# Human–computer interactions predict mental health

**MAILA — a MAchine learning framework for Inferring Latent mental states from digital Activity**

**Veith Weilnhammer**[1,2,*], **Jefferson Ortega**[3], **David Whitney**[1,3,4]

[1] Helen Wills Neuroscience Institute, University of California Berkeley, USA
[2] Max Planck UCL Centre for Computational Psychiatry and Ageing Research, London, UK
[3] Department of Psychology, University of California Berkeley, USA
[4] Vision Science Group, University of California Berkeley, USA
[*] Corresponding author. Email: veith.weilnhammer@gmail.com

Scalable assessments of mental illness remain a critical roadblock toward accessible and equitable care. Here, we show that everyday human–computer interactions encode high-dimensional information about self-reported psychological distress and wellbeing. We introduce MAILA, a MAchine-learning framework for Inferring Latent mental states from digital Activity. We trained MAILA on 18,200 cursor and touchscreen recordings labeled with 1.3 million mental-health self-reports collected from 9,500 participants. MAILA predicts dynamic mental states along 13 dimensions of distress and wellbeing, detects within-person changes over time, and resolves experimentally induced fluctuations in experienced arousal and valence. At the group level, MAILA recovers demographic and time-of-day patterns in self-reported mental health with high fidelity. MAILA also captures information only partially reflected in verbal self-report and, in a synthetic proof of concept, improves the ability of a frontier large language model to infer user mental health. By extracting signatures of psychological function that have so far remained untapped, MAILA provides the first large-scale, systematic proof of principle that human–computer interactions encode multidimensional and transferable information about mental health.

# Introduction

Mental illness is the leading cause of disability worldwide[1,2]. Despite their impact, symptoms often go undetected for years[3,4]. Delayed access to care increases the risk of poor outcomes[5].

Language, the medium through which mental health is commonly expressed and understood, is not sufficient to close the gap between symptom onset and access to care. Mental illness can make it difficult to recognize and articulate the experiences that give rise to distress[6]. Feelings of shame, stigma, and language barriers may prevent people from reaching out[7,8]. In support systems with limited resources, shared moments of communication are often difficult to achieve[9]. While fluent in conversation and, to some extent, reflective of human cognition[10], large language models still lack the contextual understanding and interpretability required for responsible deployment[11–13].

Efforts to develop more accessible and efficient mental health care are therefore expanding from language-based assessments, such as interviews and questionnaires, to non-verbal markers, including polygenic risk scores[14–16], neuroimaging[17–19], wearable sensors[20–24], sleep monitoring[25], cognitive tasks[26], and digital behaviors[27–29]. Human–computer interactions like cursor and touchscreen activity are of particular interest, because they are generated by virtually every consumer-grade device, recorded continuously at zero cost, and independent of language, introspection, and social expectations[30]. Establishing a systematic mind-body connection in these digital behaviors will allow mental states, and their changes, to be decoded every time a person uses a computer, tablet, or smartphone[28–40].

The idea that mental states are expressed in movement is supported by centuries of research on facial expression, posture, gait, and gestures[41–43]. According to motor-control theory, actions rely on internal models that are continuously shaped by ongoing affective and cognitive processes[44,45]. Human–computer interactions, like other forms of motor behavior[41–45], are therefore expected to encode signatures of mental states, including those central to mental health.

So far, however, the extent to which human–computer interactions reflect mental states remains an open question. Previous attempts have been limited by small, homogeneous samples that restrict statistical power and external validity[46]. Many have focused on narrowly defined features that may overlook the high-dimensional nature of human–computer interactions[47,48]. In addition, prior work has mostly targeted binary diagnostic traits rather than the dynamic and continuous fluctuations in mental health that matter most in psychology, medicine, and neuroscience[49,50].

Here, we introduce MAILA, a machine learning framework for inferring latent mental states from digital activity, and the MAILA dataset, a large-scale collection of human–computer interactions annotated with self-reports about mental health. Our results demonstrate that cursor movements and touchscreen activity, two ubiquitous components of human–computer interactions, carry information about the mental state of the person behind the screen. MAILA provides a large-scale, systematic proof of principle that psychological function can be decoded from everyday digital behavior, pointing to more cost-efficient and scalable ways to measure mental health in digital environments[51].

# Results

We recorded cursor and touchscreen activity during a variety of digital activities and at multiple times in 9,500 unique participants who answered 67 questions about their current psychological distress and wellbeing. We projected each participant into two spaces, one defined by patterns of human–computer interaction, and one defined by self-reported mental health, and trained MAILA to map from one to the other (Figure 1).

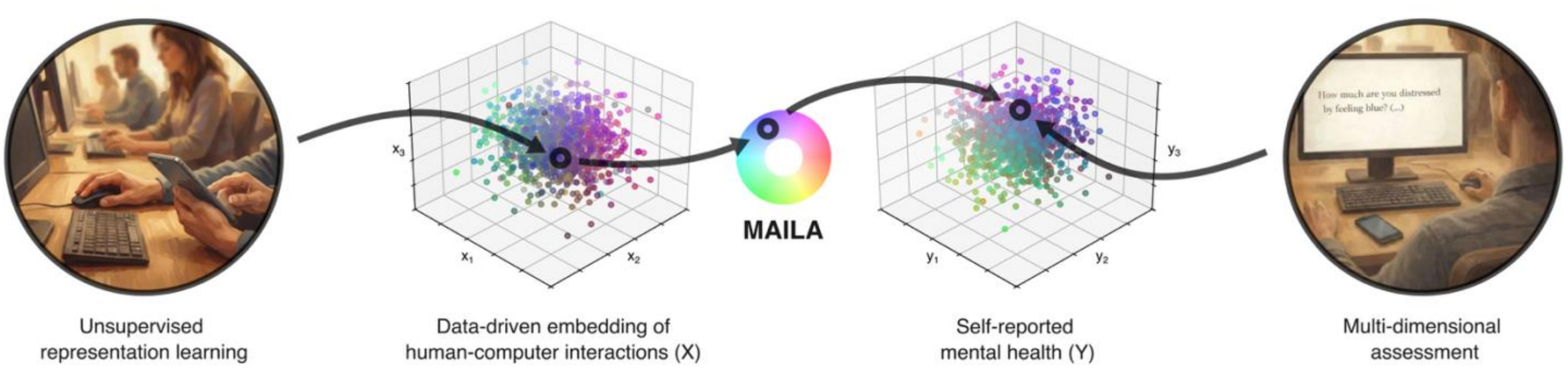

***Figure 1. Decoding mental states from digital behavior.*** *MAILA predicts mental states from cursor and touchscreen activity, two integral components of everyday interactions with computers and handheld devices. Each participant (N = 9,500 total) is represented in two spaces: (i), the space of digital behavior $X^{N \times C}$, where $C$ denotes features of human–computer interaction identified by unsupervised representation learning (illustrated for three features $x_1$–$x_3$, left), and (ii), the space of mental health $Y^{N \times Q}$, where $Q$ denotes self-reported dimensions describing mental states (illustrated here with three example items $y_1$–$y_3$, right). By training support vector regression to map from $X$ to $Y$, MAILA predicts each participant's mental-health profile from their movement fingerprint.*

## The space of human–computer interaction

We tracked cursor movements in 4,000 participants from the general population who completed a web interface designed to mimic everyday computer use. 2,000 of the 4,000 baseline participants repeated the assessment at a later time. Among the follow-up participants, 600 completed an additional non-mental-health survey, and another 600 played an interactive decision-making game. Separately, we recorded touchscreen activity in 5,500 participants who completed a creative drawing task and a mobile version of the web interface. Among these, 1,000 self-identified as living with depression, and 500 reported living with obsessive–compulsive disorder (OCD). A final validation sample of 500 participants completed the GAD-7[52] and PHQ-9[53] in their standard format in addition to the touch-based MAILA tasks.

MAILA uses unsupervised representation learning to encode each participant's cursor or touchscreen activity as a distribution over a data-driven library of movement motifs. We segmented each recording, containing on average 24,642.35 ± 462.35 screen-normalized coordinates, into partially overlapping windows of 100 consecutive samples. A long short-term memory autoencoder, pretrained on naturalistic human–computer interactions[54], transforms each segment into a 50-dimensional movement embedding. We pooled the embeddings across all participants N and grouped them into C = 500 K-means clusters, each representing a distinct, recurring pattern of human–computer interaction. MAILA then computes, for individual participants, the proportion of segments assigned to each cluster (Figures S2–S5).

This process transforms the recorded cursor or touchscreen activity into a $\mathrm{X}^{\mathrm{N} \times \mathrm{C}}$ feature matrix. Each row in X defines a participant's location in the space of digital behavior, that is, a C-dimensional distribution over recurring patterns of human–computer interaction.

## The space of mental health

We annotated all recorded human–computer interactions with self-reports of psychological distress and wellbeing as two distinct but related domains of mental health[55]. To this end, we devised a novel, continuous-response interface aligned with the Brief Symptom Inventory (BSI[56]) and the Mental Health Continuum–Short Form (MHC-SF[57]). We assessed distress on 53 items across 10 subscales, including depression, anxiety, phobic anxiety, somatization, obsession-compulsion, interpersonal sensitivity, psychoticism, paranoid ideation, hostility, and clinically relevant features. We quantified wellbeing on 14 items across 3 subscales, covering emotional, social, and psychological experiences (Table S1; Figure S6).

We organized the continuous self-reports in a $Y^{N \times Q}$ mental health matrix, where Q represents individual items, dimension-specific scores, and global questionnaire scores, scaled from 0 to 1. Each row in this matrix defines a location in the space of psychological distress and wellbeing, that is, a Q-dimensional description of the experiences that define mental health.

Our ground truth assessment achieved an internal consistency of 0.91 (Cronbach's $\alpha$), with mental health profiles that spanned the full continuum from distress to wellbeing at an average inter-quartile range of 0.49 ± 0.01. Test-retest correlations reached 0.86 for follow-up intervals shorter than one week, and declined to 0.69 after eight weeks, indicating reliable measurements with sensitivity to mental health changes that accumulated over time. Standard-format PHQ-9 and GAD-7 scores correlated with the corresponding MAILA dimensions of depression and anxiety at $R = 0.48$ ($p < 0.001$) and $R = 0.51$ ($p < 0.001$), respectively, providing external convergent validity with established self-report instruments (see Methods).

## Linking human–computer interactions and mental health

Together, $X^{N \times C}$ and $Y^{N \times Q}$ form a paired representation of digital behavior and mental health. We trained support vector regression machines to predict self-reported mental states from patterns of human–computer interaction and evaluated performance using participant-disjoint 5-fold cross-validation, within-participant transfer across time and behavioral contexts, and frozen transfer to independent datasets. Cursor and touchscreen results were obtained using modality-specific models.

## Human–computer interactions predict mental health

Experimental markers and predictive models of mental health, such as polygenic risk scores[14–16], neuroimaging[17–19], wearable sensors[20–24], sleep monitoring[25], cognitive tasks[26], and digital behaviors[27–29], have yielded correlations below $R = 0.2$ with inter-individual differences in psychological function[19,26,28] and an area under the curve (AUC) between 0.55 and 0.75 when classifying diagnoses such as depression[18,20,24,27], anxiety[22,23], or schizophrenia[14,15,21].

On held-out human–computer interaction data, MAILA yielded effect sizes within the ranges reported across these psychiatric prediction studies using signals that can be acquired at zero marginal cost from consumer-grade devices (Figure 2). Based on cursor movement alone, MAILA predicted overall levels of distress ($R = 0.26$, $p < 0.001$, relative to 1,000 randomly permuted baselines, Spearman correlation) and wellbeing ($R = 0.18$, $p < 0.001$) in held-out participants. Predictions extended to inter-individual differences on all 13 subdimensions of mental health, including depression, anxiety, phobic anxiety, somatization, obsession-compulsion, interpersonal sensitivity, psychoticism, paranoid ideation, hostility, clinically relevant features, and emotional, social, and psychological wellbeing ($R = 0.2 \pm 0.02$, $p < 0.001$, across dimensions, 4,000 participants; Figure 2A). The model achieved an equivalent level of performance when evaluated on touch-based interactions with phones and tablets ($R = 0.16 \pm 0.03$, $p < 0.001$, 3,500 participants).

MAILA's data-driven representations of digital behavior distinguished individuals who self-identified as living with depression (AUC = 0.64, 1,000 participants) and obsessive–compulsive disorder (AUC = 0.7, 500 participants) from those who did not endorse these labels[16] (3,500 participants; Figure 2B). For comparison, classifiers trained directly on all available self-reports in the MAILA dataset achieved an AUC = 0.75 for self-identified depression and 0.76 for obsessive–compulsive disorder (Figure 2B). MAILA remained sensitive to inter-individual differences even within these self-identifying subgroups (R = 0.07 ± 0.01; Figure S7).

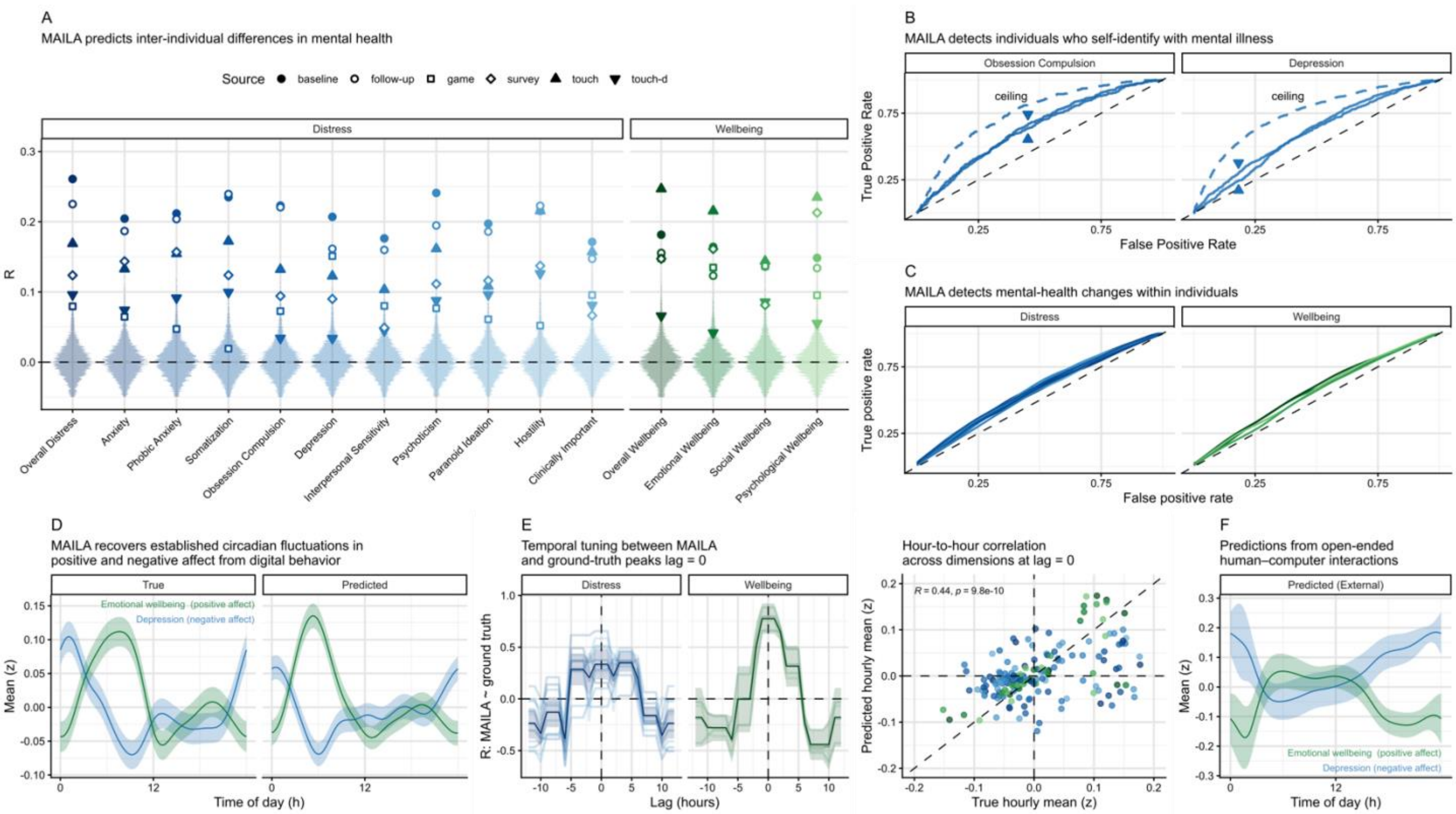

***Figure 2. Human–computer interactions predict mental health. A.*** *Correlations (R) between MAILA predictions and self-reports across 13 mental-health dimensions. Violin plots show null distributions from 1,000 label permutations; filled markers indicate observed correlations from 5-fold cross-validation (general population data; baseline: N = 4,000; touch interface: N = 3,500; touch drawing: N = 3,500); open markers indicate estimates from later or alternate-task recordings (follow-up, N = 2,000; survey, N = 600; game, N = 600).* ***B.*** *ROC curves for distinguishing participants who self-identified as living with obsessive–compulsive disorder (N = 500, left) or depression (N = 1,000, right) from those who did not endorse these labels (general population, N = 3,500). Upward and downward triangles indicate performance based on touch-based interface interactions and drawings, respectively. The dashed line shows the corresponding self-report–based classifier (using all 67 questionnaire items) as an empirical upper bound.* ***C.*** *ROC curves for within-person change detection from follow-up, survey, and game recordings using out-of-fold baseline estimates.* ***D.*** *Ground truth and predictions for depression (negative affect; blue) and emotional wellbeing (positive affect; green) across the day in the MAILA dataset (shaded: 95% confidence intervals).* ***E.*** *Lagged correlations between predicted and true diurnal profiles across all dimensions (left; thin lines: individual dimensions; thick lines: averages) peaked at 0 lag, indicating that MAILA captures shared circadian structure in mental state. Predicted versus true group means across time-of-day bins (right).* ***F.*** *Applying frozen MAILA models to open-ended cursor activity from an external dataset produced corresponding diurnal fluctuations across 19 individuals with repeated sessions at varying times of day.*

## Human–computer interactions detect changes in mental health

Detecting changes in mental health is central to early intervention and personalized care[28,58–60]. Most existing psychiatric biomarkers, however, are used as static predictors of risk or diagnostic status[14–16,18]. To address this gap, we calibrated MAILA on cursor movements from 4,000 participants recruited at baseline and applied the frozen model, without any retraining or adaptation, to 2,000 participants who repeated the experiment 5 to 76 days later. Between baseline and follow-up, participants reported median mental-health changes of 19.49% (inter-quartile range: 10.33%) relative to the full response range.

In contrast to the analyses above, which assessed psychological distress and wellbeing at a single point in time, we asked whether baseline-calibrated MAILA estimates could capture within-person changes in mental state. Using unseen follow-up cursor recordings as the only new input and baseline estimates generated out of fold, MAILA detected whether mental health improved versus worsened at an average AUC of 0.58 ± 0.01 (all 13 dimensions exceeded their permutation baselines; Figure 2C). Predicted and self-reported changes correlated at R = 0.18 ± 0.02 after controlling for baseline self-reports (partial correlation).

Models trained at baseline also predicted inter-individual differences in unseen follow-up recordings from the same participants (R = 0.18 ± 0.02, $p < 0.001$; Figure 2A), and errors did not increase with the interval between recordings ($p = 1$).

Self-reports in the MAILA dataset, grouped by the local time at which participants started the task, followed established systematic fluctuations in mental health: participants reported higher emotional wellbeing in the early morning and increasing depression-related distress as the day progressed, consistent with a morning peak in positive affect and a gradual rise in negative affect toward nightfall[61]. MAILA's predictions closely tracked time-of-day effects across all mental-health dimensions (R = 0.47, $p < 0.001$; Figure 2D). Temporal tuning curves, obtained by shifting predicted and true diurnal signals, peaked at zero lag, indicating that cursor movements capture shared circadian dynamics in mental state (Figure 2E).

To test whether naturalistic, open-ended human–computer interactions encode similar time-of-day effects, we applied frozen MAILA models to cursor movements from a public dataset of 19 users, who jointly contributed 2,550 hours and 160,000 sessions spanning web browsing, file management, office applications, coding, and entertainment[54]. Strikingly, using only these everyday digital behaviors as input, MAILA produced within-user diurnal prediction patterns that corresponded to the hour-by-hour self-report profiles observed in the MAILA cohort: higher emotional wellbeing in the morning (correlation with the hour-by-hour average self-reports in the MAILA dataset: R = 0.4, $p < 0.001$) and increasing depression-related distress toward nightfall (R = 0.42, $p < 0.001$; Figure 2F). In this external dataset, predictions evolved smoothly across successive sessions, consistent with slow natural fluctuations in mental state[62], while preserving the expected covariance structure across dimensions (e.g., higher distress co-occurring with lower wellbeing[63]; Figure S8). Because this dataset contained no concurrent mental-health reports, these analyses test temporal and structural generalization rather than person-level accuracy.

## MAILA generalizes across contexts

In everyday digital environments, users typically navigate between central content and peripheral controls along horizontal, vertical, and diagonal paths[54]. We trained MAILA on cursor and touchscreen activity recorded while participants engaged with a custom web interface that replicated such naturalistic human–computer interactions by placing self-report elements at random central locations and navigation elements at fixed corner positions (Figures S1, S9-S10).

We designed the interface so that cursor and touchscreen trajectories carried no trivial spatial information about the content of participants' responses. In the baseline and follow-up experiments, participants answered items such as "How much are you distressed by feeling fearful?" or "To what extent do you feel happy?" on a continuous scale from "Not at all" to "Very much". Questions appeared in random order, and responses were given by moving a cursor or dragging a dot onto a response line whose start and endpoint were independently randomized on every trial. MAILA received the full cursor or touchscreen trajectory as input; the model was never told which pixels corresponded to the chosen response, which item was on screen at any moment, or in what order items were answered. Control analyses confirmed that raw pointer coordinates alone could not decode self-reports above chance (Figure S11).

Having established MAILA's sensitivity to dynamic mental states within the training interface, we next evaluated the model's robustness across a diverse set of human–computer interactions labeled with mental-health self-reports. To this end, we trained MAILA on cursor movements recorded at baseline and follow-up, then applied the frozen models, without retraining or adaptation, to two non-overlapping subgroups of the follow-up cohort performing two additional tasks unrelated to mental health: 600 participants completed a non-psychological survey, and another 600 participants played an interactive decision-making game[64] (Table S2; Figures S1–S2). The survey and game analyses therefore test within-participant transfer to held-out behavioral contexts.

Responses in both tasks carried no above-chance information about the participants' mental health (Figure S12). Human–computer interactions recorded in these contexts were not predictive of the non-psychological survey ($R = 0.01 \pm 0.03$, $p = 0.28$) or gameplay behavior ($R = 0.02 \pm 0.02$, $p = 0.089$, Table S2). Any ability to decode psychological distress or wellbeing must therefore arise from how people moved the cursor in a new behavioral and cognitive context, rather than from how they responded to specific survey questions or game events.

Despite the shift in context, MAILA's decoding errors remained within the baseline distribution, increasing by 3.66 ± 1.61% when frozen models were applied to human–computer interactions recorded during non-mental-health survey completion and decreasing by 0.43 ± 3.25% for models applied to gameplay behavior (Figure S12). MAILA remained sensitive to inter-individual differences (survey: $R = 0.12 \pm 0.03$, $p < 0.001$; game: $R = 0.08 \pm 0.02$, $p < 0.001$; Figure 2A) and changes in mental health with only the out-of-context data as input (survey: AUC = 0.58 ± 0.0095; game: AUC = 0.55 ± 0.01; Figure 2C). At the time of the survey and game experiments, participants also repeated the mental-health task. MAILA estimates derived from held-out mental-health-task interactions correlated with estimates derived from held-out survey and game interactions in the same participants (survey: $R = 0.22 \pm 0.01$, $p < 0.001$; game: $R = 0.11 \pm 0.02$, $p < 0.001$).

To test whether MAILA can generalize beyond rigid user interfaces, we asked all touchscreen participants to complete a series of prompted drawings on their phones or tablets before starting the questionnaire. Each prompt, for example, "Draw a spaceship" or "Draw the digits 036", specified only what to draw, but not how (see Table S3 for all prompts and Figure S13 for example drawings). When trained and tested on these free-form, creative digital behaviors, MAILA predicted overall distress ($R = 0.1$, $p < 0.001$), wellbeing ($R = 0.07$, $p = 1 \times 10^{-3}$), and their subdimensions ($R = 0.07 \pm 0.02$, $p < 0.001$).

Despite relying on entirely different interaction modes, independent models trained on the structured touchscreen interface and the drawing behavior converged on correlated predictions for the same held-out participants ($R = 0.05 \pm 0.02$, $p < 0.001$). Errors decreased by 4.31 ± 0.18% ($p < 0.001$) when predictions from the two touchscreen recordings were combined, indicating that repeated measurements across contexts improve MAILA's accuracy[65].

Across all recorded human–computer interactions, predictions derived from random splits of each participant's data correlated at $R = 0.61 \pm 0.05$, demonstrating a level of reliability that exceeds many experimental markers of psychological function[26] (Figure S14).

Together, these results show within-participant transfer of mental-health-related movement signals across tasks and cognitive contexts, alongside convergent predictions across structured and free-form touchscreen interactions. Although predictive performance decreased under a shift in behavioral context, MAILA remained sensitive to inter-individual differences across all tested tasks and interfaces, and to within-person change over time.

## MAILA predicts mental health across demographics

Bias is a major concern when applying predictive models to people who differ in age, gender, or cultural background, since unequal performance across demographic groups is known to amplify existing disparities in care[66,67]. Figures S15-S16 summarize the demographic composition of our sample. The MAILA dataset includes participants between 18 and 89 years of age, 94 nationalities and a balanced gender distribution (48.37% female, 47.09% male), representing varied ethnicities and a range of employment and student statuses (Figures S15–S16). Demographic distributions were closely aligned between all subsets of the MAILA dataset (see Figure S16 for a visualization of age, gender, employment, and ethnicity across subsets). While no dataset can fully reflect global populations[68], this diversity provides a meaningful foundation for testing whether MAILA's predictions generalize fairly. Demographic factors modulated the magnitude of prediction errors only marginally, accounting for a median of 0.22% of residual error variance (range: 0.02–1.22%; partial $\eta^2$; Figure S15).

Beyond differential error across groups, an additional concern is that MAILA may predict mental health only indirectly via confounding traits, because characteristics such as age, gender, cultural background, and the ability to engage with an online task may correlate with both human–computer interaction patterns and mental health. MAILA's sensitivity to dynamic mental states argues against a purely trait-driven account, because static characteristics cannot explain fluctuations within the same individual over time (Figure 2). Consistent with this, the correlation between MAILA's predictions and the ground truth remained significant after residualizing both measures with respect to all available demographic covariates (age, gender, ethnicity, country of birth/residence, nationality, language, student status, and employment status) and two proxies of digital literacy (task completion time and the number of prior successfully completed tasks unrelated to MAILA). The adjustment reduced predictive performance by 3.28 ± 0.6% to $R = 0.1 \pm 0.01$ while ground-truth correlations remained significant (across dimensions and datasets, $T(89) = 17.67$, $p < 0.001$; Figure S17).

Prediction errors were likewise unrelated to identification accuracy within the baseline session ($R = 0.02$, $p = 0.21$) or across sessions at follow-up ($R = -0.03$, $p = 0.17$; Results S1), providing no evidence that participants who were easier to identify received more accurate mental-health predictions.

## MAILA tracks group-level mental health

Human–computer interactions can be collected continuously at zero marginal cost, enabling population estimates at arbitrary scale. In our data, individual-level prediction errors ranged from 17.17 to 24.86%. In simulations calibrated to MAILA's empirical error, group-level estimates approach ceiling once sample sizes exceed ~100, suggesting that MAILA may be especially well suited for population-level mental-health monitoring (Figure S18).

From human–computer interactions alone, and without access to any demographic or temporal information, MAILA recovered established demographic and environmental effects on mental health at high fidelity. This included the effects of age, with older adults reporting lower distress and higher wellbeing[69] ($R = 0.97$, $p < 0.001$; Figure 3A); of employment, with unemployed and part-time employed individuals reporting higher distress and lower wellbeing than retired or full-time employed participants[70] ($R = 0.67$, $p < 0.001$; Figure 3B); and of gender, with participants identifying as female reporting higher distress than those identifying as male[71] ($R = 0.67$, $p < 0.001$; Figure 3C).

Prior genome-wide association studies have shown strong genetic correlations between clinician-assigned diagnosis and self-reported mental illness[16]. Consistent with self-identification, MAILA predicted higher depression-related distress in participants who identified as living with depression ($b = 0.0077 \pm 0.002$, $p < 0.001$), and higher obsessive–compulsive distress in those who reported living with OCD (b =

0.01 ± 0.0027, $p < 0.001$), with strong agreement between the ground-truth and the information encoded in human–computer interactions across all dimensions of mental health ($R = 0.79$, $p < 0.001$, Figure 3D).

Together, these findings benchmark MAILA against known population-level correlates of mental health and illustrate how routine human–computer interactions can support real-time public mental-health monitoring at scale.

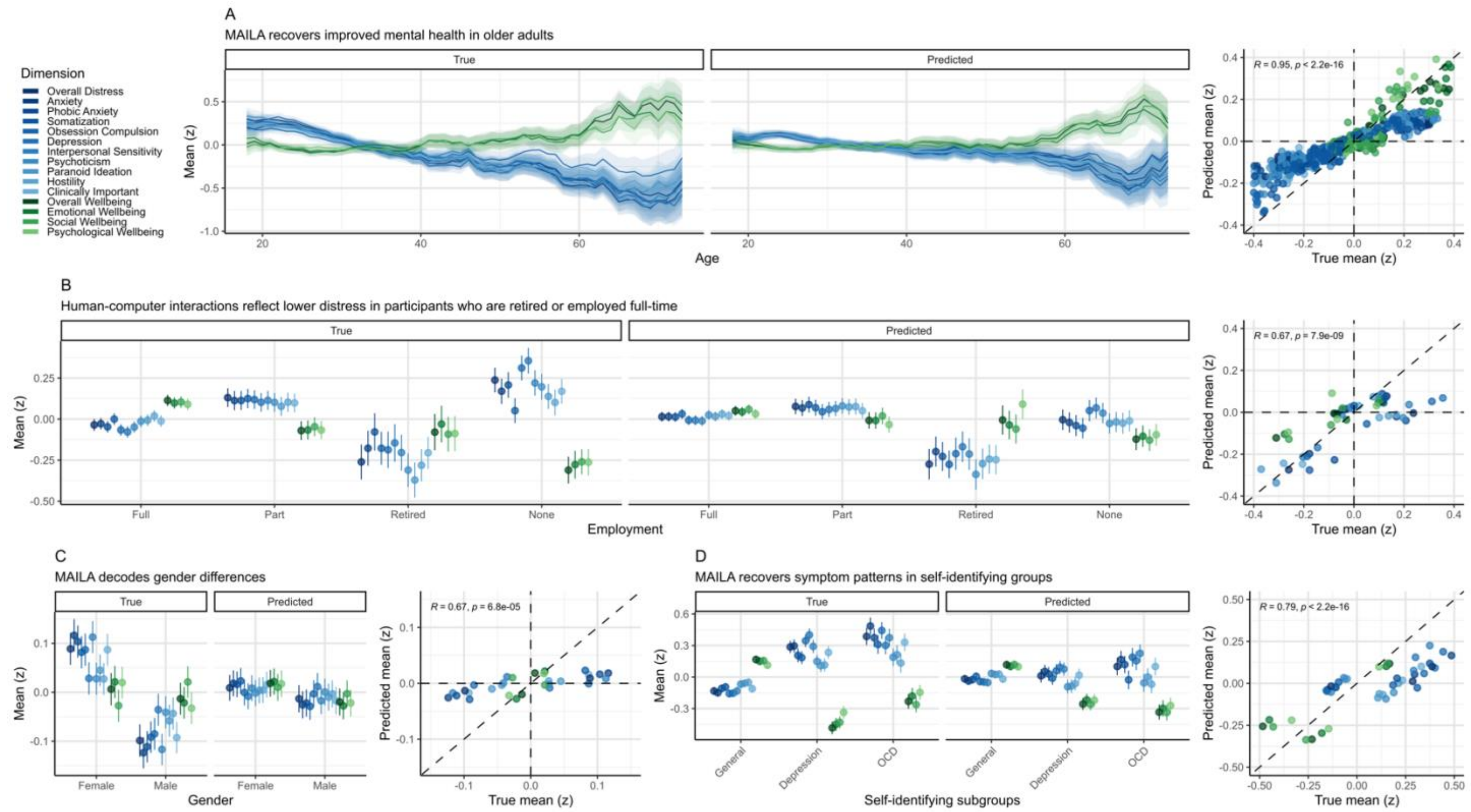

***Figure 3. MAILA tracks group-level mental health.*** *Human–computer interactions enable scalable population estimates of mental health. For each demographic factor, we compare group-level self-reported means (True) to behavior-based estimates from MAILA (Predicted). Error bars and ribbons indicate 95% confidence intervals. Scatter plots show predicted versus true group means across all 13 dimensions for age (**A**), employment (**B**), gender (**C**), and self-identified mental illness (**D**).*

## Human–computer interactions encode 3 orthogonal axes of variation in mental health

Language-based descriptions of mental health are typically interrelated[63]: low mood, for example, is frequently accompanied by social withdrawal and persistent worry. These covariations were also present in the MAILA dataset: participants who felt more distressed reported lower wellbeing (and vice versa, $R = -0.25 \pm 0.02$). Higher scores on one dimension were accompanied by higher scores on others in the same domain (correlations within distress dimensions: $R = 0.66 \pm 0.02$; wellbeing: 0.73 ± 0.09; Figure S6). Such shared variance is often taken to reflect a global factor capturing an individual's propensity toward distress[63]. A key question that follows is whether patterns of human–computer interaction encode particular thoughts and emotions that shape the content of psychological distress and wellbeing, or whether they reflect only a general tendency toward poor or good mental health.

To test whether human–computer interactions encode symptom-specific markers beyond a one-dimensional scalar of distress, we transformed the $Y^{N \times Q}$ mental health matrix into a set of orthogonal principal components (PC). The PC structure was stable across the independent cursor and touchscreen cohorts (Tucker congruence: PC1 = 0.99, PC2 = 0.98, PC3 = 0.92; three-axis Procrustes similarity = 0.97).

We trained independent SVR models to predict each PC, treating PCs as orthogonal axes of variation in self-reported mental health. Based on the recorded human–computer interactions alone, MAILA successfully predicted the location of held-out participants on the first 3 PCs of mental health, which together explained 37.91% of the variance across all datasets (Figure 4A).

PC1 reflected a general distress-to-wellbeing axis and was decoded at $R = 0.16 \pm 0.05$ ($p < 0.001$). PC2 separated depression and interpersonal sensitivity from other types of distress ($R = 0.22 \pm 0.05$, $p < 0.001$). PC3 placed somatization and hostility on one end, obsessive-compulsive symptoms and interpersonal sensitivity on the other, and anxiety, depression, psychoticism, and paranoid ideation in between ($R = 0.2 \pm 0.07$, $p < 0.001$; Figure 4B).

Using out-of-fold baseline estimates, MAILA detected the direction of within-participant change separately along PC1 (AUC = 0.61), PC2 (AUC = 0.60), and PC3 (AUC = 0.60). Follow-up MAILA estimates remained associated with self-reported change after controlling for the corresponding baseline PC score (partial correlations; PC1: $R = 0.20$; PC2: $R = 0.22$; PC3: $R = 0.20$; Figure 4C).

We found weaker mental state predictions beyond PC3 ($R = 0.01 \pm 0.0036$), suggesting that human–computer interactions capture the most dominant axes of variation in self-reported mental health (Figure S19).

Having established that PC1–PC3 organize inter-individual differences in self-reported mental health, we next asked whether the same axes also track moment-to-moment changes in subjective experience within a person[72]. To this end, we applied pretrained, frozen MAILA models beyond their supervised calibration data: to naturalistic, open-ended human–computer interactions from 19 users performing everyday computer activities[54] (Figure 4D), and to three independent emotion-tracking datasets in which a total of 298 participants used cursor movements to continuously report the arousal and valence they experienced while watching 130 emotionally evocative videos[73–75] (Figure 4E). These emotion-tracking datasets differed from MAILA in participants, stimuli, task demands, and response geometry. The entire MAILA inference pipeline was frozen before application and had never encountered these participants, trajectories, affect labels, or the continuous affect-report mapping. Arousal and valence reports were used only after predictions had been generated to group windows for comparison. The analysis therefore tests zero-shot transfer to a distinct response geometry.

With only naturalistic cursor movements as input, MAILA recovered systematic within-day trajectories in mental-health PC space. Across the day, decoded states shifted toward higher PC1 scores, consistent with increasing global distress-linked signatures, and toward lower PC2 scores, consistent with a shift away from emotional wellbeing and toward depression-linked distress (Figure 4D).

During the video tasks, MAILA also tracked moment-to-moment fluctuations in affective experience. Periods of higher experienced arousal were accompanied by higher PC1 scores (shift toward distress; $p < 0.001$) and higher PC2 scores (shift away from depression-linked distress; $p < 0.001$), as well as lower PC3 scores (shift toward somatization/hostility; $p < 0.001$; Figure 4E). More positive valence was associated with lower PC1 and higher PC2 scores ($p = 0.0091$ and $p < 0.001$, respectively), consistent with a shift toward wellbeing and away from depression-linked distress, with no additional effect on PC3 ($p = 0.88$).

Together, these results show that human–computer interactions encode not only higher-order variation in mental health across individuals, but also within-person dynamics in subjective experience. Moment-to-moment changes in arousal and valence mapped onto the same low-dimensional PC space that organized inter-individual differences in mental health, with arousal expressed most consistently along PC1 and valence along PC2.

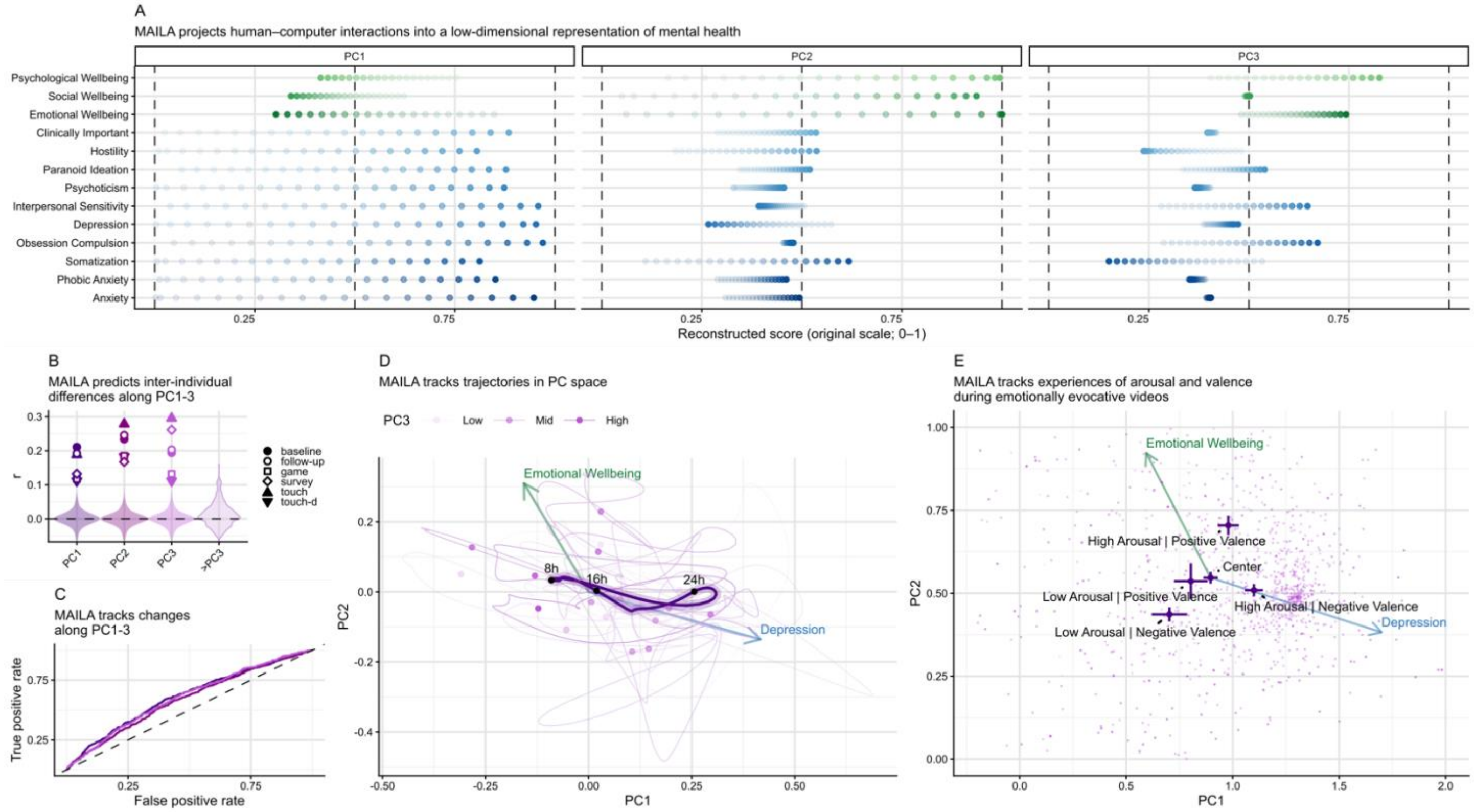

***Figure 4. Human–computer interactions predict 3 orthogonal axes of variation in mental health.*** *MAILA predicts the level (PC1) and composition (PC2–PC3) of distress and wellbeing.* ***A.*** *Reconstructed dimension scores (0–1) while sweeping each PC score from its empirical minimum (low opacity) to maximum (high opacity), holding the remaining PCs at 0.* ***B.*** *Predictive performance across PCs. Violins show null distributions; points mark observed correlation from baseline, follow-up, survey, game, and touch datasets (filled: five-fold cross-validation; unfilled: frozen-model transfer to later or alternate-task recordings).* ***C.*** *ROC curves for classifying higher versus lower PC changes at follow-up relative to out-of-fold baseline estimates.* ***D.*** *Thin lines show predicted within-user trajectories in PC space during naturalistic human–computer interactions across time of day (black dots); the thick line shows the population mean trajectory. Purple points mark each participant's centroid, with opacity indicating depth along PC3 (higher opacity = larger PC3).* ***E.*** *Participant-level PC1 and PC2 predictions are shown as points (opacity scales with absolute PC3), overlaid with group centroids (mean ± 95% CI) for the four arousal x valence states defined by within-participant median splits (high/low arousal; positive/negative valence). Centroids were computed across participants after applying frozen MAILA models to three external emotion-tracking datasets.*

## MAILA extends cognitive phenotypes of mental health

Conventional mental health assessments rely on predefined indicators of mental health, such as responses to structured interviews and questionnaire items[49]. MAILA, by contrast, learns an end-to-end mapping from human–computer interactions to mental health without specifying the underlying signals. This creates a new framework for discovering latent cognitive phenotypes that may not be exhaustively queried by self-report, yet shape everyday behavior in clinically meaningful ways[29] (Figure 5A).

We first tested whether MAILA preserved structure shared with established symptom measures beyond the questionnaire space on which it was trained (BSI and MHC-SF). Frozen MAILA models were applied to touch-based interactions from an independent validation sample of 500 participants who also completed the PHQ-9 and GAD-7. MAILA predictions remained aligned with PC1 ($R = 0.15$, $p < 0.001$), PC2 ($R = 0.19$, $p < 0.001$), and PC3 ($R = 0.21$, $p < 0.001$; averaged across touch-based interface and drawing tasks; Figure 5B).

To evaluate MAILA relative to standard symptom scales, we compared frozen MAILA predictions with questionnaire-only models based on GAD-7, PHQ-9, and PHQ-2 summary scores. GAD-7, PHQ-9, and PHQ-2 primarily measure depression and anxiety, providing a strong benchmark for the global distress axis (PC1) but a weaker benchmark for orthogonal components capturing other sources of mental-health variation. Consistent with this, MAILA reached 30.93% of the mean questionnaire-only benchmark on PC1 (0.15 versus 0.5). For PC2 (depression and interpersonal sensitivity vs. emotional wellbeing) and PC3 (somatization and hostility vs. obsession-compulsion), MAILA outperformed GAD-7, PHQ-9, and PHQ-2 by $\Delta R = 0.13$ and $\Delta R = 0.21$, respectively (Figure 5B). Adding MAILA to questionnaire-only models further improved prediction by $\Delta R = 0.02$, 0.07, and 0.14 for PC1, PC2, and PC3, respectively. These findings suggest that human–computer interactions capture mental-health information that is only partially reflected in conventional language-based measures of depression and anxiety.

We next asked whether human-computer interactions can help predict cognitive markers of mental health. Specifically, we tested whether MAILA could improve the decoding of belief instability, a latent cognitive phenotype describing how readily individuals update internal models in response to new evidence[64]. Belief instability is typically inferred from controlled behavioral experiments in which participants integrate sequential noisy observations. Rigid updating has been linked to perseveration and compulsive control, whereas overly flexible updating may manifest as volatile, inconsistent inference associated with impulsivity or disorganized behavior[64].

In the MAILA dataset, 600 participants completed a gamified beads task that measured belief updating under uncertainty. On each round, participants inferred the hidden source of a sequence of colored beads and reported their confidence after each new observation. We quantified belief instability as the entropy of confidence-weighted belief updates, operationalizing a continuum between rigid cognition (low entropy) and reactive cognition (high entropy)[76] (see Methods for details).

Our data confirmed that this reactive-to-rigid continuum mapped onto mental health[64], with positive associations with hostility and somatization, and negative associations with depression, obsession-compulsion, and interpersonal sensitivity (Figure 5C). These associations persisted when self-reports were replaced by MAILA predictions derived solely from cursor movements: MAILA recovered the group-level association between mental-health dimensions and belief instability at $R = 0.78$ ($p < 0.001$). Thus, a cognitive phenotype inferred from a controlled behavioral task aligned with mental-health signatures decoded from digital motor behavior across contexts.

Separately, within the beads-task data, MAILA's movement embeddings predicted unique variance in belief instability over and above all available self-reports (cross-validated partial correlation controlling for self-reports: $R = 0.19$, $p < 0.001$). Self-reports also carried unique, albeit weaker, information about belief instability after controlling for cursor-derived predictions ($R = 0.14$, $p < 0.001$). Combining self-reports and cursor-based features achieved the highest performance ($R = 0.28$, $p < 0.001$) and significantly reduced prediction error relative to self-reports alone ($p < 0.001$; Figure 5D).

Together, these results show that human–computer interactions encode cognitive information that is not fully captured by language-based mental-health assessments. Data-driven markers extracted from everyday digital behavior can therefore validate, refine, and extend established cognitive phenotypes of mental health, providing a step toward scalable, data-driven models of mental health.

## MAILA improves mental-health inference by a frontier AI chatbot in a synthetic proof of concept

MAILA's ability to decode mental states from non-verbal digital behavior may be particularly useful as context for frontier AI chatbots, where misunderstanding a user's latent mental state in text-based interactions can amplify risk in vulnerable interactions[13]. We therefore conducted a controlled synthetic proof of concept testing whether MAILA context improves mental-state inference from everyday assistant-facing language.

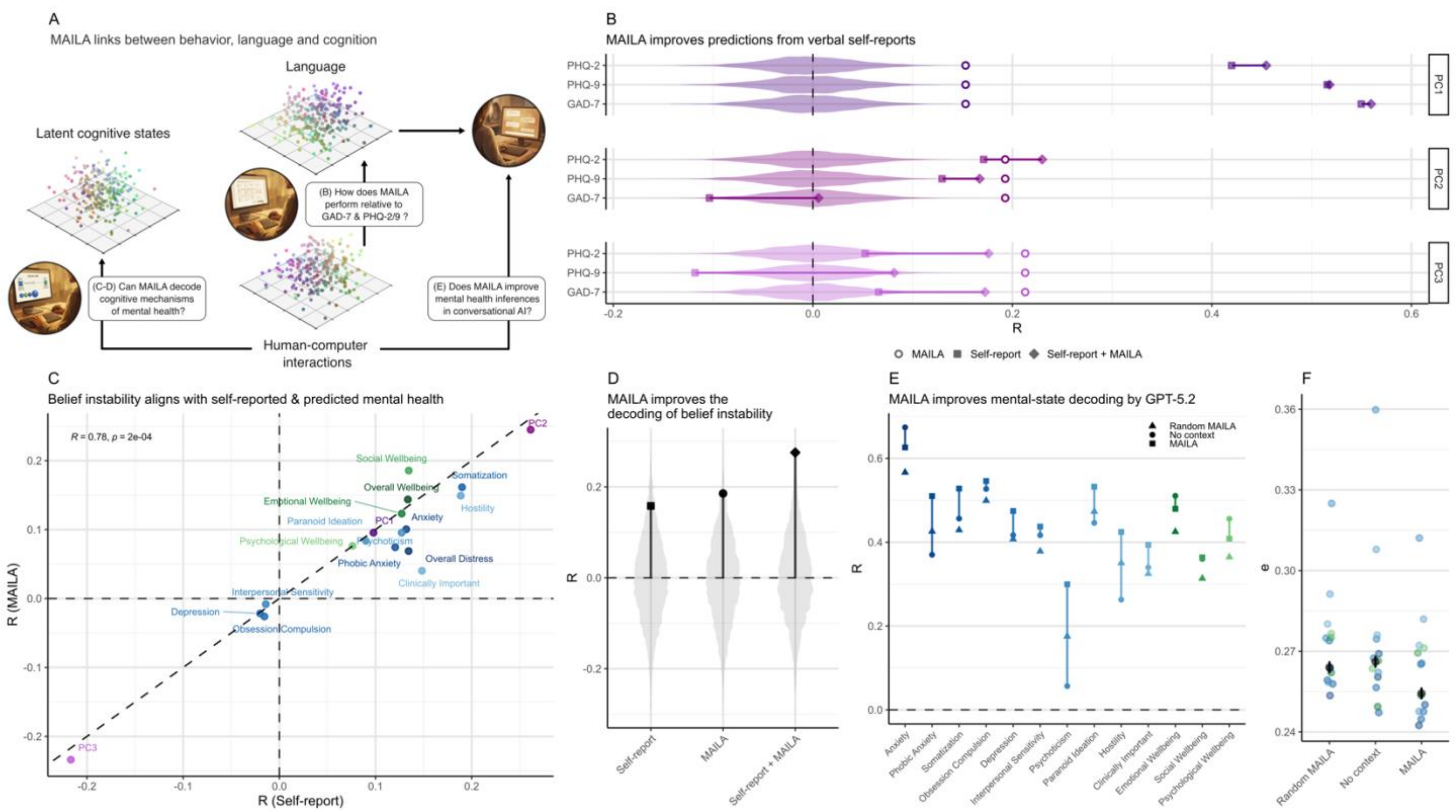

***Figure 5. MAILA extends cognitive phenotypes of mental health and improves chatbot inference in a synthetic proof of concept. A.*** *Each point is a participant shown in three interconnected spaces: human–computer interactions (behavior), self-reported mental health (language), and cognition.* ***B.*** *Performance for self-report models, MAILA, and self-report models augmented with MAILA across the first 3 mental-health principal components. Horizontal separation indicates the gain associated with adding MAILA; violins show permutation-based null distributions.* ***C.*** *Belief instability showed aligned associations with self-reported and MAILA-predicted mental-health dimensions, with positive associations with hostility and somatization, and negative associations with depression, obsession-compulsion, and interpersonal sensitivity.* ***D.*** *Prediction of belief instability from self-report measures, MAILA, and their combination. Violins indicate permutation-based null distributions; points and line ranges show observed correlations (R).* ***E.*** *Mental-state decoding by a frontier AI chatbot (GPT-5.2) from single-message assistant-facing interactions with simulated users. Points show dimension-wise R between chatbot-inferred and ground-truth mental-health profiles under three conditions: text only, text plus MAILA context, and text plus random MAILA context; vertical lines connect the text-only and MAILA conditions for each dimension.* ***F.*** *Normalized prediction error (e; root mean squared difference between inferred and ground-truth percentile ranks across the 13 dimensions, divided by the 100-point percentile range) across the three context conditions; colored points show dimension-level error, black points and line ranges show the mean and 95% CI across simulated users.*

We simulated short messages from synthetic users (user simulators, independent Claude-Sonnet-3.7 instances), each conditioned on the 13-dimensional ground-truth mental-health profile of one participant in the 4,000-person baseline cursor cohort. Each user simulator wrote a realistic message to a general AI assistant (GPT-5.2) about everyday topics such as planning, communication, decision-making, routines, social situations, work, family, household tasks, travel, health administration, finances, or situations that felt difficult to interpret. These messages were not framed as mental-health disclosures; instead, the latent profile could shape what the user wanted help with, how the situation was framed, and which concerns surfaced indirectly. GPT-5.2 inferred the user simulator's mental-health profile under three conditions: text-only (message alone), with MAILA as context (message plus out-of-fold MAILA predictions for the same participant), and with random MAILA context (message plus out-of-fold MAILA predictions sampled from a randomly selected other participant). Core prompts are summarized in Table S4; the scenario bank is listed in Table S5.

Across dimensions, profile recovery was highest when GPT-5.2 received out-of-fold MAILA cursor-based predictions as context ($R = 0.46 \pm 0.05$), intermediate from text alone ($R = 0.41 \pm 0.08$), and lowest with random MAILA context ($R = 0.4 \pm 0.05$; Figure 5E). Normalized prediction errors were lower with MAILA ($e = 0.25 \pm 0.002$) as compared to text alone ($e = 0.27 \pm 0.002$) and random MAILA context ($e = 0.26 \pm 0.002$; $F(2, 103{,}454) = 175.27$, $p < 0.001$; Figure 5F). In this controlled simulation, MAILA thus improved mental-health-profile recovery from single-message interactions with synthetic users.

## Explaining MAILA with handcrafted features

To probe which movement features underlie MAILA's predictions, we regressed its predicted scores onto a battery of handcrafted cursor and touchscreen features (Table S6). In participants from the general population, higher predicted wellbeing was characterized by higher path efficiency, whereas higher predicted distress was associated with more tortuous trajectories and greater variability in speed (Figure S20A). Relative to the general population, participants who self-identified with mental illness showed more of the reverse-engineered features associated with distress ($p < 0.001$) and less of the features linked to wellbeing ($p < 0.001$; Figure S20B). MAILA thus provides a principled way to test whether candidate handcrafted features capture meaningful inter-individual differences in mental health.

Between cursor- and touch-based interactions, we observed substantial variability in how handcrafted features related to mental states, including reversals in the direction of association for 41.11% of all features (Figure S20C). MAILA outperformed these intuitive behavioral descriptors, which struggled to distill modality-general signatures of mental states, across all benchmarks, including lower prediction errors for inter-individual mental health differences in the general population ($p < 0.001$) as well as higher accuracy in classifying individuals who self-identified as living with depression (AUC = 0.64 vs. 0.59) and obsessive–compulsive disorder (AUC = 0.7 vs. 0.6).

MAILA's accuracy decreased when we reduced the number of group-level clusters, excluded participants from the training set, limited the amount of available test data, or distorted human–computer interactions with increasing levels of random noise (Figure S21). These patterns show that behavioral diversity, realistic sample sizes, and longer recordings enhance MAILA's performance, while modest user-side scrambling can substantially reduce unwanted digital profiling.

# Discussion

Closing the gap between mental-health need and access requires measurements that are scalable, accessible, affordable, and accurate enough to guide decision-making[31,32,35]. Gold-standard tools such as clinical interviews and questionnaires provide high-quality information, but rely on limited availability, shared language, and cultural context. Experimental markers and predictive models, including polygenic risk scores[14–16], neuroimaging[17–19], wearable sensors[20–24], sleep monitoring[25], cognitive tasks[26], and digital behaviors[27–29] often encode persistent traits rather than dynamic mental states, depend on active participation, time-consuming protocols, or expensive equipment, generate data that are difficult to store and anonymize, or capture behavior in contexts far removed from everyday life. Current methods therefore occupy a narrow Pareto frontier on which no single tool achieves sufficient accuracy, cost-efficiency, and ecological validity at once[51].

MAILA provides a large-scale, systematic proof of principle that everyday human–computer interactions carry multidimensional information about psychological distress and wellbeing. Its predictions are continuous and dynamic, can be generated from both cursor and touchscreen interactions, transfer across tasks and time, and improve when multiple observations are combined. Because cursor and touchscreen activity can be acquired at near-zero marginal cost from billions of consumer devices, MAILA offers a scalable framework for studying mental health in digital environments. Aggregated across users, its predictions recover group-level patterns with high fidelity, motivating future work on population-level measurement and public-health applications. Compared with handcrafted movement features, MAILA's learned representation may be less directly interpretable, but it reduces reliance on task- and device-specific feature engineering and may therefore support transfer across interaction contexts.

Unlike many areas of medicine where diagnostic categories are anchored to well-defined causal markers, mental-health classifications remain largely symptom-based. Existing diagnostic categories are therefore heavily shaped by language, history, and culture, grouping together individuals with diverse symptoms and treatment responses[16,49,50]. As a data-driven behavioral assessment, MAILA may be sensitive to mental states that traditional language-based assessments obscure[29,77,78]. If validated in clinical and underserved populations, this approach could eventually complement language-based assessments for individuals who struggle to recognize, articulate, or report their experiences, including non-verbal individuals and those facing language barriers[7,8].

MAILA may also support more personalized and context-aware conversational AI by giving language models access to user-state information that is not fully expressed in language. Because human–computer interactions are less susceptible than self-report to deliberate distortion, impression management, and social desirability, discrepancies between movement- and language-based estimates could provide a target for future research on multimodal assessment. Data-driven predictions from human–computer interactions also provide convergent validation for computational phenotyping[28,29,77]. Because of its scalability, MAILA can help to move the science of mental health into the data-rich regime in which deep learning most effectively advances large-scale predictive models of human cognition, emotion, and behavior[79].

## Limitations

The MAILA dataset is large and demographically varied, but all participants were recruited online, limiting claims about non-English-speaking, low-resource, digitally underserved, neurodivergent, and clinical populations. MAILA was calibrated to self-reported psychological distress and wellbeing, not to structured diagnoses, clinician-rated outcomes, treatment response, or demonstrated clinical utility in care pathways. The present study evaluates statistical inference from digital behavior, not clinical deployment, and therefore does not establish diagnostic validity, safety, effectiveness, or utility in care pathways.

At the same time, MAILA's predictive signal was observed in both cursor and touchscreen interactions and generalized across tasks, time, demographic groups, and open-ended digital activities, providing a robust proof of principle that human–computer interactions carry transferable mental-health information. A central next step is prospective validation in clinical cohorts, ideally alongside ongoing treatment and in randomized trials testing whether MAILA improves decision-making and outcomes in mental health.

MAILA's central strength is also its central risk: it provides sensitive information about an individual's mental state from signals that can be obtained on any digital device[51,80,81]. Passive mental-health screening, even when well-intentioned, can produce unintended consequences. For example, individuals flagged by automated tools may experience anxiety, stigma, or confusion if results are presented outside of an ecosystem that bridges the last mile toward mental-health support[31,32,82]. There is also a danger that predictive mental-health technologies will be implemented in contexts that prioritize institutional or economic interests over individual wellbeing[51]. Without strong safeguards, MAILA could be misused in hiring decisions, insurance risk assessments, or unwanted profiling in sectors such as education, immigration, or law enforcement[80]. Our own results show that human–computer interactions can in principle be used to detect inconsistent self-reports and to partially re-identify users across sessions (Results S1). When used in any of these contexts, mental-health predictions may exacerbate discrimination and reinforce existing inequalities.

Preventing such harms requires transparent disclosure, opt-in participation, user control, privacy-preserving design, anti-misuse safeguards, continuous fairness validation, and strong normative and regulatory frameworks that limit use to beneficial contexts. To guide responsible development and deployment, predictive models should be evaluated against established standards for trustworthy AI. Table S7 summarizes MAILA's alignment with the FUTURE-AI framework, a system for assessing the fairness, universality, traceability, usability, robustness, and explainability of AI in healthcare[83]. By adopting these recommendations, we aim to advance MAILA in a way that is transparent, inclusive, and ultimately beneficial to those in need of mental health support.

# Acknowledgements

We thank Shi Chen, Maria Crespo-Ribadeneyra, Raymond Dolan, Sophocles Goulis, Janis Karan Hesse, Jochen Michely, Matthew Nour, Philipp Sterzer, Dominik Thalmeier, and Pierre Vassiliadis for their valuable feedback and support. V.W. was supported by the Leopoldina, German National Academy of Sciences, and the Brain and Behavior Research Foundation (BBRF; Young Investigator Award).

# Author Contributions

V.W. conceived the study and designed the experiments. V.W. developed the computational framework and performed the analyses. D.W. provided conceptual guidance and supervision. All authors contributed to interpretation of the results and to writing the manuscript.

# Competing Interests Statement

This work is related to a pending patent application filed by the University of California, Berkeley. The authors declare no other competing interests.

# Data Availability Statement

The group-level data supporting the main figures of this study are publicly available in the MAILA repository (https://github.com/veithweilnhammer/maila_sdk), in the data/ subfolder. The release contains de-identified, aggregated results for the main results, a manifest describing each file, and a reproduction tool (reproduce.py) that recomputes the reported results directly from the data.

# Code Availability Statement

Code supporting this study is publicly available in the MAILA repository (https://github.com/veithweilnhammer/maila_sdk). This includes the MAILA Sandbox — a browser-side software development kit (maila-recorder.js), an end-to-end demo interface (example.html), and example recordings — for recording human–computer interactions and requesting model predictions from the MAILA backend, as well as the data-reproduction utilities (reproduce.py) used to regenerate the main-figure results from the released data. The manuscript was created using the R Quarto framework, which integrates the data, analysis code, and text within a single document.

# Supplementary Information for *Human–computer interactions predict mental health*

**Veith Weilnhammer**[1,2,*], **Jefferson Ortega**[3], **David Whitney**[1,3,4]

[1] Helen Wills Neuroscience Institute, University of California Berkeley, USA
[2] Max Planck UCL Centre for Computational Psychiatry and Ageing Research, London, UK
[3] Department of Psychology, University of California Berkeley, USA
[4] Vision Science Group, University of California Berkeley, USA
[*] Corresponding author. Email: veith.weilnhammer@gmail.com

**This PDF file includes:**

# Materials and Methods

## The MAILA Dataset

The MAILA dataset is a large-scale dataset linking mental-health self-reports with digital behavior that can be acquired at zero marginal cost. It comprises approximately 18,200 recordings of cursor and touchscreen activity collected between August 2024 and July 2025 (Figure S1), except for an independent validation sample collected in March and April 2026. Each behavioral recording was paired with the participant's corresponding 67-item mental-health profile, yielding approximately 1.3 million recording-level item labels.

**Participants:** We recruited participants through the online research platform Prolific® (www.prolific.com). All participants provided informed consent prior to participation. The study was approved by the Institutional Review Board of the University of California, Berkeley (IRB2011-01-2701), and conducted in accordance with the Declaration of Helsinki. We pre-screened participants for English proficiency and for willingness to answer mental-health questions, including sensitive topics such as self-harm or suicidality. The final analytic sample comprised 9,500 participants, including 4,000 desktop/laptop users for cursor-based recordings and 5,500 smartphone/tablet users for touch-based recordings. During data collection, submissions were rejected and replaced to reach these target sample sizes if participants (i) left more than 10 mental-health items unanswered, (ii) omitted required items needed to compute at least one dimension score, or (iii) completed the study on the wrong device type (e.g., using a desktop/laptop for the smartphone/tablet experiment, or vice versa). All accepted submissions were entered into the analysis pipeline, with no further exclusion criteria. Participants were compensated at an average rate of $10 per hour.

Figure S1 outlines the structure of the MAILA dataset. The 4000 cursor-based participants came from the general population (no filters applied except the hardware filter). For test–retest assessment, all 4,000 cursor-based participants were re-invited via Prolific®, and the first 2,000 participants who completed the follow-up session were included in the longitudinal subset (i.e., follow-up participation was voluntary and determined by response to re-invitation). The follow-up interval (delay between baseline and retest completion) ranged from 5 to 76 days.

Among the follow-up participants, we recruited two mutually exclusive additional subsets of 600 participants each: one subset completed an additional non-mental-health survey, and a separate, non-overlapping subset completed an interactive decision-making game. Assignment to these two subsets was randomized at the time of re-invitation (i.e., eligible follow-up participants were randomly allocated to one of the two tasks until each subset reached N = 600). The 5,500 participants in the touch-based datasets were recruited once. This touchscreen sample comprised 3,500 participants recruited once from the general population (no Prolific prescreening filters beyond the device/hardware filter), plus two mutually exclusive pre-screened groups: 1,000 participants who self-identified as living with depression and 500 participants who self-identified as living with obsessive–compulsive disorder. A final validation sample of 500 general population participants completed the GAD-7 and PHQ-9 in addition to the touch-based MAILA experiments.

Figures S15-S16 summarize the demographic composition of our sample. The MAILA dataset includes participants between 18 and 89 years of age, 94 nationalities, ethnicities, 6 types of employment, and a balanced gender distribution (48.37% female, 47.09% male). Demographic distributions were closely aligned between all subsets of the MAILA dataset (see Figure S16 for a visualization of age, gender, employment, and ethnicity across subsets).

**Recordings:** The recordings had a median duration of 13.22 minutes. Because the experiment was conducted on the participants' own devices, the interface's sampling rate was determined by the participant's hardware and browser implementation. The median sampling rate was 60Hz for cursor and 60.6Hz for touchscreen activity, with 74.45% of data falling within a ±5 Hz window around the median. To preserve ecological validity, we did not re-sample or exclude any movement data, requiring all downstream analyses to account for natural variation in hardware at the user end. This design supports the generalizability of the framework to uncontrolled, real-world settings.

**Randomized response interface:** We developed a custom web-based questionnaire interface in JavaScript that allowed us to collect self-reports while eliciting cursor movements and touchscreen gestures characteristic of everyday digital activity (Figures S1). In the interface, a randomized response mapping dissociated the observed human–computer interactions from the semantic content of the responses provided by participants. We used the interface to collect mental health self-reports, survey responses unrelated to mental health, and confidence reports in a gamified decision-making experiment.

At the start of the interface, participants received standardized on-screen instructions and completed a brief training trial. Each trial guided participants through a self-paced two-step question-response loop for a single, randomly selected item. On the question display, a question appeared in large font at the center of the screen (e.g., "How much are you distressed by feeling blue?"). In the cursor-based interface, participants proceeded by clicking on a circle randomly positioned in one of the screen's four corners. The response display appeared after a short delay (250 ms). The same item was displayed again in smaller font at the top, and a response line appeared at a randomly generated position, length, and orientation. The two endpoints of the line were marked with a green and a blue circle. A reference displayed at the bottom of the screen on every trial explained that the green end corresponded to "Not at all" and the blue end to "Very much".

Response positions were mapped to a continuous scale from 0 to 1, where 0 corresponded to the green end and 1 to the blue end. For instance, if a participant answered "How much are you distressed by feeling blue?" by clicking one-third of the way from green to blue, the recorded response would be 0.33, indicating mild distress. Clicks were only registered within the diameter spanned by the endpoints. As participants may not click exactly on the response line, the relative distance between the green and blue endpoints was used to compute their response. We randomized the position, orientation, and length of the response line on every trial (length range: 15-50% of screen height). As a result, the same response (e.g., "Not at all") could be associated with any absolute screen location. For example, one participant might click in the lower-left quadrant to indicate a distress level of 0.33, while another might click near the center for the exact same distress. Both the question order and the response mappings were independently randomized across participants and items. In the gamified decision-making experiments, the questions were replaced by information about the current outcome of the game (see below for details). Figure S11 illustrates how this design ensured that the location of the pointer on the screen was orthogonal to the underlying mental health self-reports.

The touch-based version differed from the cursor-based interface in three ways: first, participants viewed and responded to each item on a single screen; second, participants advanced the questionnaire by pressing a centrally located button at the bottom of the display; third, instead of clicking directly on the response line, participants dragged a response dot, initially placed at random in one of the four corners of the screen, onto a randomly positioned response line. These adjustments accommodated smaller screens and transformed the interaction into a continuous dragging gesture, providing a touchscreen analogue to the continuous cursor movements.

**Mental health assessments:** We used the interface to assess the participants' mental states using a novel self-report instrument that captured current distress and wellbeing as two complementary domains of mental health[55]. By adapting 67 items from established clinical and positive psychology questionnaires (BSI[56] and MHC-SF[57]), we mapped mental health across a spectrum of negative and positive states[55] (Table S1). The distress domain consisted of 53 items grouped into the subdimensions of anxiety, phobic anxiety, somatization, obsession-compulsion, depression, interpersonal sensitivity, psychoticism, paranoid ideation, hostility, and items of clinical relevance. The wellbeing domain comprised 14 items spanning emotional, social, and psychological wellbeing. All items were reworded to fit a digital, continuous-response format (Table S1). Rather than using a Likert scale, participants reported their experiences on a continuous scale ranging from 0 ("Not at all") to 1 ("Very much"). Distress and wellbeing items were intermixed and presented in randomized order, such that each participant experienced a different order of items.

We computed global scores for distress and wellbeing, as well as subdimension scores, by averaging the available responses across the respective items (person-level available-case scoring). This yielded a mental health matrix $\mathrm{Y}^{\mathrm{N}\times\mathrm{Q}}$, where N is the number of participants and Q the number of mental health features (items, subdimensions, global scores). Each row in this matrix represents an individual's location in a high-dimensional space of mental health, without reference to clinical thresholds or normative cutoffs. By decomposing the mental health matrix $\mathrm{Y}^{\mathrm{N}\times\mathrm{Q}}$ into orthogonal principal components (PCs), we derived independent axes of variation in self-reported mental health across participants.

We evaluated the psychometric properties of the questionnaire interface in terms of internal consistency, item structure, test-retest reliability, and external validity. We first assessed the reliability of our mental health assessments using Cronbach's $\alpha$, which was high for both distress (0.96) and wellbeing (0.86), indicating strong coherence among items within each scale. The average correlation between each item and the corresponding global score fell within the expected range of well-functioning items (distress: 0.55 ± 0.01; wellbeing: 0.51 ± 0.06). Mean inter-item correlations confirmed that the items within each domain were related but not redundant (distress: 0.31 ± 0.003; wellbeing: 0.3 ± 0.02). The subdimensions of distress and wellbeing each captured coherent and interpretable variance, as reflected by their correlations with the respective global scores, which ranged from 0.77 to 0.89.

To assess temporal stability, we correlated baseline and follow-up responses for intervals that ranged from 5 to 76 days. For follow-up intervals shorter than a week, test-retest correlations were high for overall distress (0.88) and wellbeing (0.84). After eight weeks, test-retest correlations declined to 0.69 for distress and 0.69 for wellbeing. A linear mixed-effects model with random intercepts for each item revealed that changes in self-reported mental health increased significantly with longer follow-up intervals ($p = 0.0077$).

Finally, we examined external convergent validity in an independent validation sample of 500 participants who completed the standard GAD-7 and PHQ-9 in addition to the touch-based MAILA tasks. For this analysis, we focused on the MAILA dimensions most closely corresponding to anxiety and depression, as assessed through the response interface for BSI[56] and MHC-SF[57]. GAD-7 and PHQ-9 summary scores were computed in the standard way by summing item responses coded from 0 to 3 ("not at all", "several days", "more than half the days", "nearly every day"), yielding total scores from 0 to 21 for the GAD-7 and from 0 to 27 for the PHQ-9, with higher scores indicating greater symptom severity. The anxiety dimension correlated with GAD-7 total score at $R = 0.51$ ($p < 0.001$), and the depression dimension correlated with PHQ-9 total score at $R = 0.48$ ($p < 0.001$). These results indicate that the randomized MAILA questionnaire interface recovers variance that converges with established self-report measures of anxiety and depression.

Together, these validation results indicate that the questionnaire interface provided stable, consistent, interpretable and valid estimates of mental health, while remaining sensitive to real-world variation in psychological state over time.

**Generalization experiments:** The response interface dissociated the content of self-reported mental health (that is, to what degree a participant endorsed a specific item of the assessment) from cursor movements and touchscreen activity recorded during questionnaire completion. This calibration procedure minimized the amount of data required to link motor behavior to mental health. At the same time, it recorded human–computer interactions in the cognitive context of self-reflection about mental health. Whether this context constrains or enhances generalization remains an open question: on the one hand, cursor and touch dynamics elicited during introspection may differ from those in everyday digital activity; on the other hand, activating a mental health context may amplify behavioral signatures that are diagnostic across settings.

To assess the robustness of models calibrated in this way, we evaluated MAILA across datasets that varied in content, task structure, and cognitive context. Within the MAILA dataset, the generalization structure was nested: all cursor-based recordings originated from a baseline assessment of 4,000 participants from the general population. Of these, 2,000 completed a follow-up session. At follow-up, participants completed the same 67-item mental-health assessment as at baseline with re-randomized item order. Within the follow-up group, two additional subsets of 600 participants each completed (i) a non-psychological survey and (ii) a gamified decision-making task.

For the inter-individual cross-context analyses, we maximized the available calibration data by fitting MAILA in a shared, baseline-defined movement-motif space using all baseline and follow-up mental-health-task recordings. The resulting models were frozen before application to survey and game movements from two non-overlapping subgroups of the follow-up cohort; no alternate-task movements or responses entered model fitting. We further tested frozen MAILA models on an external public dataset of open-ended human–computer interactions, including web browsing, file management, office applications, coding, and entertainment, with multiple sessions per participant[54].

**Non-psychological survey:** We recorded cursor movements while participants answered general survey questions unrelated to mental health (Table S2). The task interface and randomized response mapping were identical to the calibration paradigm, isolating the effect of content while keeping the motor context constant.

**Gamified decision-making task:** Participants completed a gamified version of the beads task, a probabilistic reasoning paradigm used as a transdiagnostic marker of altered decision-making in computational psychiatry[64]. At the beginning of each of six rounds, one of two jars was selected at random: a “blue jar” containing mostly blue beads and some green, or a “green jar” containing mostly green beads and some blue. The majority–minority ratio was set to 90%/10%, 75%/25%, or 60%/40%, and was displayed, but the identity of the majority color was hidden. Each participant completed all three ratios twice (6 rounds total), with the ratio–round assignment fully randomized in order independently for each participant. Each round consisted of eight sequential bead draws. After each draw, participants viewed the bead and an updated count of blue and green draws in the current round. They then indicated their certainty about which jar the beads were coming from on a continuous scale ranging from “100% certain: green jar” to “100% certain: blue jar.” The report interface was identical to the questionnaire interface outlined above. Across the six rounds, participants observed 48 bead draws and provided 48 certainty judgments. Cursor movements were recorded throughout the entire game. With the same interface logic and randomized response mapping as the calibration dataset, this task extended MAILA from survey completion to a novel interactive context involving sequential evidence accumulation, probabilistic reasoning, and gamification.

**Touchscreen drawing task:** Before starting the touchscreen questionnaire, participants completed a brief creative drawing task on their phone or tablet. Each participant drew 20 prompted and self-paced sketches (Table S3) in a randomized order (independently randomized per participant). Drawings were made on a canvas whose pixel dimensions were determined by the participant's device viewport (i.e., 90% of the maximum available drawing area on that screen). Drawing and touchscreen-interface recordings were represented and modeled independently using modality-specific clustering and SVR pipelines, yielding two behaviorally distinct estimates of the same participant-level mental-health profile. Because the drawing task contained no structured response interface, successful prediction from drawing tested whether the movement–mental-health association extended beyond questionnaire-response behavior.

**Benchmarking against GAD-7 and PHQ-9:** To benchmark the predictive value of MAILA against established brief symptom scales, we applied the frozen MAILA models to an independent validation sample of 500 participants who completed the touch-based tasks as well as standard-format PHQ-9 and GAD-7 questionnaires. We then asked how well the target variables in this validation cohort could be predicted from: (i) MAILA alone, using the pretrained frozen models transferred without refitting; (ii) PHQ-9, PHQ-2, or GAD-7 summary scores alone; and (iii) a combined model integrating questionnaire summary scores with MAILA predictions. Because the questionnaire-based benchmarks were fit within the validation cohort, their performance was estimated using 5-fold cross-validation, whereas MAILA performance reflected direct out-of-sample transfer to a fully independent dataset.

This analysis allowed us to distinguish two related questions: first, whether behavioral predictions from MAILA generalized to unseen participants and a new sample; and second, whether those predictions captured information comparable to, or additive over, widely used brief self-report instruments. PHQ-9 and GAD-7 provided strong benchmark predictors for depressive and anxiety-related variance, whereas PHQ-2 provided a more stringent ultra-brief comparator.

## External datasets

**Naturalistic cursor movements:** We applied MAILA, without retraining, to naturalistic cursor movements from the Boğaziçi dataset[54], downloaded 07/01/2024. This dataset comprises continuous recordings of cursor activity from 24 individuals, totaling approximately 2,550 hours of active computer use. Cursor movements were logged via a custom Python application that continuously captured mouse actions, timestamps, window titles, and contextual details of user interactions. Following the authors' protocol, we analyzed data from 19 participants who contributed sufficient training and testing data. The LSTM autoencoder was pretrained on these naturalistic trajectories, whereas the movement clusters and supervised mental-health mapping were calibrated only on the MAILA data.

**Emotion tracking:** To test whether MAILA's latent mental-health components track moment-to-moment affect, we analyzed three independent emotion-tracking datasets in which participants used cursor movements to continuously report arousal and valence while viewing emotionally evocative clips (Bayes dataset: 115 participants, 12 videos; Beat dataset: 81 participants, 83 videos; Context dataset: 102 participants, 35 videos; total: 298 participants, 130 videos; see[73–75] for dataset-specific sample characteristics, stimulus sets, and acquisition details). We applied pretrained, frozen MAILA models (trained on cursor movements labeled with mental-health self-reports in the MAILA dataset) to these external datasets to generate out-of-sample PC1–PC3 predictions from the recorded human–computer interactions. No component of the inference pipeline was trained or refitted on the affect-tracking data or their response mapping; affect reports were used only to stratify predictions generated by the frozen models. Affect labels therefore entered the analysis only after prediction and could not influence representation learning, clustering, or the mental-health mapping, making this a zero-shot transfer test from a separately calibrated task. Cursor trajectories were converted to MAILA inputs using the same segmentation scheme as in the MAILA dataset.

For each participant, we defined affective states based on participant-specific medians of arousal and valence and assigned each window to one of four arousal–valence combinations (high arousal/positive valence, high arousal/negative valence, low arousal/positive valence, low arousal/negative valence). We averaged predicted PC scores across windows within each affective state. For each participant and PC, we then computed high-minus-low contrasts separately for arousal and valence, such that each participant served as their own control. Participant-level contrasts were pooled across the three datasets and tested against zero using two-sided one-sample t-tests. The joint arousal–valence pattern was visualized using the four within-participant state combinations.

## MAILA

To model the relationship between human–computer interaction and mental states, we developed MAILA, a machine learning framework that transforms raw cursor and touchscreen activity into a data-driven movement feature matrix $\mathrm{X}^{\mathrm{N\times C}}$ (N = number of participants, C = number of movement features) and predicts the associated self-report matrix $\mathrm{Y}^{\mathrm{N\times Q}}$ (Q = number of mental health features).

**Inputs:** We segmented screen-normalized cursor and touchscreen positions $(\mathrm{a_t}, \mathrm{b_t}) \in [0,1]^2$ (min–max normalization per recording, applied independently to screen width and height) with a sliding window of fixed length L = 100 and stride $\delta$ = 10 samples. For each participant n, this yields $\mathrm{S_n}$ segments $X_i^{L\times 2}$:

$$X_i = \begin{bmatrix} a_{i,1} & b_{i,1} \\ \vdots & \vdots \\ a_{i,L} & b_{i,L} \end{bmatrix}$$

**Autoencoder:** Each movement segment $X_i^{L\times 2}$ consists of a sequence of L normalized 2D cursor positions $\{x_{i,t}\}_{t=1}^{L}$, with $x_{i,t} \in [0,1]^2$. An LSTM autoencoder transforms each segment into a single low-dimensional latent vector that summarizes its movement dynamics.

**Encoder.** The encoder LSTM (hidden dimension H = 64) processes each movement segment $X_i$ as a sequence of 2D cursor positions $\{x_{i,t}\}_{t=1}^{L}$, producing a hidden state $h_t$ at each time step. The final hidden state $h_L$ summarizes the full trajectory and is projected into a single latent vector $z_i^{1\times E}$ of dimension E = 50:

$$h_t = \mathrm{LSTM_{enc}}(x_{i,t},\, h_{t-1}), \qquad z_i^{1\times E} = \sigma(W_z h_L + b_z).$$

Here, $x_{i,t} \in [0,1]^2$ denotes the cursor position at time t, $h_t$ is the recurrent hidden state of the encoder, and $h_L$ is the final hidden state summarizing the entire trajectory. The latent vector $z_i$ is obtained by linearly projecting $h_L$ through $(W_z, b_z)$ followed by a sigmoid activation that bounds its values to (0,1). Thus, each trajectory segment (100 positions) produces exactly one latent vector $z_i$ , which serves as MAILA's movement embedding for downstream analysis.

**Decoder.** To reconstruct the original sequence during training, the decoder LSTM (hidden dimension H = 64) is conditioned on the latent code $z_i$ and generates a sequence of predicted 2D cursor positions:

$$\hat{h}_t = \mathrm{LSTM_{dec}}(z_i, \hat{h}_{t-1}), \qquad \hat{x}_{i,t} = \sigma(W_o \hat{h}_t + b_o), \quad \hat{x}_{i,t} \in [0,1]^2.$$

The same architecture allows extraction of latent features through the encoder or reconstruction from any latent vector using the decoder. The sigmoid output ensures that all predicted coordinates remain within the normalized input range. We used the decoder only for reconstruction during training. The latent vectors $z_i$ are the only quantities used downstream as MAILA's movement features.

We trained the autoencoder on an independent public cursor tracking dataset[54] for 100 epochs, using a batch size of 128 and a learning rate of 0.001. Training minimized the mean squared reconstruction error:

$$\mathcal{L}_{\text{recon}} = \frac{1}{L}\sum_{t=1}^{L} \| x_{i,t} - \hat{x}_{i,t} \|^2.$$

The final validation loss after training was 0.000052. Figure S4 shows examples of original and reconstructed cursor movements from the MAILA dataset.

**Movement feature representation:** MAILA pools all segment embeddings $z_i^{1\times E}$ across participants and clusters them using K-means into $C = 500$ discrete clusters:

$$\mathcal{C} = \{c_1, \dots, c_C\}, \qquad z_i \mapsto \underset{j}{\operatorname{argmin}} \| z_i - c_j \|.$$

Each cluster represents a recurring movement motif as captured in the latent space of the autoencoder at the group level. For each participant n, MAILA computes the proportion of their $\mathrm{S_n}$ segments assigned to each cluster, resulting in a movement feature vector $\mathrm{m}_n^{1\times C} \in [0,1]^C$ that sums to 1:

$$m_{n,j} = \frac{1}{S_n}\sum_{i=1}^{S_n} \mathbb{I}\,[\, z_i \in \mathcal{C}_j \,].$$

Stacking these feature vectors yields the movement matrix $\mathrm{X^{N\times C}}$, where each row describes a participant's distribution over recurring patterns of human–computer interaction. To define a stable, label-free movement vocabulary, K-means clustering was performed once on the complete set of unlabeled embeddings within each calibration modality. The resulting 500 centroids were held fixed throughout participant-disjoint supervised cross-validation. For transfer analyses, baseline cursor centroids were applied unchanged to follow-up, survey, game, naturalistic, and affect-tracking recordings; touchscreen-interface and drawing analyses used their respective modality-specific motif dictionaries.

All MAILA parameters were prespecified before data collection; sensitivity analyses varying cluster number, training-set size, test-data availability, and input noise are presented in Figure S21.

## Prediction of mental health

MAILA uses the movement feature matrix $\mathrm{X}^{N\times C}$ to predict participants' self-reported mental health features from the matrix $\mathrm{Y}^{N\times Q}$. MAILA approximates the decoding function $f(X;\delta)\colon X \to Y$, which maps cursor or touchscreen activity to latent mental states as indicated by the questionnaire responses in $\mathrm{Y}^{N\times Q}$.

To implement $f(X;\delta)$, we trained one support vector regressor (SVR, radial basis function kernel, $C = 1.0$, $\epsilon = 0.1$, automatic kernel scaling) per mental-health feature:

$$\hat{y}_{n,q} = f_q(x_n;\delta_q) = \mathrm{SVR}_q(x_n), \qquad q = 1, \dots, Q.$$

Here, $\delta_q$ denotes the SVR parameters for feature $q$, and $x_n$ is the movement feature vector of participant n. We evaluated model performance in two complementary settings. First, we assessed predictive accuracy using 5-fold cross-validation with non-overlapping participant IDs, ensuring that all data from a given participant appeared in a single fold. Second, to assess transfer across time and context, we trained models on the relevant calibration data and applied them without training on the target recordings to later, alternate-task, or independent test datasets. In all transfer analyses, both the calibration-derived motif dictionary and the supervised mapping were fixed before application to the target recordings.

After training independent SVR models for each target item, we averaged held-out predictions and targets by participant and dimension (depression, anxiety, phobic anxiety, somatization, obsession-compulsion, interpersonal sensitivity, psychoticism, paranoid ideation, hostility, clinically relevant features, emotional, psychological, and social wellbeing) to obtain per-participant dimensional estimates, including overall distress (mean across all distress dimensions) and overall wellbeing (mean across all wellbeing dimensions).

Model performance was quantified using three complementary metrics. First, Spearman's rank correlation coefficient (R) captured the rank-order correspondence between predicted and observed values. Second, the normalized root mean squared error (e) measured absolute deviations while accounting for the outcome's scale, defined as the square root of the mean squared error normalized by the range of the outcome variable. Third, the area under the receiver operating characteristic curve (AUC) assessed discriminative performance. To compute AUC, continuous SVR outputs were treated as ranking scores and ground-truth responses were binarized at the 10th, 25th, 50th, 75th, and 90th percentiles of the empirical outcome distribution. This procedure enabled evaluation of model sensitivity across multiple cut-offs along the mental-health continuum. Together, these metrics provide complementary assessments of prediction accuracy, error magnitude, and classification performance.

**Principal component analysis:** To test specificity to orthogonal axes of variation, we applied principal component analysis (PCA) to the 67 ground-truth self-reports (i.e., the $N \times Q$ matrix of all assessed mental health items, excluding summary scores for each dimension) after centering and scaling each item to zero mean and unit variance. To establish a common coordinate system for mental-health variation across datasets, we fitted a reference PCA once to the pooled, standardized 67-item profiles from the baseline, follow-up, and general touchscreen cohorts. The resulting standardization parameters and loadings were then fixed for all projections and PC-target models. Supervised prediction of PC scores was evaluated using participant-disjoint folds. We trained independent SVR models directly on the PC scores as targets (rather than computing PC estimates by post-hoc transforming the original item- or dimension-level SVR predictions into PC space). Because PC2 and PC3 were orthogonal to the first global component and encoded variation in symptom composition, above-chance prediction of these axes would argue against an account based solely on a single global response-position or response-magnitude signal.

To assess stability across cohorts and devices, we fitted PCA separately to the standardized 67-item profiles from the baseline cursor and independently recruited touchscreen cohorts. We matched component order among the first six PCs by maximizing absolute Tucker congruence and aligned component signs before comparison. We quantified the stability of PC1–PC3 using Tucker congruence and the similarity of their joint subspace using orthogonal Procrustes analysis (mean principal cosine).

**T-SNE visualization:** To visualize multivariate structure, we computed dimension-wise correlation matrices for ground-truth scores and for MAILA's predictions, converted them to dissimilarities (D = 1 − R), and embedded the resulting matrices using t-SNE (perplexity = 6), plotting the first two embedding dimensions (Figure S6).

**Significance testing:** To assess statistical significance, we compared observed model performance to null distributions obtained from 1,000 iterations with randomly shuffled target values. Permutation-based p-values quantify the proportion of permuted scores that were equal to or more extreme than the empirical metric. Unless otherwise indicated, we used t- and F-test statistics for group-level inferences. Linear mixed-effects models were applied in analyses where repeated measurements or hierarchical data structures required explicit modeling of dependency. Unless otherwise specified, summary statistics are reported as the mean ± 95% confidence interval. P-values reported alongside cross-dimensional means are from one-sample t-tests against zero (or the corresponding permutation baseline).

## Control analyses

We used two complementary controls to test whether MAILA's performance could be explained by direct spatial or task-response information. First, we repeated prediction using only the participant-level mean x and y coordinates of response endpoints and, separately, endpoint eccentricity. These controls were evaluated using 5-fold participant-disjoint cross-validation against null distributions generated from 1,000 within-item permutations. Second, we tested whether movements recorded during alternate tasks directly encoded the responses made in those tasks by predicting non-mental-health survey responses and bead-task confidence reports using 5-fold participant-disjoint cross-validation.

## Internal reliability

To assess the internal reliability of MAILA, we quantified the consistency of its predictions across two independent subsets of human–computer interaction data from the same individuals. For each participant, we randomly divided the LSTM-derived movement embeddings into two non-overlapping halves (50/50 splits), with split boundaries randomized independently to avoid systematic alignment across participants. The clustering step was performed on one split only, and the resulting cluster centroids were then transferred to the other split to construct its corresponding feature matrix. This yielded two independent feature matrices $X_{1/2}^{N\times C}$, each associated with the same mental health response matrix $Y^{N\times Q}$.

For each mental-health item, we trained independent MAILA models on the self-reports from one split and applied them to the opposite split. The procedure was then repeated in the reverse direction, resulting in two independent prediction vectors per participant. The correlation between these cross-split predictions served as an estimate of model reliability relative to a null distribution generated by random permutation of one prediction vector.

## Prediction of changes in mental health

To evaluate whether human–computer interactions track within-person changes in mental health, we calibrated MAILA exclusively on baseline movement features $X_{\text{baseline}}$ and the corresponding self-reports $y_{\text{baseline}}$. Both the baseline-derived K-means centroids and the supervised mapping were then fixed before the pipeline was applied to follow-up, non-mental-health survey, and game recordings. Thus, later or alternate-task behavior was the only new input used to estimate change relative to baseline. We correlated the difference between the ground-truth reports,

$$y_{\text{followup}} - y_{\text{baseline}}$$

with the predicted difference,

$$\hat{y}_{\text{followup/survey/game}} - \hat{y}_{\text{baseline}}$$

For the primary item- and dimension-level change analysis, $\hat{y}_{\text{baseline}}$ was generated out of fold using 5-fold participant-disjoint cross-validation, so the model estimating a participant's baseline score had not been trained on that participant's baseline data or label. Predictions for later or alternate-task recordings were generated by the frozen baseline-calibrated models described above. AUC quantified MAILA's ability to distinguish participants whose self-reported mental health improved versus worsened relative to baseline. To address regression to the mean in change detection, we additionally computed, separately for each mental-health dimension and PC, the partial Spearman correlation between the follow-up MAILA estimate and observed self-report change while controlling for the corresponding baseline score.

## Demographics

To test whether MAILA's predictive performance was consistent across demographic groups, we analyzed participant-level normalized prediction errors across PC1–PC3 and all datasets using a joint linear mixed-effects model. Mental-health component, age, gender, ethnicity, country of birth and residence, nationality, language, student status, and employment status were entered simultaneously as fixed effects, with participant included as a random intercept. Type III ANOVAs and partial $\eta^2$ quantified the unique prediction-error variance associated with each demographic characteristic while accounting for all other model terms and repeated estimates from the same participant. Factor-specific tests and category-level prediction errors are reported in Figure S15.

We further tested whether MAILA's predictions remained associated with self-reported mental health after accounting for demographic and digital-engagement covariates. Separately within each dataset and mental-health dimension, we residualized both observed and predicted scores with respect to all available demographic characteristics, task completion time, and the participant's number of prior approved Prolific submissions. Covariates available and variable within each subset were entered simultaneously, using complete cases. We then computed Spearman correlations between the residualized predictions and observations, quantifying their association after removing variance explained by all measured demographic and digital-engagement factors.

## Group-level mental health

To estimate how MAILA's prediction errors translate into performance in larger groups, we conducted simulations of population-level monitoring under realistic noise levels (Figure S18). We first generated synthetic "true" mental-health scores by sampling from a Gaussian distribution with fixed mean and variance. We then created corresponding "predicted" scores by adding noise such that the normalized root mean squared error between true and predicted values matched the range observed for MAILA in our empirical analyses. For each simulated population and group size, we computed group-level means and variances for both the true and predicted scores, and quantified alignment by correlating the true group-level statistics with their predicted counterparts.

To evaluate whether MAILA can support population-level mental health monitoring, we tested its ability to reproduce established demographic and temporal patterns in mental health from human–computer interactions alone. For each mental-health dimension, ground-truth questionnaire scores and MAILA predictions were z-scored independently before aggregation. We then computed group-level means by binning participants according to age (years), employment status (full-time, part-time, retired, job seeking), gender, and local time-of-day.

## Time-of-day analyses

Local time-of-day was defined as a continuous hour-of-day variable. For the primary analysis, we assigned participants to one-hour time bins (24 bins) and computed bin-wise mean and 95% confidence intervals for both ground truth and predictions. For each grouping variable (including local time-of-day), alignment between MAILA and ground truth was quantified using Spearman's rank correlation between true and predicted group-level means across bins. We also evaluated lagged correspondence by circularly shifting the predicted circadian profile relative to the ground-truth profile (integer-hour lags within ± 12 hours; shift in bins determined by the bin size) and recomputing Pearson correlations to identify the peak alignment.

To visualize circadian structure, we additionally estimated smooth time-of-day profiles on the 24-hour clock using circular Gaussian smoothing. Profiles were evaluated on a dense time grid (15-minute resolution), using a kernel width of 1.5 hours and circular wrap-around distance (i.e., the minimum of $|h -$ $t|$ and $24-|h-t|$). We computed weighted mean and approximate 95% confidence intervals for both ground-truth and predicted z-scores at each grid time point.

To test whether within-participant temporal structure in mental health could be inferred from open-ended human–computer interaction, we applied the frozen models to an external dataset[54] and, for each participant, aggregated MAILA's predictions by local hour of day using the same binning and smoothing procedures. Local time for the external dataset was obtained from the dataset's timestamp metadata.

## Self-identified mental illness

To evaluate MAILA in people who experience more extreme mental states, we analyzed touchscreen data from 1,000 participants who self-identified as living with depression and 500 who self-identified as living with obsessive–compulsive disorder. We trained support vector classifiers (SVC; radial basis function kernel, C = 1.0, automatic kernel scaling, balanced weights) to distinguish each self-identifying group from the general-population cohort using movement features $\mathrm{X}^{\mathrm{N}\times\mathrm{C}}$. Classification performance was estimated using 5-fold participant-disjoint cross-validation, with AUC computed from pooled out-of-fold class probabilities. As an empirical self-report benchmark, we replaced the movement feature matrix $\mathrm{X}^{\mathrm{N}\times\mathrm{C}}$ with the mental-health matrix $\mathrm{Y}^{\mathrm{N}\times\mathrm{Q}}$ and evaluated an independent SVC using the same procedure.

To assess sensitivity to inter-individual differences within self-identifying groups, we trained models on the full sample (including both general and self-identifying participants) using 5-fold cross-validation and computed Spearman's rank correlations between ground truth and predicted scores within the depression and OCD groups. We applied linear mixed-effects models to test whether MAILA predicted mental health profiles aligned with expected group characteristics, that is, higher depression scores among participants who self-identified with a history of depression and higher OCD symptoms among those who self-identified with a history of OCD.

## Decoding belief instability from human–computer interaction

To derive a behavioral marker of belief instability, we analyzed trial-by-trial confidence updates from the gamified decision-making task. In each round, participants observed eight sequential draws from one of two jars with known bead ratios but unknown majority color. After each draw, they provided a confidence-weighted belief about which jar generated the sequence. For each round $r$, we quantified the magnitude-weighted entropy of belief-update directions as:

$$\Delta b_{r,t} = b_{r,t} - b_{r,t-1}, \qquad T_r = \sum_t |\,\Delta b_{r,t}|,$$

$$p_{+,r} = \frac{\sum_{\Delta b_{r,t}>0} |\,\Delta b_{r,t}|}{T_r}, \qquad p_{-,r} = 1 - p_{+,r},$$

$$H_r = -p_{+,r}\log p_{+,r} - p_{-,r}\log p_{-,r}.$$

Here, $b_{r,t} \in [0,1]$ denotes the belief at trial $t$ of round $r$, $T_r$ is the total magnitude of belief change within the round, and $p_{+,r}$ and $p_{-,r}$ denote the proportions assigned to positive and negative updates, with $0\log 0$ defined as 0. The measure operationalizes belief instability as variability in update direction: high entropy indicates bidirectional, reversal-prone updating and was interpreted as reactive cognition, whereas low entropy indicates predominantly unidirectional updating and was interpreted as rigid cognition. It is invariant to total update magnitude and therefore does not quantify update size or reversal frequency. For each participant, we averaged $H_r$ across valid rounds to obtain a single belief-instability score.

To examine how belief instability relates to mental health, we correlated belief-instability scores with self-reported mental-health dimensions, and repeated the same analysis using MAILA's predicted mental-health scores (derived from cursor movements in held-out participants). Alignment between ground-truth and MAILA-derived mental-health correlations was quantified using Spearman's R, computed by correlating the vector of true correlations with the corresponding vector obtained from predictions across mental-health dimensions.

To test whether belief instability could be predicted directly from human–computer interaction, we trained SVR models under 5-fold cross-validation using three feature sets: (i) MAILA's movement feature matrix $\mathrm{X}^{\mathrm{N}\times\mathrm{C}}$, (ii) the self-report matrix $\mathrm{Y}^{\mathrm{N}\times\mathrm{Q}}$, and (iii) the concatenation $[\mathrm{X};\mathrm{Y}]$. For each fold, the model predicted belief instability for unseen participants, and predictive accuracy was quantified using Spearman's R and the normalized root mean squared error (e).

To quantify the unique association between cursor-derived predictions and belief instability beyond self-reports (and vice versa), we computed partial correlations using participant-level belief-instability scores ($y$), out-of-fold SVR predictions from the cursor model ($x_{\mathrm{cursor}}$), and out-of-fold SVR predictions from the self-report model ($x_{\mathrm{MH}}$). The partial correlation $r_{xy\cdot z}$ was defined as the Pearson correlation between residuals obtained after linearly regressing both $x$ and $y$ on the control variable $z$ (equivalently, $r_{xy\cdot z} = \mathrm{cor}(\mathrm{resid}(x \sim z),\ \mathrm{resid}(y \sim z))$). Specifically, we report (i) $r_{x_{\mathrm{cursor}},y\cdot x_{\mathrm{MH}}}$ (cursor controlling for self-reports) and (ii) $r_{x_{\mathrm{MH}},y\cdot x_{\mathrm{cursor}}}$ (self-reports controlling for cursor). Partial correlations were computed on participant-level cross-validated (out-of-fold) predictions, ensuring that each prediction was generated by a model not trained on that participant.

## LLM assistant-facing mental-health inference simulation

We tested how accurately mental-health profiles could be inferred from single-message interactions with a frontier AI chatbot in a controlled agentic simulation. The source profiles comprised all 4,000 participants in the baseline cursor cohort. We ran one simulation per participant in each of three context conditions, with one message per simulation. Ground-truth profiles and MAILA predictions were represented as population percentile ranks from 0 to 100 on each of the 13 dimensions.

The MAILA predictions supplied as context were out-of-fold predictions of the baseline cursor model, taken from the 5-fold participant-level cross-validation described above, in which all recordings from a given participant were assigned to a single fold. The context provided for a participant was therefore always generated by a model that had not been trained on that participant's data. No model was refitted for this simulation.

Messages and scenarios were drawn independently in each condition. The user simulator received the ground-truth profile but never MAILA predictions, whereas the profile-inference agent received the synthetic message and condition-specific MAILA context but never the ground-truth profile. Both agents were accessed through the OpenRouter chat-completions API in June 2026. The user simulator (anthropic/claude-3.7-sonnet; temperature = 1.0; maximum output = 900 tokens) received structured definitions of the 13 dimensions and one seeded scenario sampled from 120 broad, non-diagnostic assistant-use situations. It generated a short first-person request in which the profile could appear indirectly through content, framing, and tone, but was prohibited from mentioning mental health, diagnoses, questionnaires, items, scores, percentiles, or the simulation. Core prompts are summarized in Table S4; the full scenario bank is listed in Table S5.

The profile-inference agent (openai/gpt-5.2; temperature = 0.0; maximum output = 1,400 tokens) received the message and dimension definitions together with either (i) no non-verbal context, (ii) MAILA predictions from the same participant, or (iii) MAILA predictions from another randomly selected participant. It returned percentile-rank estimates for all 13 dimensions through a forced tool call. For both agents, sampling settings were top_p = 1.0, top_k = 0, frequency penalty = 0, presence penalty = 0, repetition penalty = 1.0, min_p = 0, and top_a = 0, with no stop sequence, no sampling seed, and default OpenRouter provider routing.

For each dimension, profile recovery was quantified using Spearman correlations between inferred and ground-truth percentile ranks across participants. Normalized prediction error was computed per participant $i$ as the root mean squared difference between inferred ($\hat{y}$) and ground-truth ($y$) percentile ranks across the 13 dimensions, divided by the 100-point range of the percentile scale,

$$e_i = \sqrt{\frac{1}{13}\sum_{d=1}^{13}\left(\frac{\hat{y}_{id} - y_{id}}{100}\right)^2},$$

and summarized across participants as the mean and 95% CI. Error was thus normalized by the fixed range of the outcome scale, as for the normalized root mean squared error reported elsewhere in this study, and not by the standard deviation or the variance of the outcome; $e$ is dimensionless and bounded in $[0,1]$. Absolute errors were compared across conditions using a linear mixed-effects model on the dimension-level absolute percentile error, with condition as a fixed effect and random intercepts for participant and participant-by-dimension, followed by a type III F-test with Satterthwaite degrees of freedom.

## Human-interpretable features

For each participant, we computed a set of 12 human-interpretable movement features from the cursor and touchscreen trajectories. Features were computed per movement segment (i.e., for each trajectory chunk used in our analysis pipeline) and then aggregated to the participant level by averaging across all segments belonging to that participant, yielding one feature vector per participant. Starting from each segment's screen-normalized 2D position trace, we derived instantaneous velocity, speed, and heading changes and summarized these dynamics with features capturing movement amplitude, intermittency, efficiency, turning behavior, directional bias, and temporal variability: average speed, speed kurtosis, jerk (mean absolute change in speed), movement area, relative idle time (fraction of samples below a stationary threshold $\tau = 0.001$), path efficiency, average turn angle, tortuosity, turn rate per distance, horizontal–vertical bias, speed entropy (FFT-based spectral entropy of the speed trace), and speed fluctuation rate (zero-crossings around mean speed). The resulting feature matrix did not include absolute screen coordinates; above-chance prediction from this representation would therefore argue against an account based solely on absolute response location. Formal operational definitions and implementation details for all 12 features are provided in Table S6.

After aggregating features within each participant, we z-scored each feature across participants and estimated linear regression coefficients by regressing each feature independently onto mental health, using both the true and MAILA-predicted scores. This analysis provided an interpretable mapping between low-level motor statistics, MAILA's predictions, and ground-truth mental health.

To directly compare the predictive utility of handcrafted features against MAILA, we replaced the original feature matrix $\mathrm{X}^{\mathrm{N}\times\mathrm{C}}$ with the handcrafted feature matrix $\mathrm{X}'^{\mathrm{N}\times\mathrm{F}}$, where F is the number of human-interpretable features. We trained SVRs to map from $\mathrm{X}'^{\mathrm{N}\times\mathrm{F}}$ to the mental-health matrix $\mathrm{Y}^{\mathrm{N}\times\mathrm{Q}}$ using the same participant-disjoint cross-validation procedure as MAILA. This yielded paired held-out predictions from learned and handcrafted feature spaces for the same participants and targets. Direct statistical comparisons used participant-by-dimension absolute prediction errors normalized by the empirical range of the corresponding outcome; overall performance was additionally compared in terms of e, R, and AUC.

## Information loss

We assessed MAILA's robustness to information loss by training and testing support vector regression models to predict scores on each dimension of distress and wellbeing while implementing four types of data degradation: (1) noise injection, where a proportion of the low-level movement embeddings of the test sets were interpolated with random values sampled from a uniform distribution; (2) within-participant data loss, where a contiguous segment of each participant's recording was deleted in the test datasets, simulating shorter recordings while preserving temporal coherence; (3) training set reduction, where we progressively decreased the number of unique participants used to train the model; and (4) cluster reduction, where we gradually reduced the number of recurring patterns used to construct the movement feature matrix $\mathrm{X}^{\mathrm{N}\times\mathrm{C}}$ (Figure S21).

Degradation was increased systematically in 5-percentage-point increments through 95%; the cluster-reduction analysis spanned movement representations from 500 to 2 motifs. At each level, we quantified the correlation between predicted and ground-truth scores separately for cursor-based interface data, touch-based interface data, and free-form touchscreen drawings.

## Detecting inconsistencies between self-report and behavior

To test whether human–computer interactions can identify inconsistent or false self-reports, we quantified how reliably mismatches between verbal reports and behavioural data could be detected as a function of the magnitude of distortion applied to otherwise valid mental-health profiles. $\mathrm{P_i}$ and $\mathrm{T_i}$ denote participant $i$'s MAILA-predicted and true mental-health profiles (z-scored). For each distortion level $\sigma$ (in SD units), we created distorted profiles by adding independent Gaussian noise to each dimension,

$$T_i^{\text{fake}} = T_i + \varepsilon_i, \qquad \varepsilon_i \sim \mathcal{N}(0, \sigma^2),$$

and clipped each dimension to the empirical range $[z_{\min,d}, z_{\max,d}]$ observed in $T$. For each $\sigma$ and repetition, we computed mismatch scores as the negative Euclidean distance between predictions and profiles,

$$s_i = -\|P_i - T_i\|_2, \qquad s_i^{\text{fake}} = -\left\|P_i - T_i^{\text{fake}}\right\|_2,$$

and used these to compute the AUC for discriminating true from distorted profiles. We repeated this procedure 1,000 times per distortion level and report the mean AUC and 95% confidence intervals across repetitions as a function of $\sigma$.

## Identification analyses

To test whether human–computer interactions carry personally identifiable information, we trained a SVC to predict the identity of 4,000 participants based on their movement feature matrix $\mathrm{X}_{\mathrm{baseline}}^{\mathrm{N}\times\mathrm{C}}$ from the baseline cursor-tracking dataset. Movement features at follow-up, $\mathrm{X}_{\mathrm{followup}}^{\mathrm{N}\times\mathrm{C}}$, were derived using K-means clustering defined exclusively on segment embeddings from $\mathrm{X}_{\mathrm{baseline}}^{\mathrm{N}\times\mathrm{C}}$. This fixed clustering ensured that no information from the follow-up dataset influenced feature construction. To assess statistical significance, we retrained the SVC on randomly permuted training labels over 1,000 iterations and compared the empirical results to the resulting null distribution.

To assess whether cursor movements contain sufficient information to track the continuity of identity across time, we trained a second SVC to distinguish whether two feature vectors $\mathrm{X}_{\mathrm{baseline}}^{1\times\mathrm{C}}$ and $\mathrm{X}_{\mathrm{followup}}^{1\times\mathrm{C}}$ belonged to the same or to different individuals. For each participant in the follow-up dataset, we paired their movement features with those from the same individual in the training set (positive pairs), as well as with features from randomly selected individuals (negative pairs). Each pair was represented by the concatenated movement features from the two sessions. The classifier was trained using 5-fold stratified cross-validation and evaluated based on its ability to discriminate between positive and negative pairs. Statistical significance was assessed using a permutation test with 1,000 iterations. We used logistic regression to examine whether classifier performance was influenced by the time elapsed between recordings or by the precision of the mental health predictions at follow-up. To test whether personally identifying movement information accounted for mental-health prediction, we correlated participant-level identification accuracy with mean mental-health prediction error at baseline and follow-up using Spearman's rank correlation.

## Results S1

Cursor movements and touchscreen gestures underlie nearly all human–computer interactions. Here, we explore two non-mental-health applications of MAILA that have broad implications for trust and accountability in digital environments: detecting inconsistencies between self-report and behavior and user identification.

**Detecting inconsistencies between self-report and behavior:** To approximate how well MAILA can detect inconsistent self-reports directly from human–computer interaction, we simulated systematically distorted questionnaire profiles by adding Gaussian noise to each participant's true mental-health profile. We then quantified the mismatch between MAILA's predictions and these distorted profiles using the negative Euclidean distance. MAILA's ability to distinguish true from distorted profiles increased monotonically with the magnitude of distortion, rising from an AUC of 0.73 when the added noise had a standard deviation equal to the original item-level variability, to 0.89 when the noise standard deviation was doubled. Increasingly inconsistent self-reports thus became progressively easier to detect from cursor and touchscreen behavior alone.

**Decoding identity:** To test whether human–computer interactions contain personally identifiable information, we asked whether a support vector classifier could predict participant identity from digital behavior alone. The model was trained on data from 4,000 baseline participants and evaluated on 2,000 follow-up participants. Classification was above chance but weak (accuracy = 1.75 ± 0.58%, chance level = 0.05%, $p < 0.001$). With cursor movements as the only available signal, MAILA's ability to identify individuals in large populations was therefore greater than zero, but limited. Identification accuracy was unrelated to mental-health prediction error both within the baseline session (participant-disjoint mental-health predictions ~ split-half identification: $R = 0.02$, $p = 0.21$) and across sessions (frozen baseline models applied at follow-up ~ baseline-to-follow-up identification: $R = -0.03$, $p = 0.17$). Thus, participants whose movement patterns were easier to identify did not receive systematically more accurate mental-health predictions.

**Tracking the continuity of identity:** In many real-world settings, the user's identity is already known, because they are logged into an account, use a recognized device, or continue a previous session. In these cases, the relevant question is not who is using the device, but whether the same person is still using it. To test whether MAILA can track the continuity of identity over time, we trained a second classifier to determine whether two recordings, one from the baseline dataset, the other from the follow-up dataset, originated from the same or a different participant. The classifier tracked the continuity of identity with an AUC of 0.58 ($p < 0.001$). Its performance remained unaffected by the delay between recordings ($p = 0.85$), the degree of change in mental health between sessions ($p = 0.62$), and MAILA's ability to predict self-reports at follow-up ($p = 0.7$).

Together, these results suggest that human–computer interactions can be used for detecting inconsistencies between self-report and behavior as well as for user identification. The ability to decode information that people may not want to share underscores the urgent need for safeguards against the unintended or unauthorized use of digital biomarkers embedded in cursor and touchscreen activity.

## Figure S1

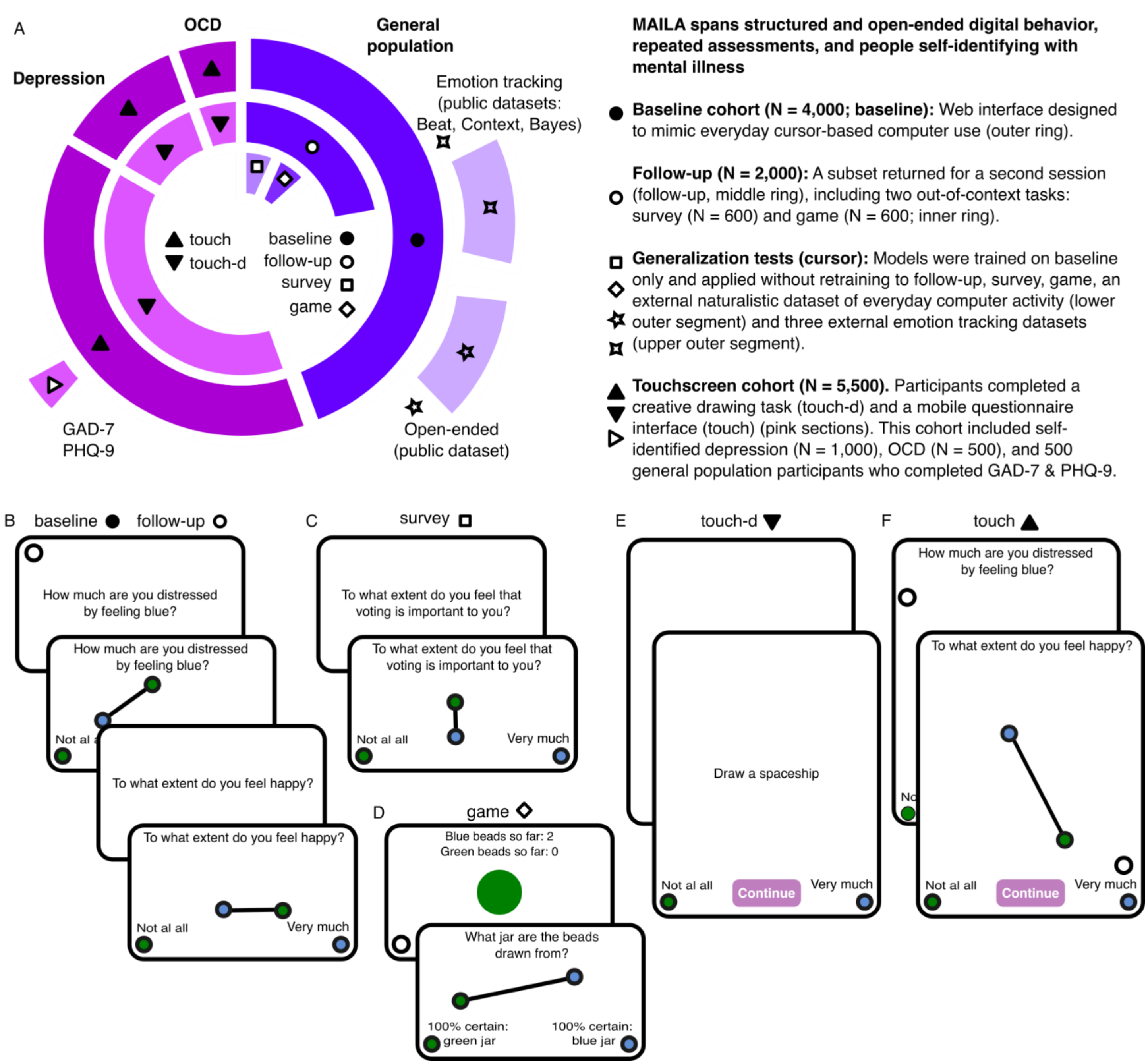

***Figure S1. The MAILA experiments.***

***A. The MAILA dataset.*** *We evaluated MAILA across structured and open-ended digital behavior, repeated assessments, and participants who self-identified as living with depression or obsessive–compulsive disorder. This figure summarizes the structure of the dataset. Sections with filled markers indicate subsets used for model training and testing (5-fold cross-validation over participants), whereas hollow markers denote subsets on which we assessed the generalizability of trained models.*

***B. Mental-health questionnaire.*** *Participants reported on their current mental health using a randomized response interface that dissociated cursor trajectories from the semantic content of their answers. Each trial began with a question screen; participants advanced by clicking a circle that appeared in one of the four screen corners at random. On the subsequent response screen, the same item reappeared and participants indicated their answer by clicking a randomly positioned and randomly oriented response line. Cursor trajectories were logged continuously from start to finish. Please note that items were presented in an order randomized per participant.*

***C. Non-mental-health survey.*** *Participants answered general survey questions unrelated to mental health using the same interface and randomized-response mapping as in **A**, isolating the effect of semantic content while holding the motor context constant.*

***D. Gamified decision-making task (beads task).*** *Participants played an interactive decision-making game using the same randomized-response mapping as in **A**, isolating the effect of semantic content beyond surveys while holding the motor context constant.*

***E. Touchscreen drawing.*** *Participants produced prompted, freehand drawings on their touchscreen device. Each prompt disappeared at first touch, and participants advanced by pressing a "Continue" button at the bottom center of the screen.*

***F. Touchscreen interface.*** *After the drawing task, participants completed a touchscreen version of the randomized-response interface from **A**, adapted for vertical screens. Instead of clicking, participants dragged a response dot, initially placed at random in one of the four corners of the screen, onto a randomly positioned response line.*

## Figure S2

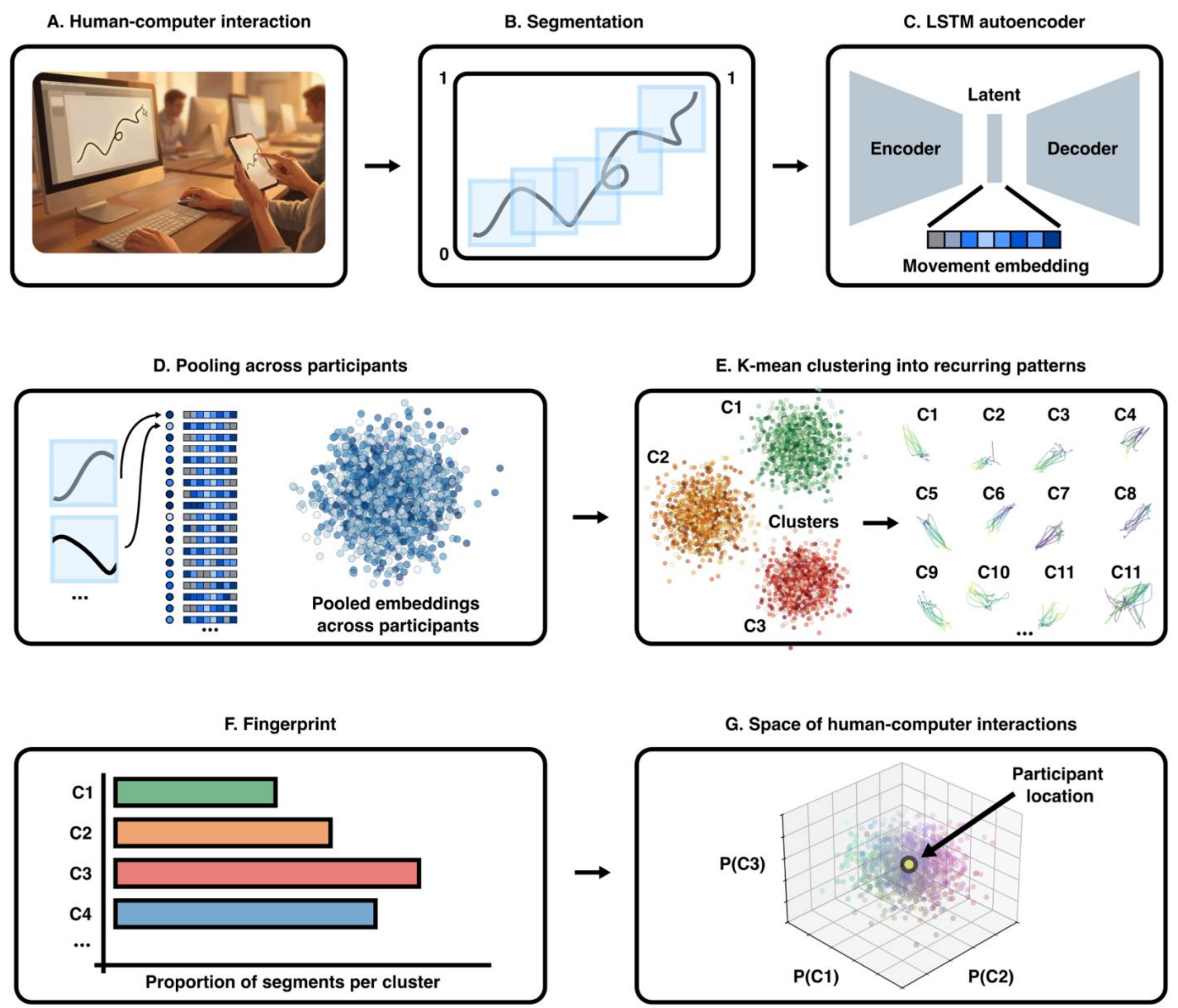

***Figure S2. MAILA transforms cursor/touchscreen activity into participant-level features.*** *This schematic summarizes the representation-learning pipeline that converts raw human–computer interactions into a participant-by-motif feature matrix* $X^{N \times C}$.

***A. Human–computer interactions.*** *Continuous cursor or touchscreen trajectories are recorded and transformed into screen-normalized coordinates.*

***B. Segmentation****. Each recording (on average 24,642.35 ± 462.35 samples) is divided into partially overlapping windows of 100 consecutive samples, yielding short trajectory segments that capture local movement dynamics.*

***C. LSTM autoencoder.*** *A long short-term memory autoencoder, pretrained on naturalistic human–computer interactions[54], maps each segment to a 50-dimensional movement embedding (encoder bottleneck) and reconstructs the input trajectory (decoder), producing a compact representation of segment-level movement structure.*

***D. Pooling across participants.*** *Segment embeddings from all participants are pooled into a shared embedding space to define a common vocabulary of recurring interaction patterns.*

***E. K-means clustering into motifs.*** *Pooled embeddings are grouped into C = 500 clusters, each representing a distinct, recurring movement motif (a "motif library") that captures stereotyped patterns of human–computer interaction.*

***F. Participant fingerprint****. For each participant, MAILA computes a fingerprint by estimating the proportion of segments assigned to each motif (distribution over C patterns).*

***G. Space of human–computer interaction.*** *This yields the feature matrix* $X^{N\times C}$*, where each row places a participant at a location in the space of digital behavior and can be used for downstream prediction of mental-health outcomes.*

---

# Figure S3

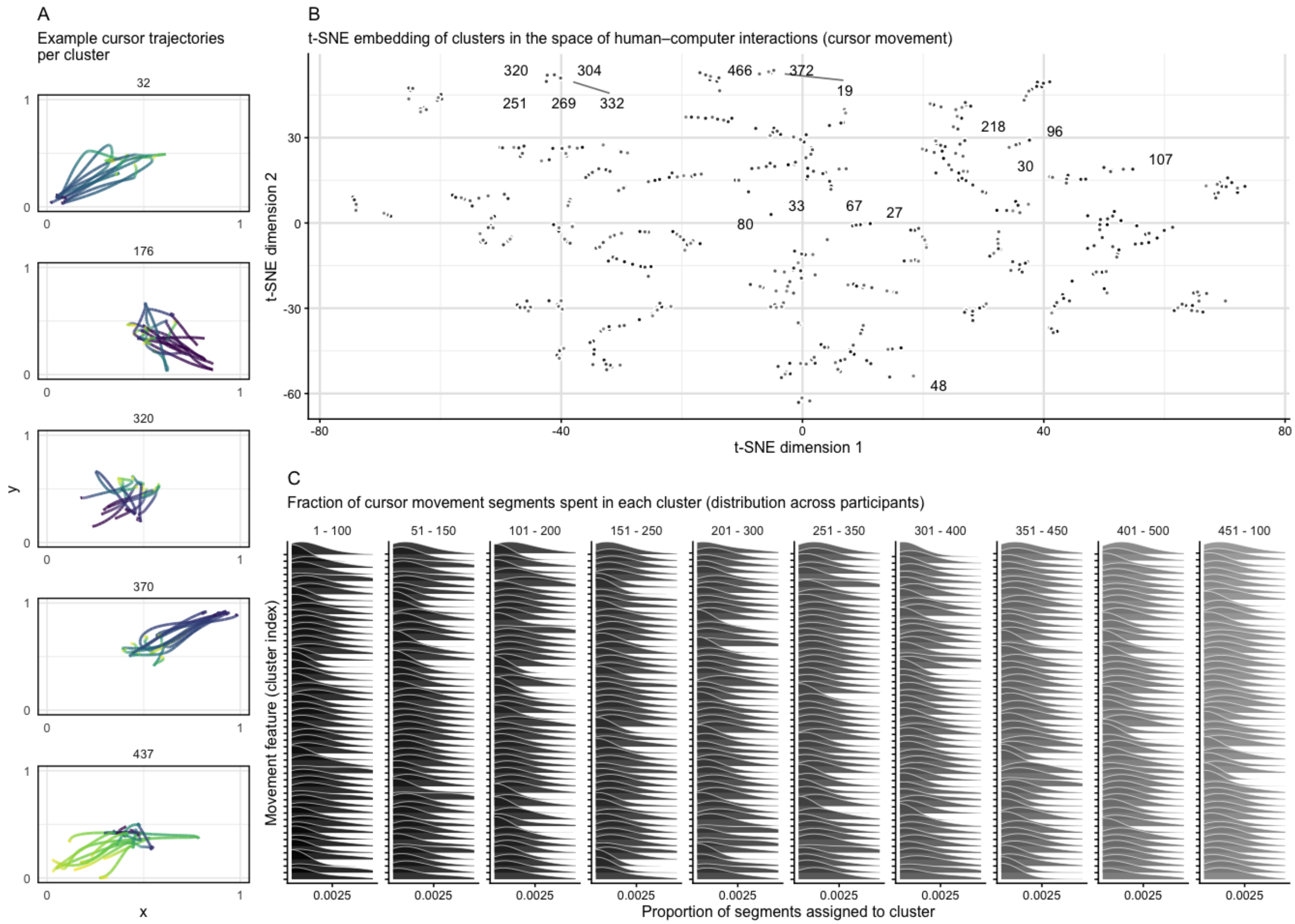

***Figure S3. The space of human–computer interaction.***

***A. Clustered cursor movements.*** *MAILA transforms segments of cursor and touchscreen movement into low-dimensional embeddings. By clustering the embeddings across participants, MAILA discovers recurring patterns of human–computer interaction at the group-level. Here, we show five exemplary clusters derived from cursor movement (time progresses from darker to lighter colors, number indicates cluster ID). Figure S5 shows examples from all 500 cursor clusters. While qualitatively similar patterns emerge for touchscreen data (not shown here), the specific clusters differ depending on the behavioral context of the interaction.*

***B. Structure of cursor movement features.*** *Each dot represents a cursor movement cluster, positioned in t-SNE space based on its similarity to other clusters in the space of human–computer interactions* $X^{N \times C}$.

***C. The cursor movement feature matrix.*** *MAILA computes the per-participant (N) fraction of segments assigned each of the C = 500 clusters, resulting in a* $X^{N \times C}$ *movement feature matrix. Each row encodes a participant's location in a space of cursor movement patterns, derived from raw trajectories that are segmented, autoencoded, and assigned to discrete clusters defined at the group-level. Plots show the distribution of features across participants for the 500 cursor movement clusters. The shape of the cluster distributions are qualitatively identical for touchscreen interactions (not shown here).*

# Figure S4

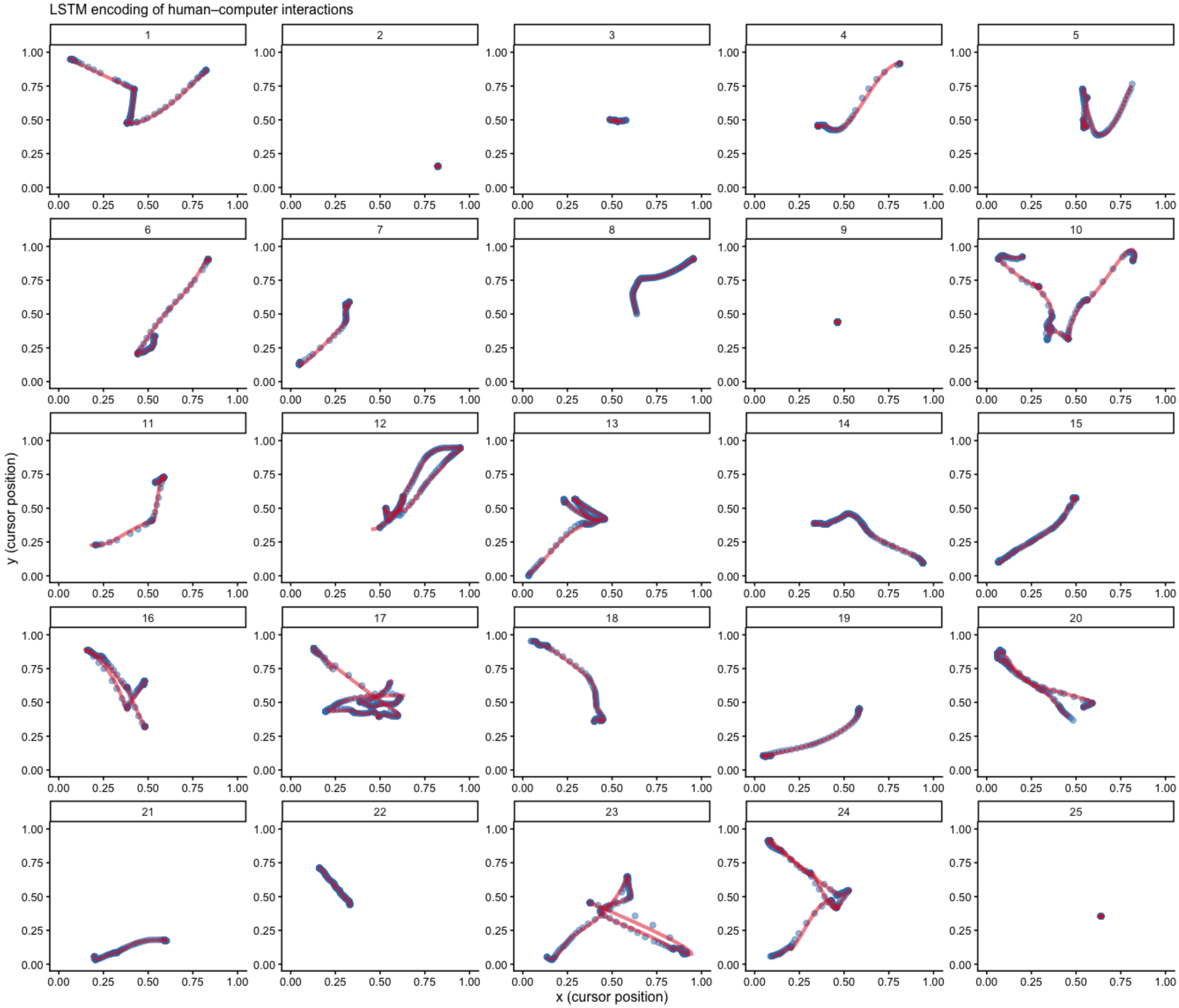

***Figure S4. Autoencoded cursor movements.*** *Each subplot represents one of 25 example cursor movements from our experiment, with original cursor positions shown in blue and reconstructed trajectories in red. MAILA's LSTM autoencoder was trained on human–computer interaction from a public dataset of naturalistic computer use[54], frozen, and applied to the MAILA dataset, where it achieved an average reconstruction loss of 0.000077 (relative to the participant's screen resolution). This confirms that MAILA captured and reconstructed human–computer interaction with high precision.*

# Figure S5

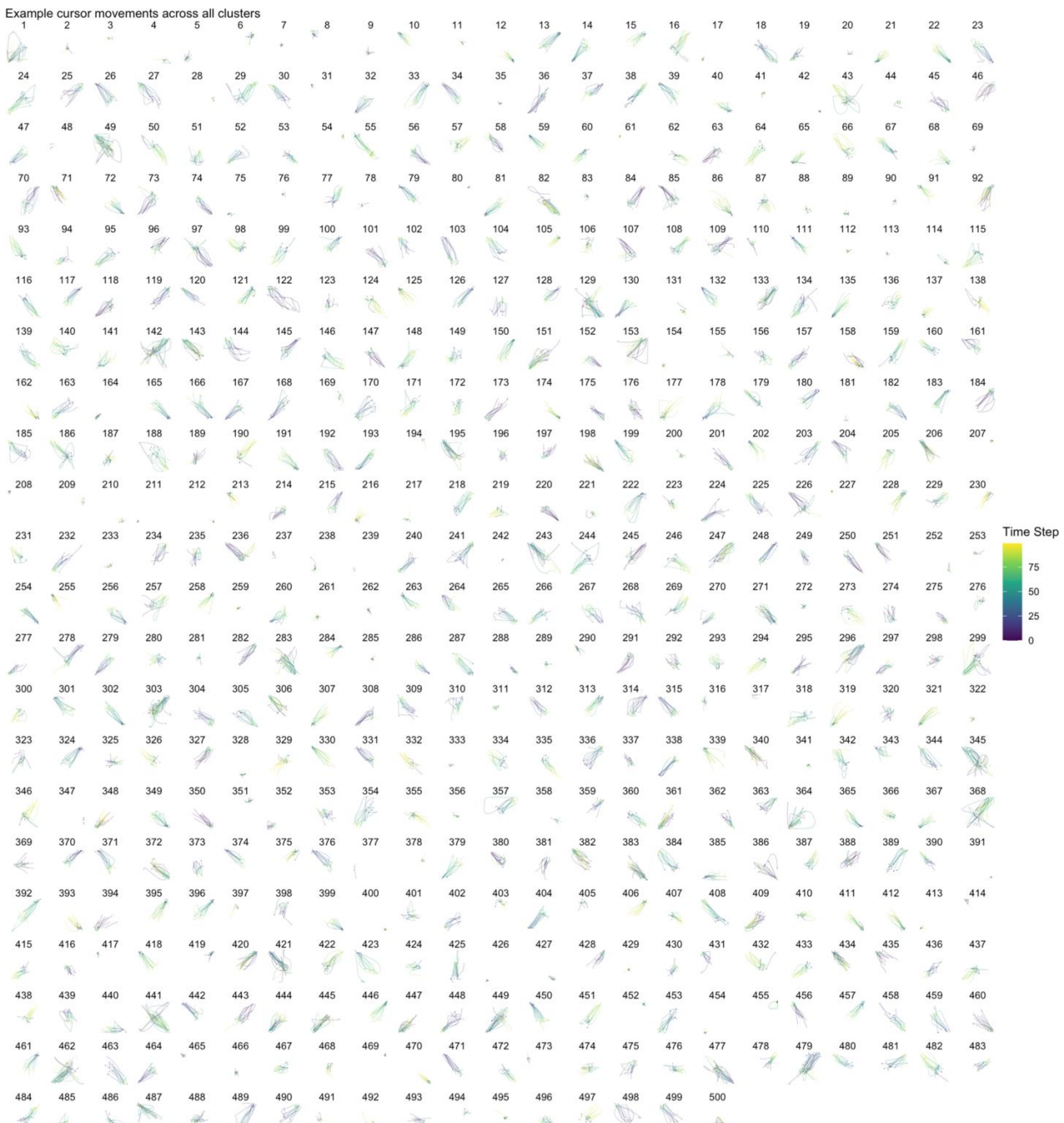

***Figure S5. Clusters of cursor movement.*** *Each subplot represents 10 example trajectories from one of the $C$ = 500 cursor movement clusters. Lighter colors indicate later time steps within each trajectory. Each cluster represents one distinct cursor movement motif observed in our experiment. For each of the N participants, we computed the fraction of embeddings assigned to each cluster C. This results in a participant-by-cluster $X^{N \times C}$ movement feature matrix. The clusters are assigned in a data-driven way, that is, without any hypotheses about what movement features are meaningful for the downstream task of predicting mental states from human–computer interaction. Different movement clusters emerge when MAILA is calibrated to structured cursor movement (shown here),*

*structured touchscreen activity (not shown here), or free-form touchscreen activity (not shown here). When assessing MAILA's ability to generalize, we performed clustering only on the training data, and transferred the frozen k-means centroids to the new datasets.*

# Figure S6

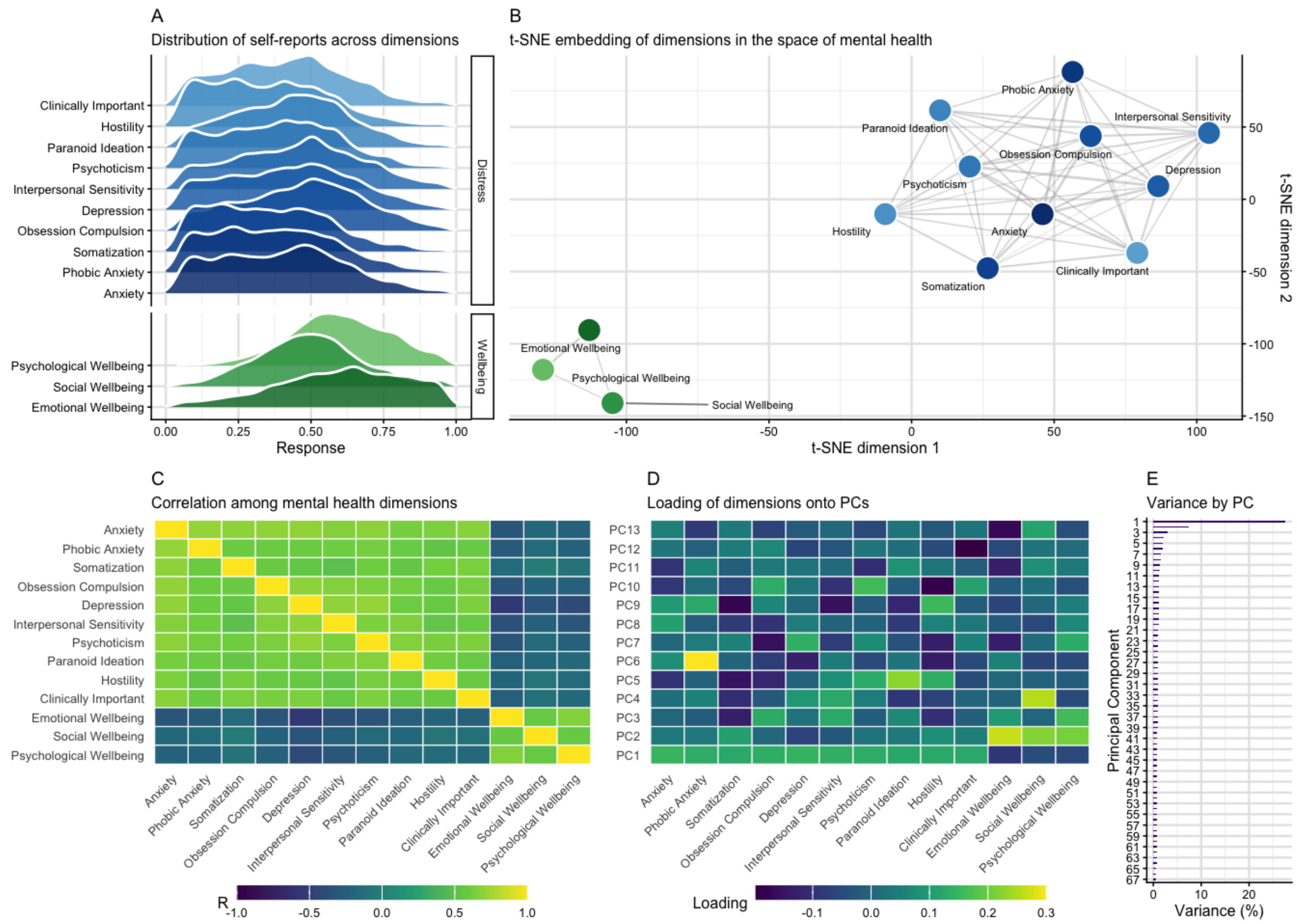

***Figure S6. The space of mental health.***

***A. The mental health matrix.*** *The mental health matrix* $Y^{N \times Q}$ *comprises self-reports for 67 questionnaire items, each belonging to an overarching dimension of distress and wellbeing. Distress distributions are shown in blue, wellbeing distributions in green, pooled across all participants from the general population in the MAILA dataset.*

***B. Structure and correlation of self-reported mental health.*** *Each point represents a mental health dimension, positioned in t-SNE space based on its similarity to other dimensions. Line thickness corresponds to the strength of positive correlations (negative correlations not shown).*

***C. Correlation matrix.*** *Correlations between self-reports in the distress and wellbeing domain, pooled across all participants from the general population in the MAILA dataset. Colors indicate the correlation strength and direction. Responses were negatively correlated between the domains of distress and wellbeing (R = -0.25 ± 0.02) and positively correlated between the dimensions of each domain (e.g., anxiety to depression, or emotional to psychological wellbeing). Average correlations reached R = 0.66 ± 0.02 between distress dimensions and R = 0.73 ± 0.09 within wellbeing dimensions.*

***D. PCA loadings.*** *Loading of each mental health dimension onto the first 13 principal components (PCs) of mental health self-reports (across participants). Colors indicate the strength and direction of the loading on the respective PC. For example, a positive loading for depression on PC1 means that, when a participant experiences increasing depressive symptoms, their score on PC1 will increase.*

***E. Variance explained.*** *The proportion of mental health variance explained by each principal component. Bars indicate the variance explained per component. Together, the first 3 components accounted for 37.91% of the variance.*

## Figure S7

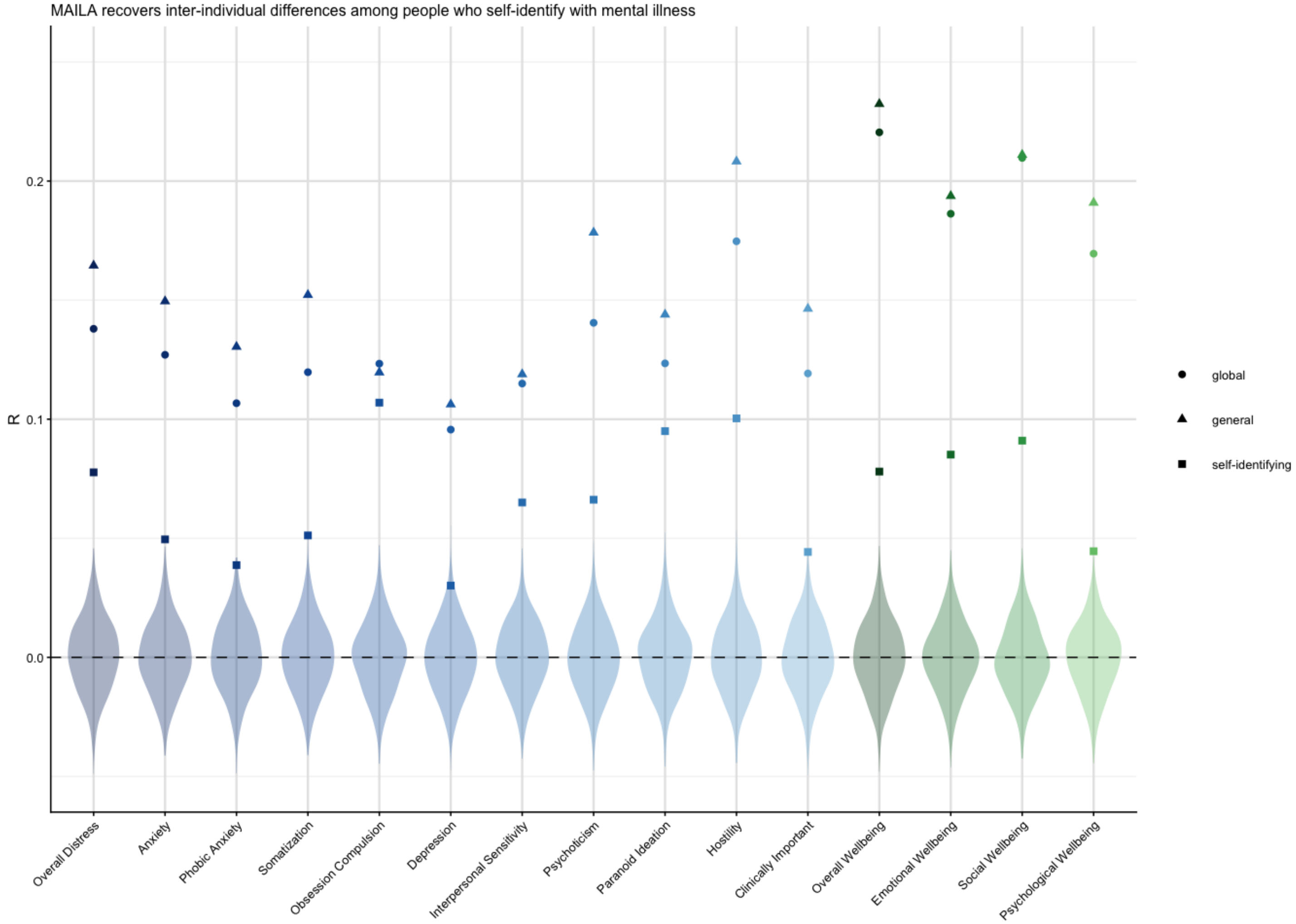

***Figure S7. MAILA predicts inter-individual differences among participants who self-identify with mental illness.*** *Models were trained on all participants and evaluated with 5-fold cross-validation. Markers denote within-group correlations of MAILA's predictions within the general sample, the self-identifying samples (depression and OCD), and the pooled (global) sample; violin plots show null distributions. MAILA captured inter-individual variation across dimensions within the self-identifying sample (R = 0.07 ± 0.01), despite broader and shifted symptom distributions relative to the general sample. This pattern supports a dimensional view in which self-identified depression and OCD reflect quantitative differences along shared mental-health dimensions rather than qualitatively distinct categories[50,55].*

## Figure S8

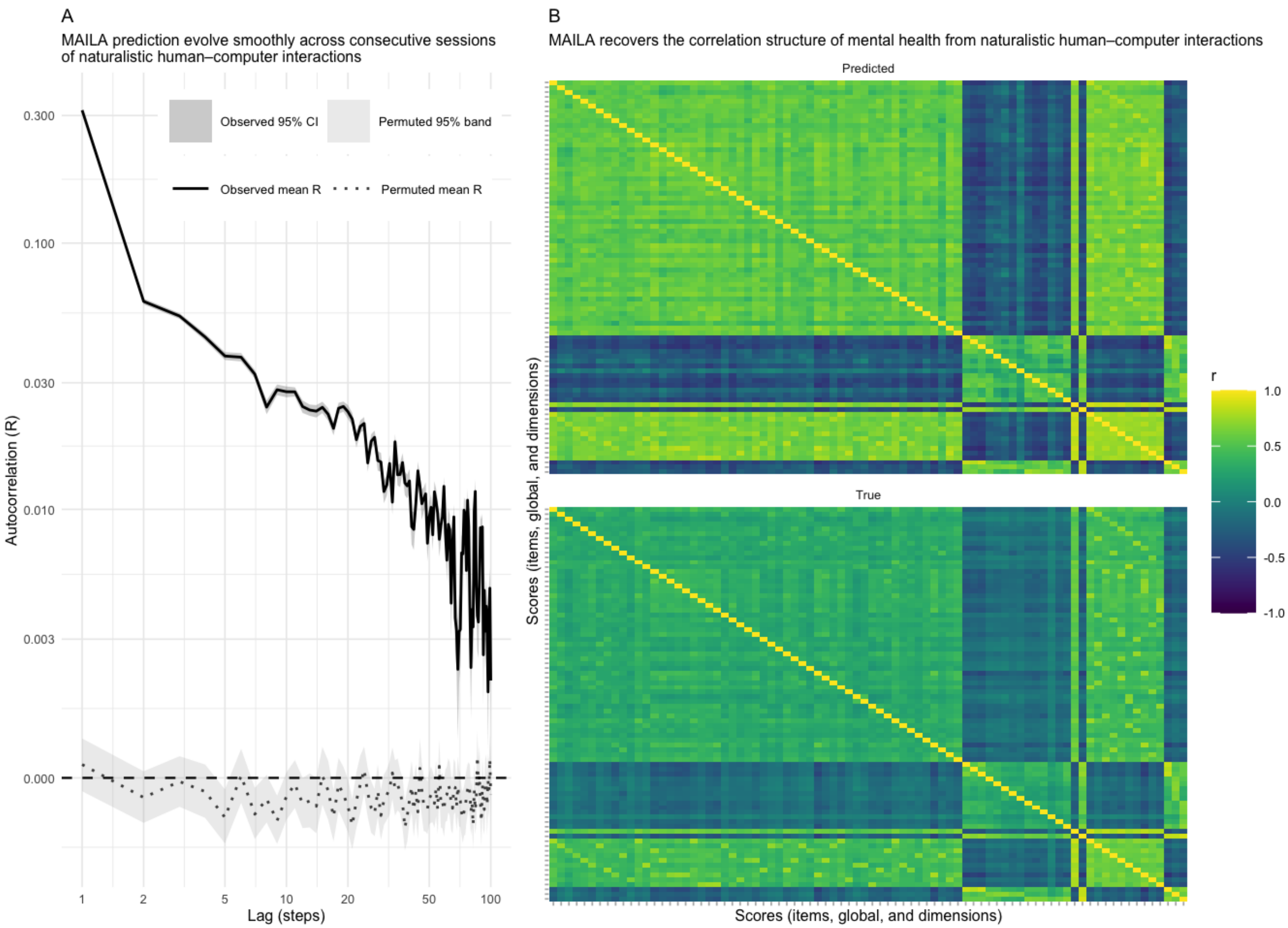

***Figure S8. Temporal and structural generalization to naturalistic cursor movement.*** *MAILA was trained on cursor movements from the baseline and follow-up datasets and applied, without retraining, to naturalistic cursor activity from 19 individuals, each contributing multiple sessions recorded across an extended period of time and at multiple times of the day[54]. While these analyses rely on unlabeled data, the temporal continuity (**A**) and structural consistency (**B**) of MAILA's predictions provide an indirect validation for the embedding of meaningful, generalizable dimensions of mental states in everyday human–computer interaction. In addition, **Figure 2F** shows that MAILA recovers known circadian fluctuations in mental state from this external dataset (higher positive affect in the morning and a rise in negative affect toward nightfall). These diurnal patterns have been independently reported before[61] and are also present in the MAILA dataset (ground truth and prediction, **Figure 2D-E**), providing external evidence that MAILA's predictions preserve expected temporal and covariance structure during open-ended computer use.*

***A. Autocorrelation of predicted mental health.*** *Predicted mental-health scores (pooled across participants and items) exhibited a significant positive autocorrelation that decayed monotonically with increasing lag on a log-scaled x-axis. The observed mean (solid line) remained above the participant-wise time-shuffled null (dashed line), with non-overlapping 95% confidence intervals at short lags and convergence toward zero at longer lags. This pattern supports the interpretation that naturalistic cursor movements reflect temporally coherent, slowly evolving latent mental states. The autocorrelation of MAILA's predictions extracted from naturalistic cursor movements closely mirrored that of self-reported scores in the MAILA dataset, where test-retest correlations declined from R = 0.88 for distress and R = 0.84 for wellbeing after one week to R = 0.69 and R = 0.69, respectively, after eight weeks.*

***B. Correlation structure of true and predicted mental health.*** *To compare the internal structure of mental health across datasets, we z-scored each item separately for true and predicted scores and computed the pairwise correlations between items. The resulting correlation matrices were highly similar (R = 0.95), indicating that predictions derived from naturalistic cursor movements preserved the inter-item structure of mental health observed in the MAILA dataset.*

# Figure S9

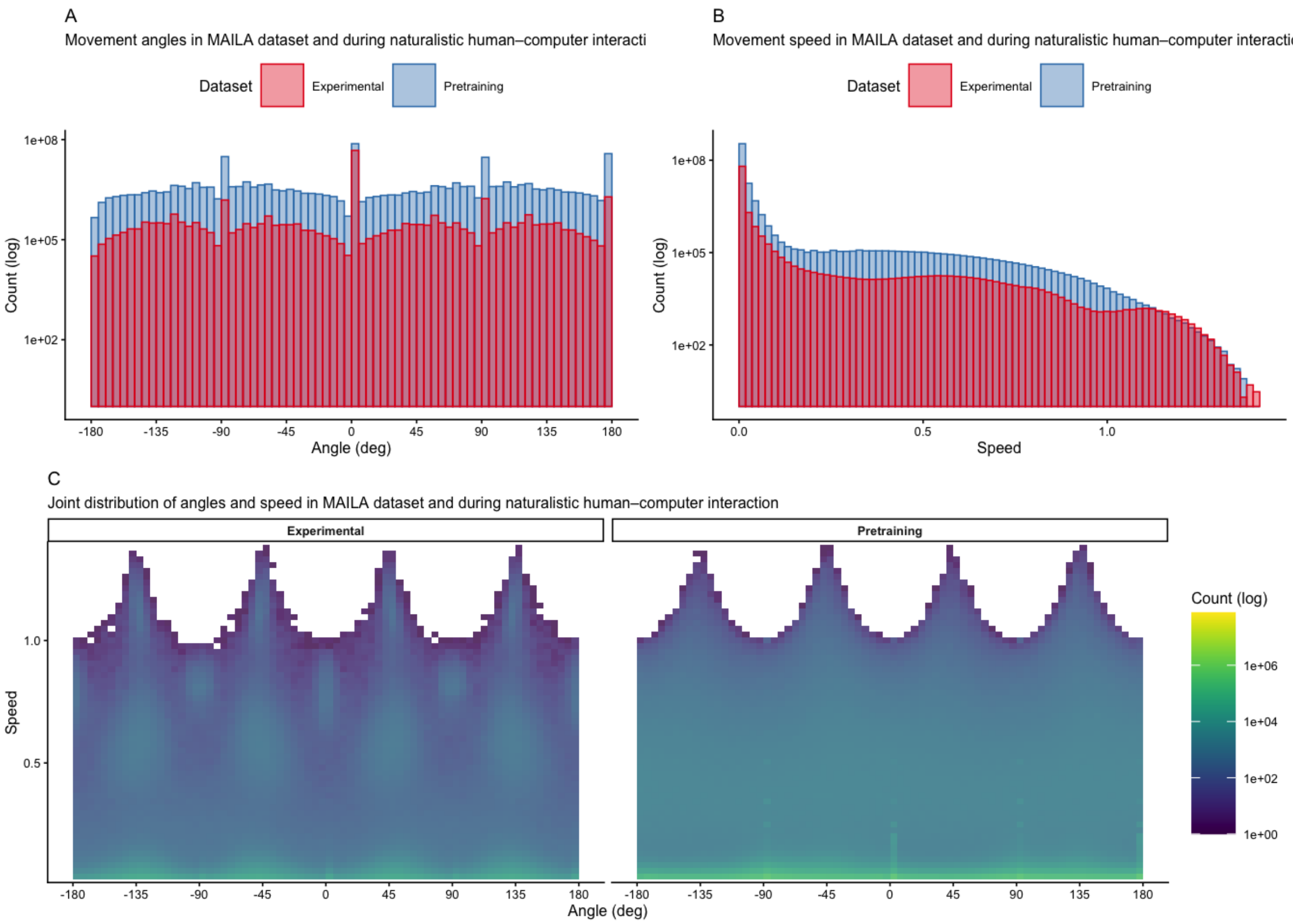

***Figure S9. Distribution of cursor movement angles and speeds in the MAILA dataset and everyday cursor movements.***

***A. Angles.*** *Histograms show the distribution of cursor movement directions (in degrees from -180° to 180°). The MAILA dataset (red) is compared to everyday cursor movements[54] (blue). The overlapping distributions indicate that MAILA captures the natural range of movement directions typically observed during everyday computer use.*

***B. Speeds.*** *Histograms show the distribution of cursor speed (log-scaled y-axis). The MAILA dataset (red) and everyday cursor movements (blue) exhibit highly similar profiles, suggesting comparable dynamics of cursor motion speed across experimental and naturalistic settings.*

***C. Joint distribution of angle and speed.*** *Heatmaps show the logarithmic density of cursor movements as a function of direction and speed for both datasets. Color intensity reflects the frequency of a specific combination of direction and speed (log-scaled). The similarity in structure across datasets indicates that MAILA's response interface reproduces core features of natural cursor trajectories.*

# Figure S10

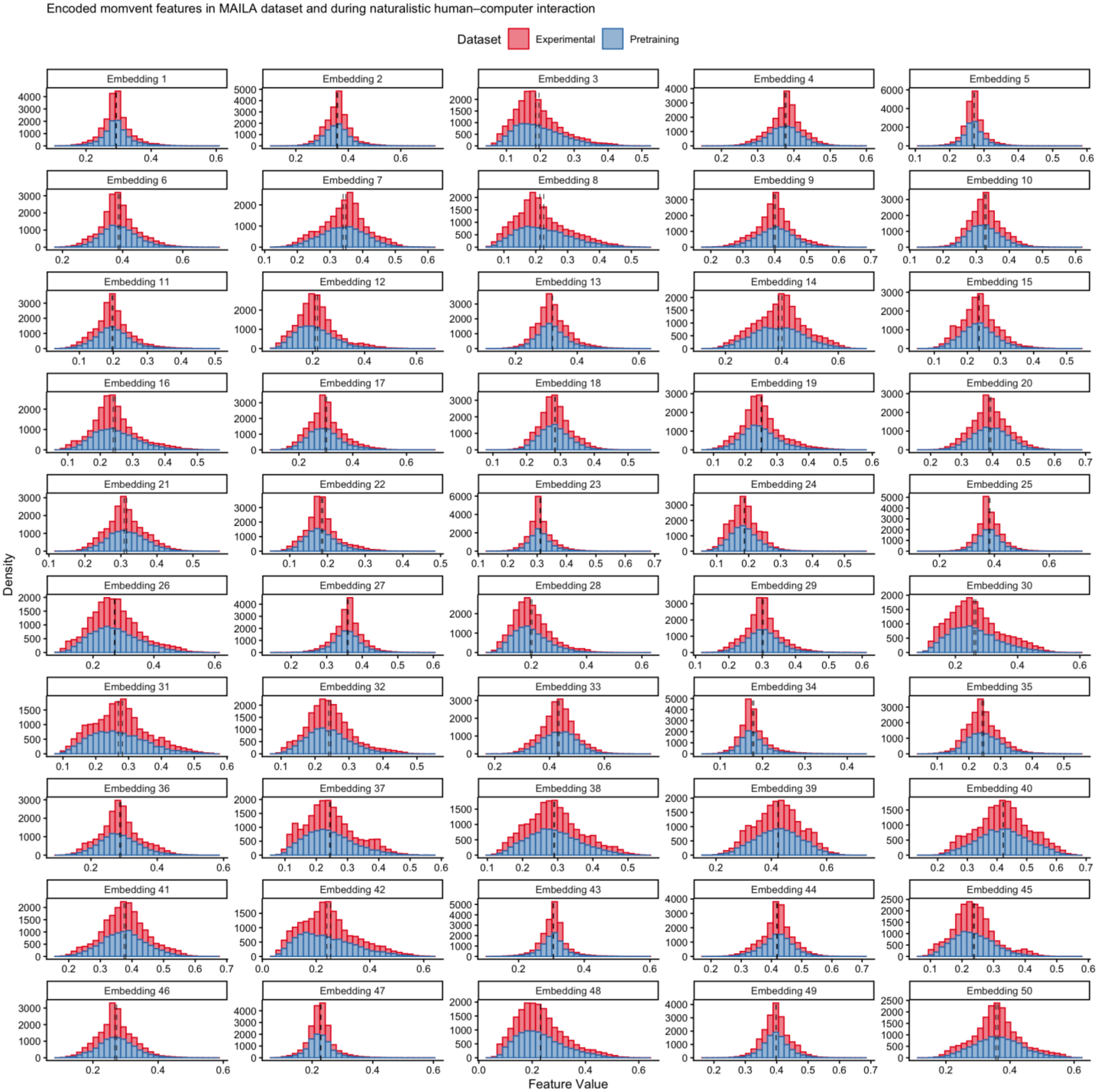

***Figure S10. Feature space similarity between the MAILA dataset and everyday cursor movements.*** *We compared LSTM embeddings from the MAILA dataset (experimentally induced cursor movements, red) to those from a public dataset of everyday computer use[54] (blue). Across all features, the distributions showed substantial overlap. On average, MAILA embeddings differed by only 0.94 ± 0.67% of the respective feature range, and 99.97 ± 0.05% of MAILA embeddings fell within the bounds of the pretraining distribution. Variance was slightly lower in the MAILA dataset compared to the naturalistic dataset ($\Delta_{var}$ = -0.00022 ± 0.00049). Together, these results suggest that, despite our dataset being experimental, it remained broadly representative of everyday computer use.*

# Figure S11

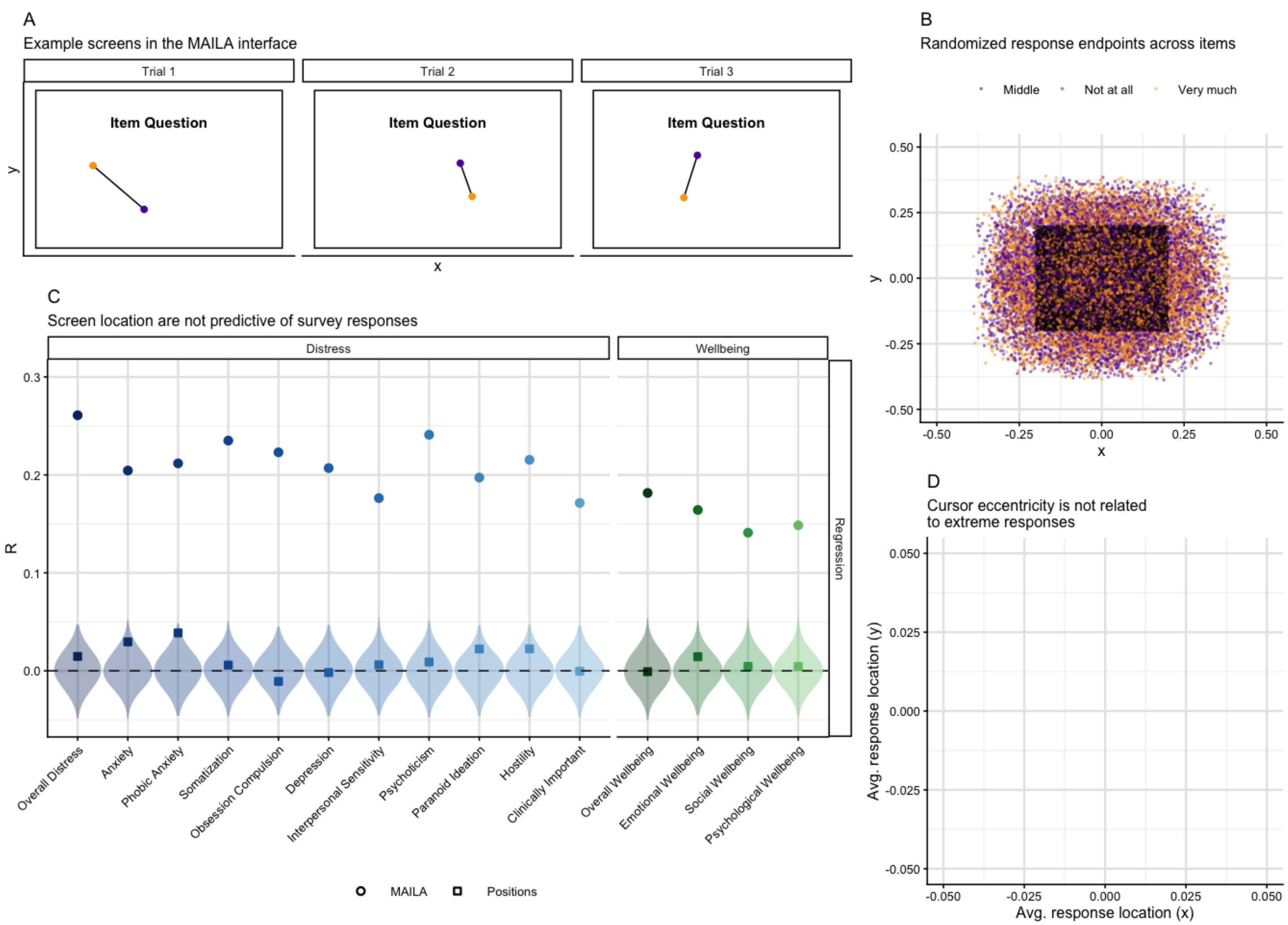

***Figure S11. Questionnaire paradigm.***

***A. Task interface and randomized response mapping.*** *Example response screens from the web-based questionnaire paradigm. Each panel shows one trial in which the response line appears at a random location and orientation. The line is flanked by two color-coded anchors: "Not at all" (illustrated here in purple) and "Very much" (illustrated here in orange). The item prompt (e.g., "How much are you distressed by feeling blue?") appears at the top of the screen. The spatial position, orientation, and length of the response line were independently randomized on every trial and for every participant, ensuring that motor behavior could not trivially encode the intended response. Note: In the actual experiment, the anchors were shown in green ("Not at all") and blue ("Very much").*

***B. Spatial distribution of response endpoints.*** *Screen coordinates of response-line midpoints and endpoints across 10,000 simulated trials. Endpoints labeled "Not at all" (purple) and "Very much" (orange) are symmetrically arranged around randomized center positions; midpoints (never displayed) are shown in black. This randomized spatial encoding prevents raw pointer coordinates from carrying systematic information about the meaning of participants' responses.*

***C. Regression analysis.*** *To further confirm that screen positions did not permit trivial decoding of mental health, even when considering human–computer interactions recorded during survey completion, we trained support vector regression models to predict self-reported mental health from x and y screen coordinates. Their cross-validated performance did not exceed permuted baselines and remained well below MAILA's movement-based features ($R = 0.01 \pm 0.0073$; round versus square markers). These controls complement analyses where we applied frozen MAILA models to non-mental-health settings (non-psychological survey and gamified decision-making* *experiment) and to free-form digital activity without any link to self-reports (Figure 2), confirming that successful decoding of mental health relied on movement features rather than trivial spatial cues.*

***D. Self-reports versus screen positions.*** *The plot shows average x and y response cursor positions per participant, colored by average of the associated self-reports. The uniform color distribution indicates that eccentricity was not correlated with the self-reports ($R = -0.0015$, $p = 0.93$). Please note that, as additional safeguards against trivial decoding, MAILA received the entirety of the cursor or touch trajectory as its input, without any labeling of the screen position of the response, or at what point in the recording a specific mental health item was presented (random order of intermixed distress and wellbeing items).*

# Figure S12

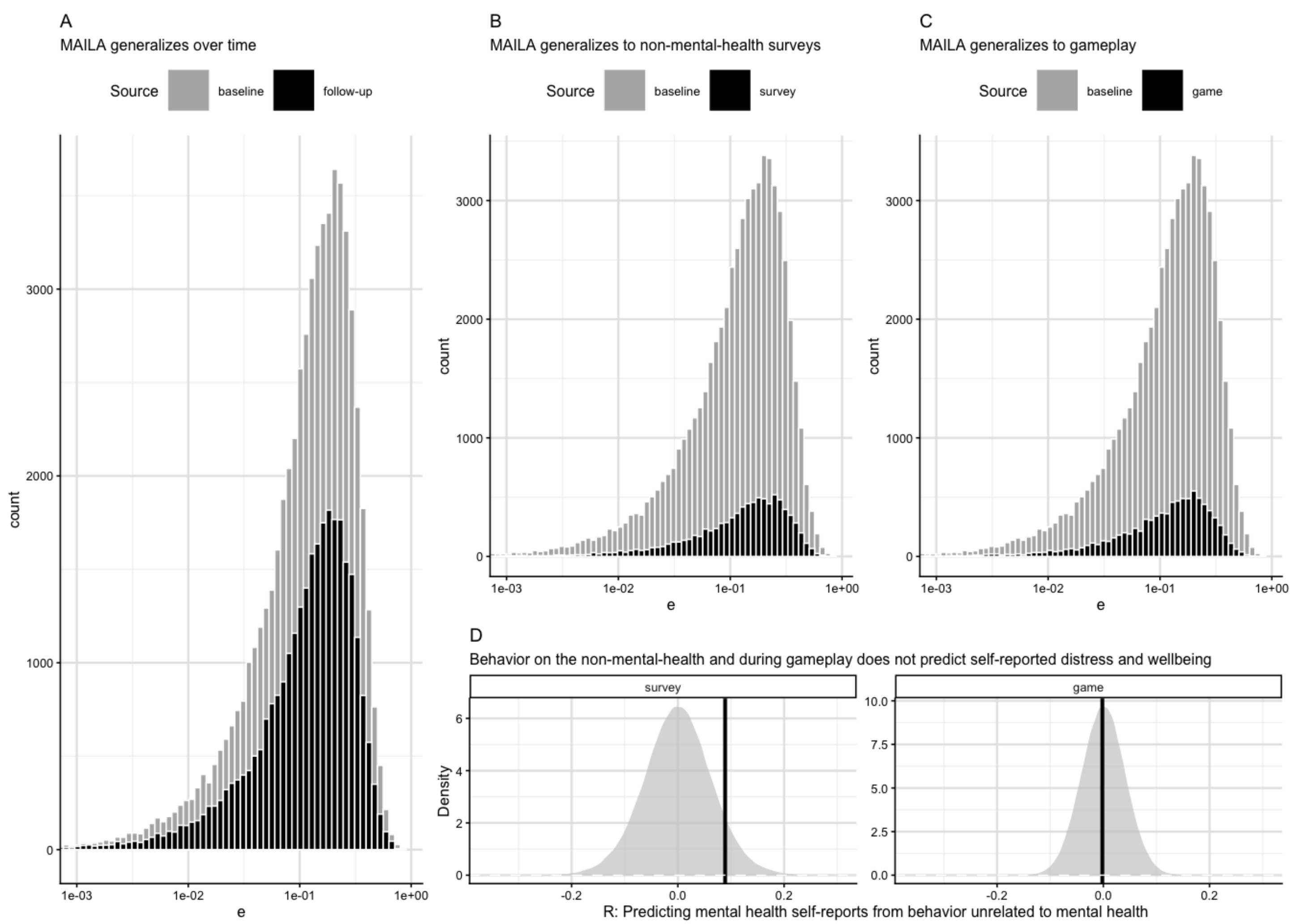

***Figure S12. Cursor-based mental health predictions generalize across contexts that are not related to mental health.***

***A.-C. Error distributions across contexts.*** *Distributions of normalized root mean squared errors (e) are shown on a logarithmic scale for the baseline (grey) and three generalization datasets (black). Baseline-trained models were applied without retraining to follow-up recordings (**A**, N = 2,000). For the cross-context analyses (**B–C**), models calibrated on baseline and follow-up mental-health-task recordings were applied, without training on the alternate-task data, to cursor movements recorded during a non-psychological survey and an interactive game completed by two non-overlapping subgroups of 600 follow-up participants each. Compared to baseline cross-validation, prediction errors decreased by 0.67 ± 1.16% when frozen MAILA models were applied to the follow-up data ($p = 0.26$). Prediction errors increased by 3.66 ± 1.61% when mental health was inferred from cursor movements during survey completion ($p < 0.001$), and decreased by 0.43 ± 3.25% for predictions based on gameplay ($p = 0.69$). The overlapping error distributions indicate that prediction errors remained similar across time, task, and behavioral context.*

***D. Predicting mental-health self-reports from responses in the non-mental health survey and gameplay****. In this control analysis, we confirmed that items from the generalization experiments (non-mental health survey and game) did not carry any above-chance information about the participants' mental-health self-reports at follow-up. We trained linear models to predict each of the 67 mental-health items from each non-mental-health item using 5-fold cross-validation, and quantified prediction accuracy using the Spearman correlation between predicted and observed responses. The grey density curves show the permuted null distributions; solid vertical lines indicate the empirical cross-validated correlations, averaged across all predictor-target pairs (survey: $R = 0.09$, $p = 0.081$; game:*

*-0.0019, p = 0.52). For both the non-mental health survey (left) and the game (right), empirical correlations did not exceed the permuted null, indicating that responses to non-mental-health items did not provide above-chance information about the participants' mental-health self-reports.*

***Figure S11** and **Table S2** highlight two additional safeguards indicating that MAILA's generalization performance relied on movement patterns that transferred across contexts rather than any trivial association with task content. First, the randomized response mapping eliminated any direct relationship between pointer coordinates and participants' answers (Figure S11). Second, MAILA did not learn to predict the non–mental-health survey items themselves ($R = 0.01 \pm 0.03$, $p = 0.28$; Table S2) or gameplay behavior ($R = 0.02 \pm 0.02$, $p = 0.089$).*

# Figure S13

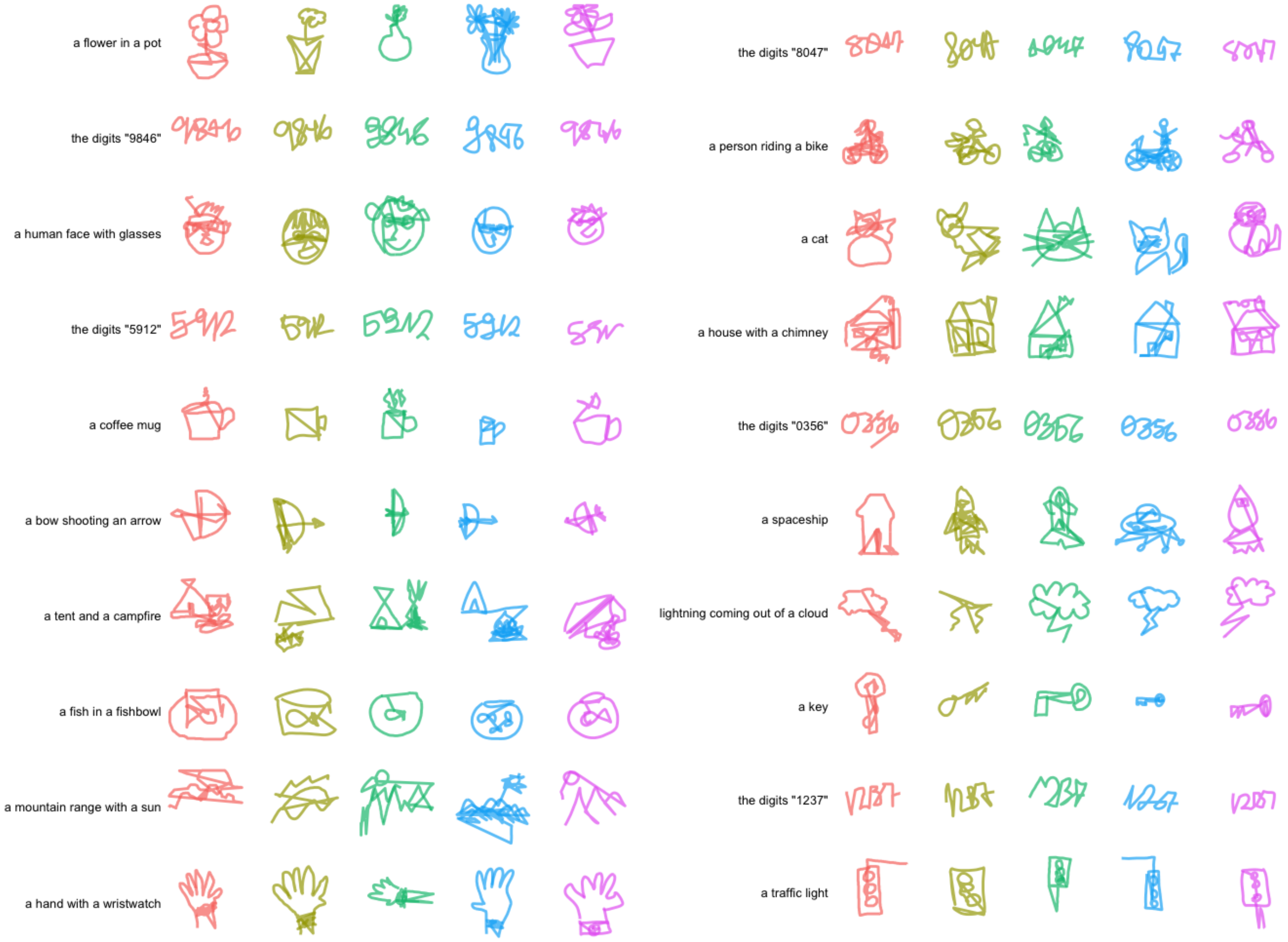

***Figure S13. Example drawings.*** *Free-form touchscreen drawings from five randomly selected participants, with prompts displayed to the left. See Table S3 for a list of prompts.*

# Figure S14

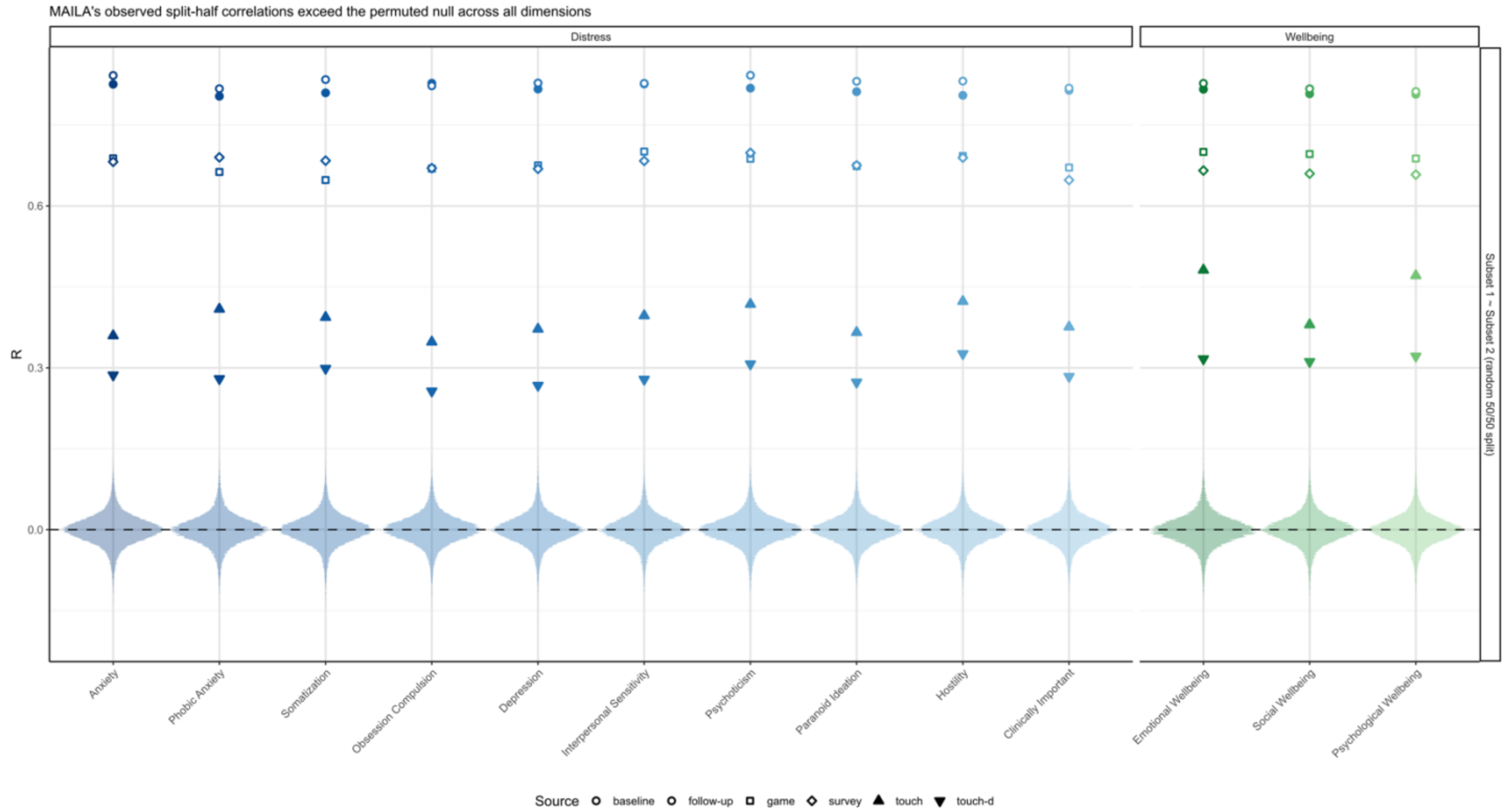

***Figure S14. Split-half reliability of MAILA across datasets and dimensions.*** *To assess the internal reliability of MAILA, we divided each participant's cursor or touchscreen trajectories into randomized 50/50 subsets and trained support vector regression models on one half to predict self-reported mental health in the other. We repeated this procedure in the reverse direction, yielding two independent prediction vectors per participant. Violin plots depict permutation-based null distributions obtained by shuffling one split. Filled markers denote correlation coefficients from unseen participants in 5-fold cross-validation (baseline, touch, and touch-d); unfilled markers denote correlation coefficients when frozen MAILA models were applied to later or alternate-task recordings from the same participants (follow-up, survey, and game). MAILA's predictions were highly consistent across randomized halves (R = 0.61 ± 0.05, across all dimensions and datasets), demonstrating a level of reliability that exceeded most behavioral and neuroimaging markers of mental health.*

# Figure S15

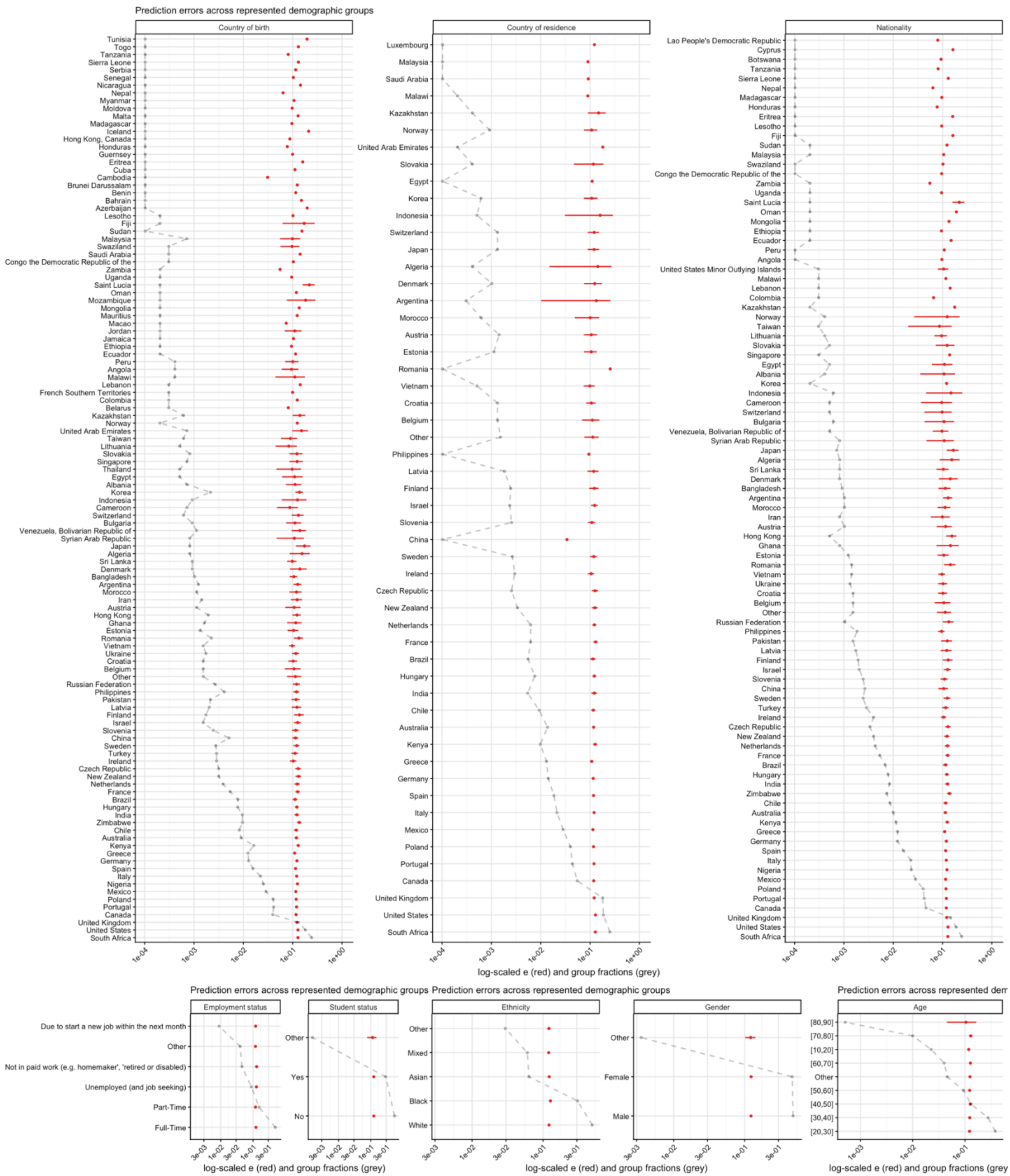

***Figure S15. Prediction errors across demographic groups.*** *MAILA's prediction errors pooled over the first three principal components of mental health (log-scaled normalized root mean squared error, e), shown as mean ± 95% confidence intervals across participants. Errors are shown in red, and the log-scaled fraction of individuals within*

*each demographic category is overlaid in grey. To assess demographic effects jointly, we fitted a linear mixed-effects model to participant-level prediction errors across PC1–PC3, entering mental-health component and all demographic characteristics simultaneously as fixed effects and participant as a random intercept. Type III ANOVAs and partial $\eta^2$ quantified the unique prediction-error variance associated with each demographic characteristic conditional on all other model terms. There was no significant effect of gender ($F = 0.63$, $p = 0.67$; partial $\eta^2 = 0.03\%$), ethnicity ($F = 0.32$, $p = 0.92$; partial $\eta^2 = 0.02\%$), country of birth ($F = 1.1$, $p = 0.23$; partial $\eta^2 = 1.22\%$), country of residence ($F = 1.04$, $p = 0.4$; partial $\eta^2 = 0.53\%$), nationality ($F = 1.06$, $p = 0.35$; partial $\eta^2 = 0.75\%$), student status ($F = 0.91$, $p = 0.4$; partial $\eta^2 = 0.02\%$), or age ($F = 1.66$, $p = 0.11$; partial $\eta^2 = 0.12\%$). MAILA's prediction errors varied significantly with employment status ($F = 3.71$, $p = 0.0011$; partial $\eta^2 = 0.22\%$), and there was a borderline significant effect of language ($F = 1.34$, $p = 0.051$; partial $\eta^2 = 0.67\%$).*

# Figure S16

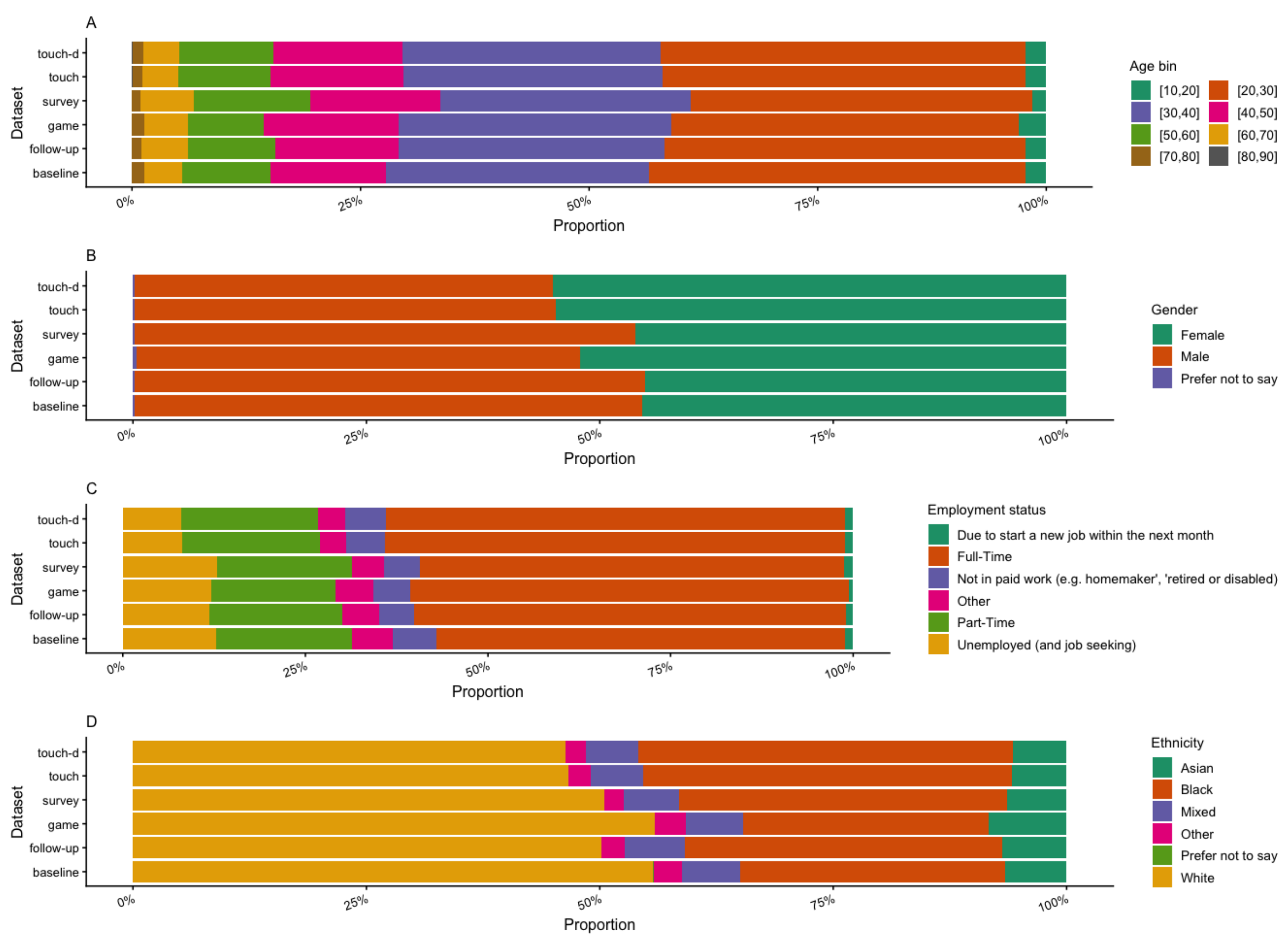

***Figure S16. Age, gender, employment, and ethnicity composition between MAILA's datasets.*** *Stacked bar plots show the proportional distribution of participants across (**A**) age bins, (**B**) gender, (**C**) employment status, and ethnicity (**D**), separately for MAILA's dataset.*

# Figure S17

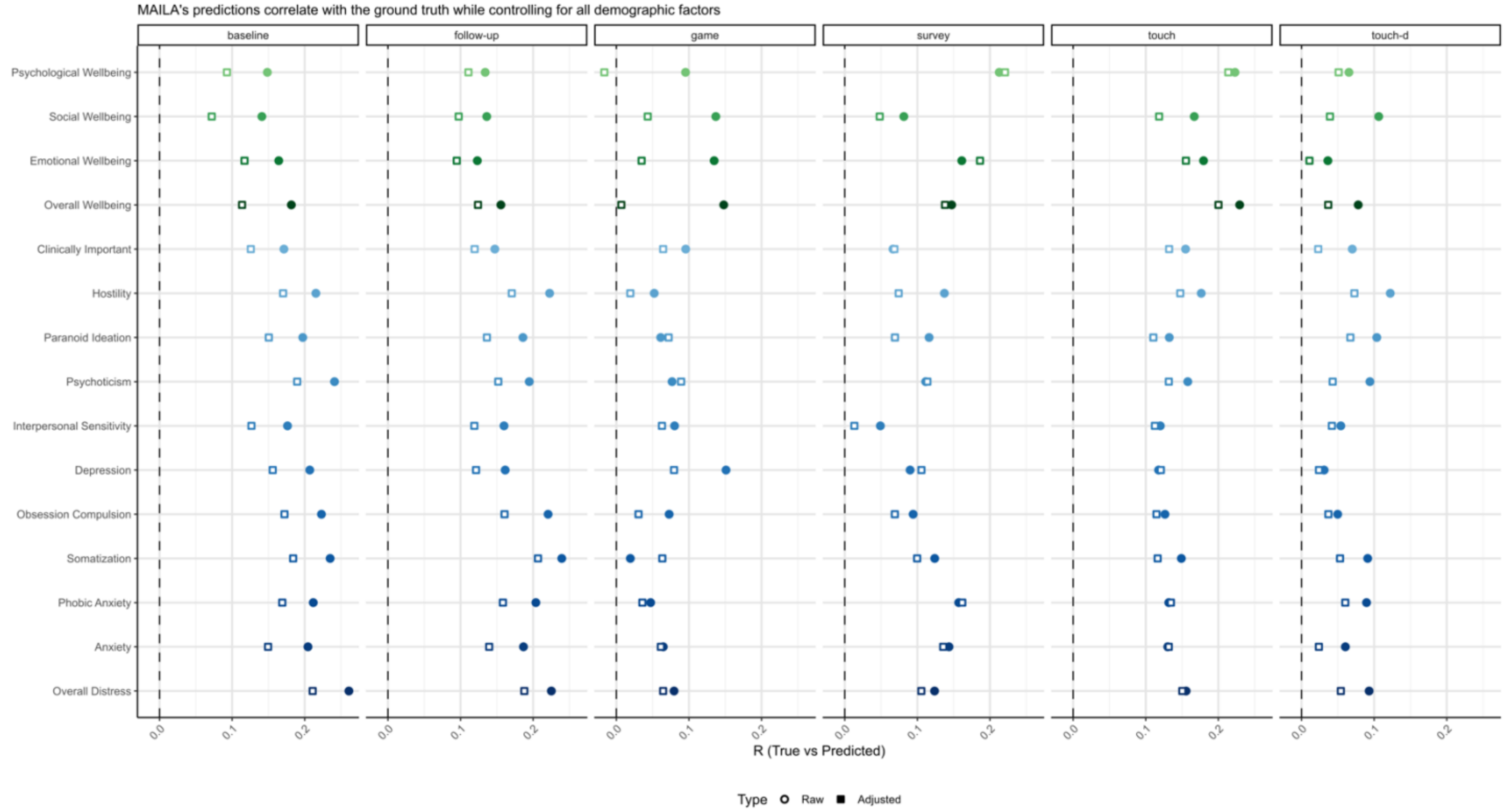

***Figure S17. Demographic confound control for MAILA's mental-health predictions.*** *Points show correlations (R) between MAILA predictions and ground-truth self-reports across mental-health dimensions, shown separately for each dataset. Filled markers show raw correlations; hollow markers show correlations after residualizing both predicted and true scores with respect to all available demographic covariates (partial correlation controlling for age, gender, ethnicity, country of birth and residence, nationality, language, student and employment status, task completion time, and the number of prior approved Prolific submissions). Across datasets and dimensions, adjusted correlations remained greater than zero (mean R = 0.1 ± 0.01, T(89) = 17.67, p < 0.001), showing that the predictive association remained after adjustment for all measured demographic and digital-engagement covariates.*

# Figure S18

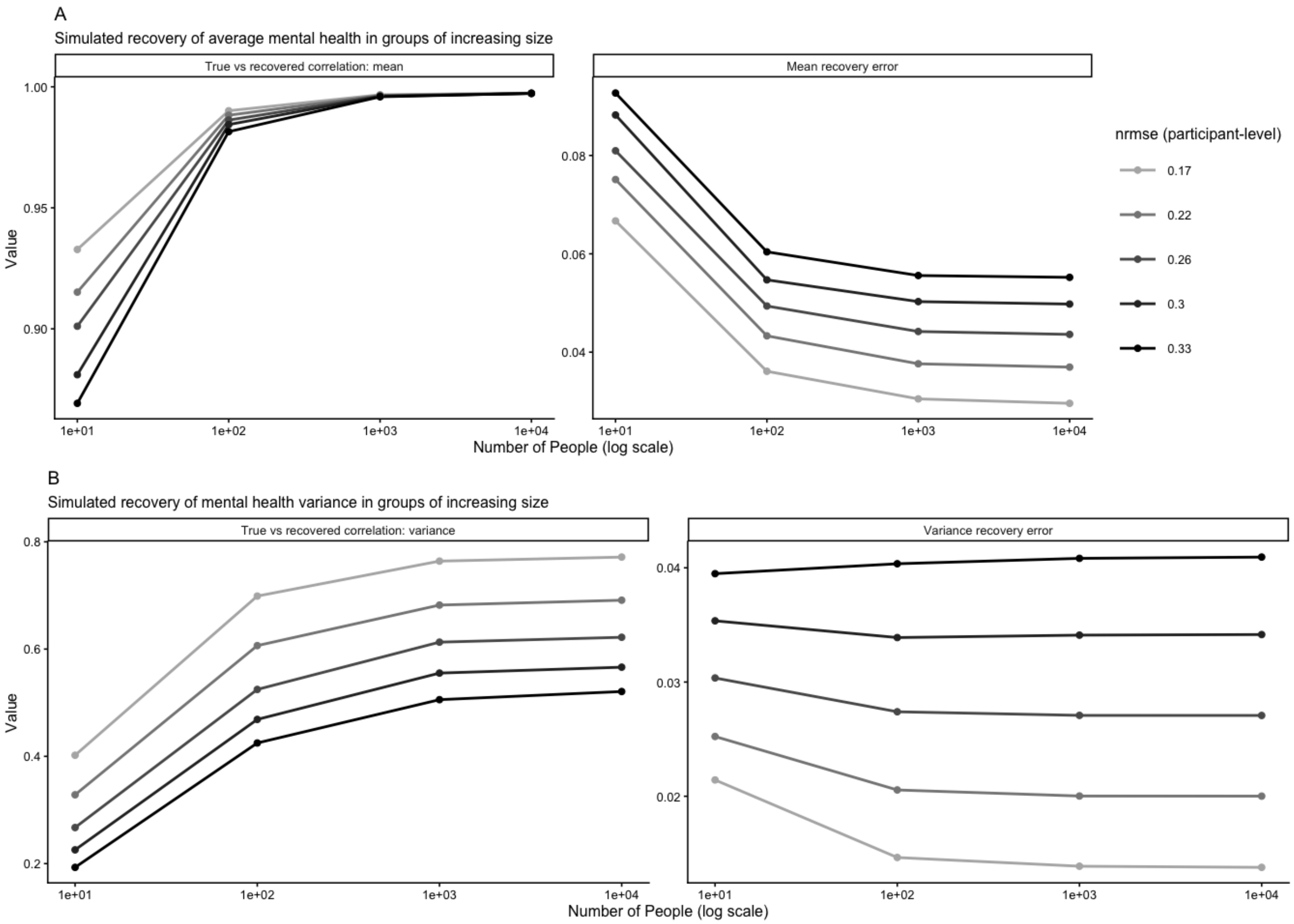

***Figure S18. Simulated recovery of population-level mental health.*** *To assess whether human–computer interactions encode information about group-level mental health, we simulated populations of group size N, each defined by a specific mean and variance of a ground truth mental-health feature. We then simulated MAILA's predictions, with noise levels defined by normalized mean squared errors (e) ranging from 17% to 35% (matching MAILA's prediction errors). The resulting mental health predictions were clipped to the unit interval to ensure they remained within the bounds of mental health scores. We then recovered the population means (**A**) and variances (**B**) from the noisy participant-level predictions and compared them to the ground truth of the simulated populations. The law of large numbers predicts that accuracy improves as a function of population size and the inverse of the prediction error.*

***A. Recovery of the group-level mean.*** *The left panel shows the correlation between the true and recovered population means. The right panel presents the corresponding absolute recovery error. The x-axis represents population size on a logarithmic scale, illustrating how larger sample sizes improve group-level accuracy. Shades of grey represent different error levels, which were informed by the range of prediction errors observed for items, dimensions, and global scores of distress and wellbeing in the MAILA dataset.*

***B. Recovery of the group-level variance.*** *The left panel shows the correlation between the true and recovered population variances. The right panel presents the corresponding absolute error when recovering the variance.*

## Figure S19

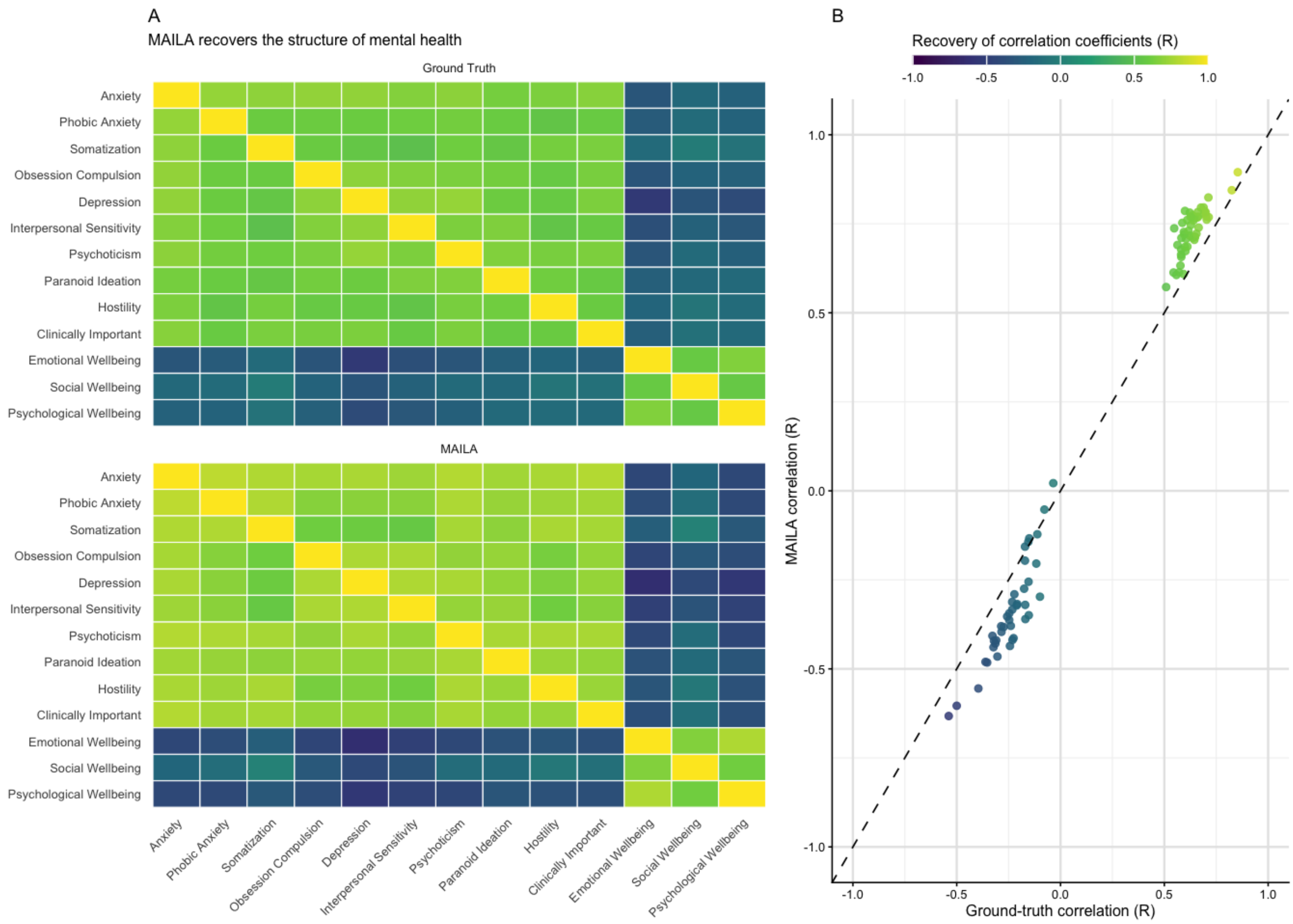

***Figure S19. Overlapping correlation structure between MAILA and the ground truth.***

***A. MAILA recovers the correlation structure of ground-truth mental health.*** *Heatmaps show pairwise correlations among mental-health dimensions in the ground truth (left) and in MAILA's predictions (right). Although MAILA was trained with separate support-vector regressions for each dimension, its predictions closely reproduced the correlation structure of the true self-reports ($R = 0.97$, $p < 0.001$). Correlation coefficients deviated from the ground-truth structure by only 5.32% of the possible range ($p < 0.001$). This indicates that human–computer interactions encode shared latent dimensions of mental health.*

***B. Agreement between ground-truth and MAILA correlations.*** *Scatter plot comparing the pairwise correlation coefficients between mental-health dimensions derived from ground-truth data (x-axis) and MAILA's predictions (y-axis). Each point represents one unique pair of dimensions (e.g., anxiety ~ depression), colored by the corresponding ground-truth correlation strength. The diagonal marks perfect agreement. Points for negative ground-truth correlations lie mostly below the diagonal and positive ones mostly above, indicating that MAILA tends to overestimate the magnitude of inter-dimensional correlations overall (paired t-test on $|r|$: $p < 0.001$).*

# Figure S20

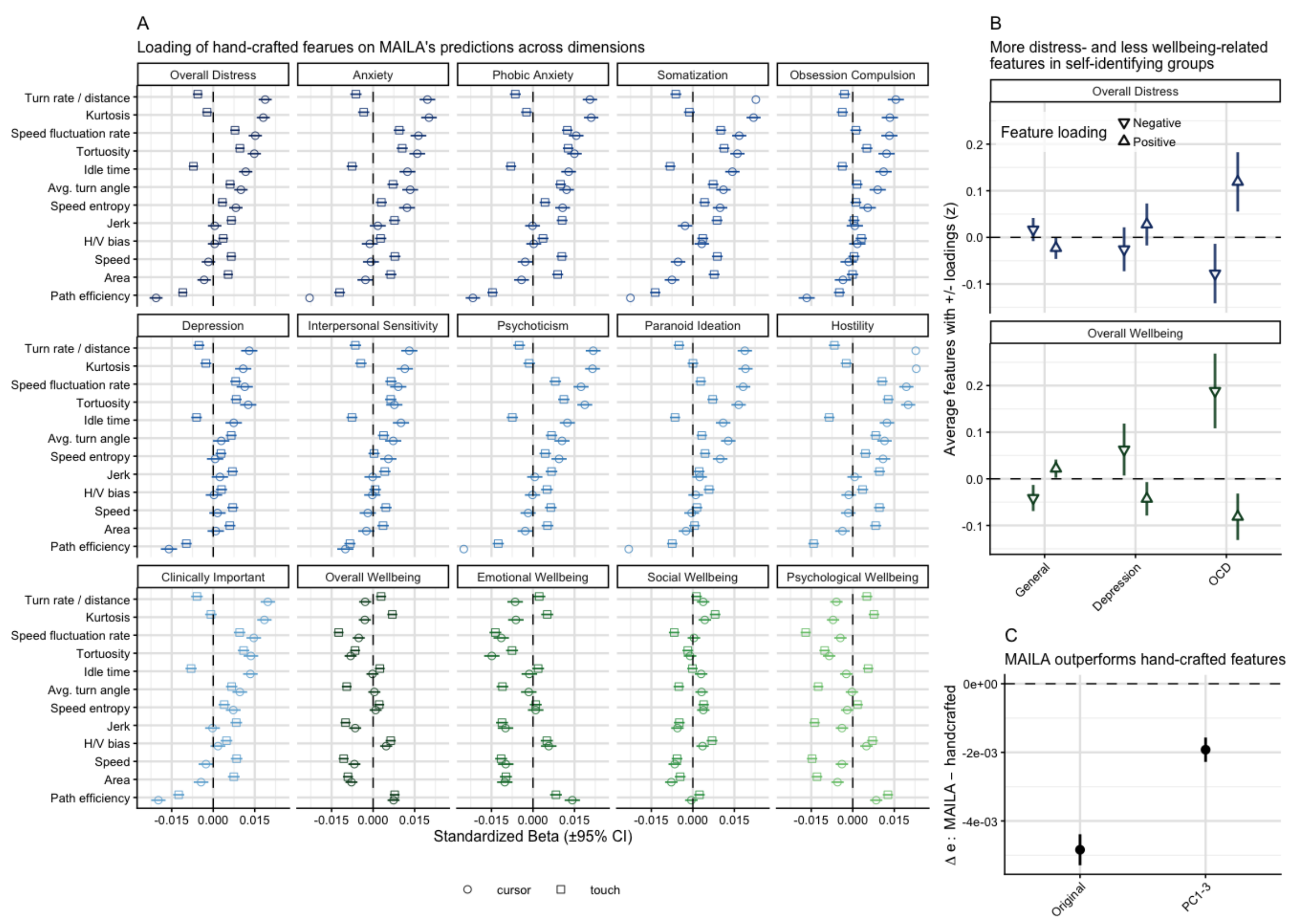

***Figure S20. Interpreting MAILA.*** *Like many data-driven models, MAILA learns predictive features that do not have immediate verbal interpretations. To explain its performance in terms of intuitive descriptors, we computed 12 established metrics of cursor and touchscreen activity and regressed them onto MAILA's predictions.*

***A. Associations between handcrafted movement features and MAILA's predictions in the general population.*** *We correlated participant-level handcrafted movement features with MAILA's predictions, shown here for overall distress in blue and overall wellbeing in green. Markers show standardized regression coefficients estimated separately for cursor- and touch-based datasets (circles vs. squares), with horizontal bars indicating 95% confidence intervals across datasets. Across modalities, higher distress and lower wellbeing were associated with more tortuous trajectories and greater variability in speed, whereas higher path efficiency predicted greater wellbeing and lower distress. Despite these broad consistencies, several handcrafted metrics showed substantial cross-modal heterogeneity: only 58.89% of features loaded onto predicted mental health in the same direction across cursor and touchscreen data.*

***B. Expression of features in self-identifying versus general participants.*** *For each population (participants who did or did not self-identify with depression and OCD), handcrafted features were grouped according to whether they loaded positively or negatively on predicted wellbeing or distress. Mean z-scored feature values (±95% CI) are plotted for positively loading (upward arrow) and negatively loading (downward arrow) feature groups. In the general population, features associated with lower distress and higher wellbeing were more strongly expressed. In participants who self-identified with depression or OCD, this pattern reversed for both distress ($p < 0.001$) and wellbeing ($p < 0.001$), indicating that interpretable aspects of human–computer interaction systematically tracked higher distress and lower wellbeing in people who endorsed living with mental illness.*

***C. Predictive advantage of MAILA over handcrafted feature models.*** *MAILA outperformed models built from handcrafted features across all benchmarks. In the original symptom space, MAILA achieved lower prediction errors for inter-individual differences in mental health in the general population ($p < 0.001$). Along PC1-3, which capture the level and specific causes of distress and wellbeing (**Figure 4**), MAILA also outperformed handcrafted models ($p < 0.001$). Together, these results demonstrate that MAILA provides more accurate and specific predictions of mental health than models based solely on handcrafted metrics.*

# Figure S21

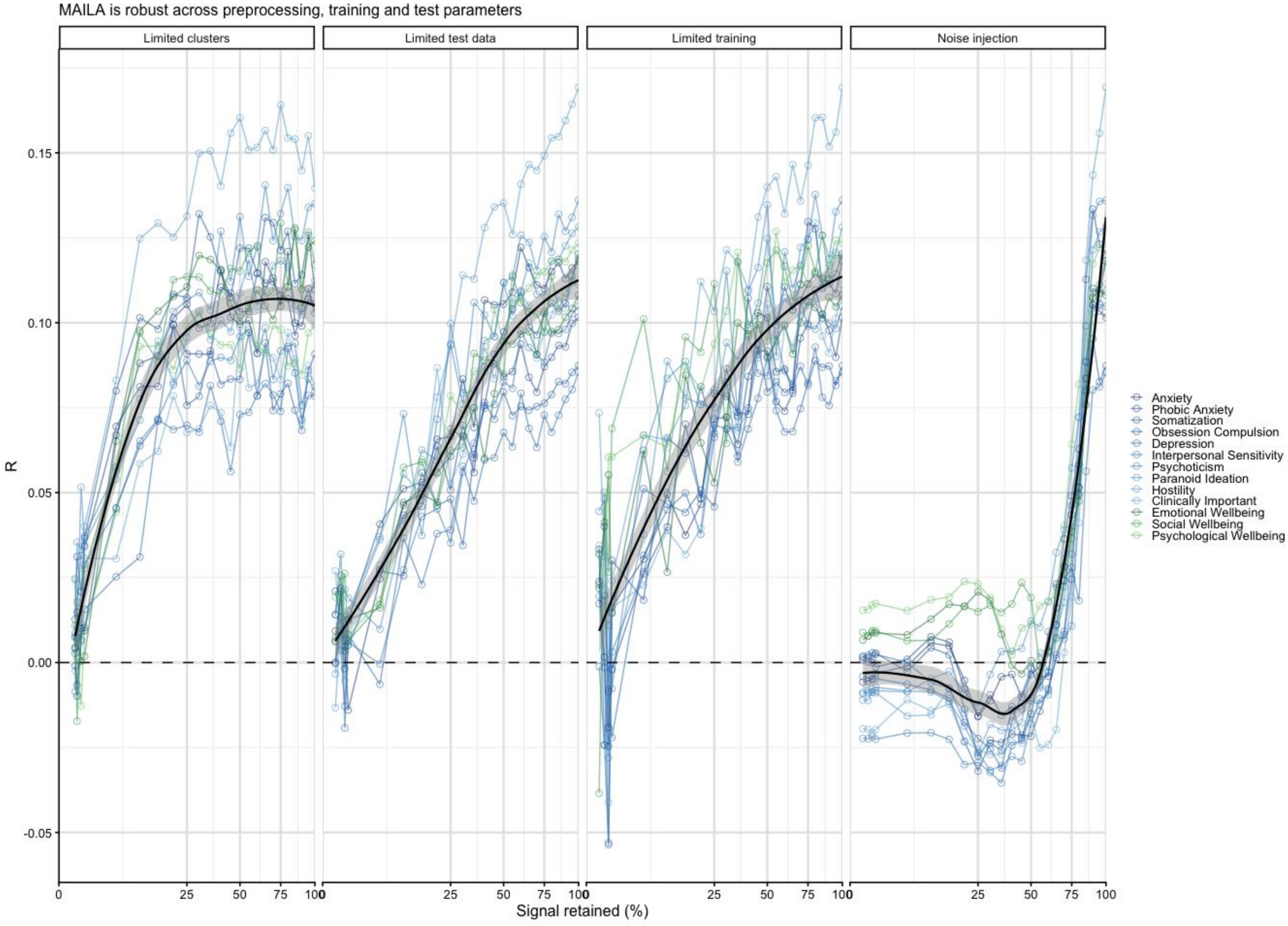

***Figure S21. Robustness of MAILA to information loss.*** *We systematically degraded MAILA in four ways (from left to right panel): (i) we limited the number of K-means clusters used to construct the movement feature matrix* $X^{N \times C}$*, simulating reduced behavioral diversity; (ii) we reduced the amount of human–computer interaction available per participant in the test folds by removing contiguous segments from each trajectory, simulating inferences from shorter cursor or touch recordings; (iii) we subsampled the number of participants in the training folds, simulating the effect of smaller calibration datasets; (iv) we corrupted cursor/touch trajectories by linearly mixing true samples with random values drawn from a uniform distribution. In all cases, correlations declined smoothly with increasing degradation, indicating predictable performance loss under impoverished behavior, limited test data, limited training data, or injected noise.*

## Table S1

| | MAILA's correlation to ground truth (R) | | | | | | |
|---|---|---|---|---|---|---|---|
| Question | Mean ± 95% CI (IQR) | baseline | follow -up | survey | game | touch-q | touch-d |
| **Distress - Anxiety** | | | | | | | |
| How much are you distressed by nervousness or shakiness inside? | 0.41 ± 0.02 (0.52) | 0.16 | 0.11 | 0.13 | -0.01 | 0.08 | 0.05 |
| How much are you distressed by suddenly being scared for no reason? | 0.36 ± 0.02 (0.48) | 0.12 | 0.15 | 0.12 | 0.06 | 0.12 | 0.06 |
| How much are you distressed by feeling fearful? | 0.41 ± 0.02 (0.49) | 0.13 | 0.11 | 0.07 | 0.07 | 0.08 | 0.04 |
| How much are you distressed by feeling tense or keyed up? | 0.45 ± 0.02 (0.49) | 0.13 | 0.10 | 0.07 | 0.08 | 0.09 | 0.03 |
| How much are you distressed by spells of terror or panic? | 0.35 ± 0.02 (0.47) | 0.13 | 0.15 | 0.13 | 0.10 | 0.08 | 0.06 |
| How much are you distressed by feeling so restless you could not sit still? | 0.36 ± 0.02 (0.48) | 0.14 | 0.14 | 0.11 | 0.06 | 0.11 | 0.05 |
| **Distress - Clinically Important** | | | | | | | |
| How much are you distressed by poor appetite? | 0.32 ± 0.02 (0.45) | 0.10 | 0.14 | 0.10 | 0.07 | 0.14 | 0.06 |
| How much are you distressed by trouble falling asleep? | 0.45 ± 0.02 (0.59) | 0.07 | 0.07 | 0.04 | 0.08 | 0.05 | 0.02 |
| How much are you distressed by thoughts of death or dying? | 0.42 ± 0.02 (0.56) | 0.06 | 0.11 | 0.01 | 0.02 | 0.11 | 0.00 |
| How much are you distressed by feeling of guilt? | 0.41 ± 0.02 (0.53) | 0.13 | 0.12 | 0.06 | 0.11 | 0.09 | 0.06 |
| **Distress - Depression** | | | | | | | |
| How much are you distressed by thoughts of ending your life? | 0.27 ± 0.02 (0.39) | 0.13 | 0.15 | 0.10 | 0.05 | 0.15 | 0.02 |
| How much are you distressed by feeling lonely? | 0.44 ± 0.02 (0.56) | 0.12 | 0.11 | 0.05 | 0.04 | 0.07 | 0.04 |
| How much are you distressed by feeling blue? | 0.45 ± 0.02 (0.55) | 0.12 | 0.11 | 0.14 | 0.04 | 0.10 | 0.00 |
| How much are you distressed by feeling no interest in things? | 0.43 ± 0.02 (0.54) | 0.13 | 0.13 | 0.02 | 0.08 | 0.05 | 0.02 |
| How much are you distressed by feeling hopeless about the future? | 0.49 ± 0.02 (0.57) | 0.14 | 0.10 | 0.04 | 0.20 | 0.06 | 0.00 |
| How much are you distressed by feelings of worthlessness? | 0.42 ± 0.02 (0.58) | 0.12 | 0.14 | 0.04 | 0.12 | 0.11 | 0.03 |

| | | MAILA's correlation to ground truth (R) | | | | | |
|---|---|---|---|---|---|---|---|
| **Question** | **Mean ± 95% CI (IQR)** | **baseline** | **follow -up** | **survey** | **game** | **touch-q** | **touch-d** |
| **Distress - Hostility** | | | | | | | |
| How much are you distressed by feeling easily annoyed or irritated? | 0.47 ± 0.02 (0.52) | 0.13 | 0.13 | 0.02 | 0.03 | 0.09 | 0.04 |
| How much are you distressed by temper outbursts that you could not control? | 0.35 ± 0.02 (0.48) | 0.13 | 0.18 | 0.12 | 0.03 | 0.09 | 0.06 |
| How much are you distressed by having urges to beat, injure, or harm someone? | 0.25 ± 0.01 (0.31) | 0.13 | 0.16 | 0.14 | 0.07 | 0.20 | 0.12 |
| How much are you distressed by having urges to break or smash things? | 0.27 ± 0.01 (0.36) | 0.14 | 0.22 | 0.15 | 0.03 | 0.17 | 0.06 |
| How much are you distressed by getting into frequent arguments? | 0.38 ± 0.02 (0.51) | 0.14 | 0.09 | 0.12 | -0.03 | 0.15 | 0.07 |
| **Distress - Interpersonal Sensitivity** | | | | | | | |
| How much are you distressed by your feelings being easily hurt? | 0.48 ± 0.02 (0.55) | 0.09 | 0.14 | 0.11 | 0.03 | 0.08 | 0.09 |
| How much are you distressed by feeling that people are unfriendly or dislike you? | 0.39 ± 0.02 (0.49) | 0.12 | 0.12 | 0.05 | 0.09 | 0.11 | 0.01 |
| How much are you distressed by feeling inferior to others? | 0.41 ± 0.02 (0.53) | 0.11 | 0.14 | 0.06 | 0.02 | 0.11 | 0.02 |
| How much are you distressed by feeling very self-conscious with others? | 0.50 ± 0.02 (0.49) | 0.11 | 0.09 | 0.04 | 0.04 | 0.03 | 0.01 |
| **Distress - Obsession-Compulsion** | | | | | | | |
| How much are you distressed by trouble remembering things? | 0.45 ± 0.02 (0.56) | 0.12 | 0.16 | 0.10 | 0.05 | 0.09 | -0.01 |
| How much are you distressed by feeling blocked in getting things done? | 0.51 ± 0.02 (0.53) | 0.13 | 0.13 | 0.13 | 0.03 | 0.07 | 0.03 |
| How much are you distressed by having to check and double check what you do? | 0.49 ± 0.02 (0.52) | 0.12 | 0.11 | 0.11 | 0.01 | 0.09 | 0.02 |
| How much are you distressed by difficulty making decisions? | 0.47 ± 0.02 (0.52) | 0.16 | 0.15 | 0.01 | 0.04 | 0.09 | 0.06 |
| How much are you distressed by your mind going blank? | 0.40 ± 0.02 (0.51) | 0.13 | 0.12 | 0.03 | -0.04 | 0.08 | 0.05 |
| How much are you distressed by trouble concentrating? | 0.47 ± 0.02 (0.54) | 0.16 | 0.21 | 0.05 | 0.10 | 0.10 | 0.03 |

| | MAILA's correlation to ground truth (R) | | | | | | |
|---|---|---|---|---|---|---|---|
| **Question** | **Mean ± 95% CI (IQR)** | **baseline** | **follow -up** | **survey** | **game** | **touch-q** | **touch-d** |
| **Distress - Paranoid Ideation** | | | | | | | |
| How much are you distressed by feeling others are to blame for most of your troubles? | 0.33 ± 0.01 (0.40) | 0.12 | 0.17 | 0.11 | 0.09 | 0.13 | 0.02 |
| How much are you distressed by feeling that most people cannot be trusted? | 0.50 ± 0.02 (0.48) | 0.09 | 0.12 | 0.08 | 0.08 | 0.07 | 0.08 |
| How much are you distressed by feeling that you are watched or talked about by others? | 0.38 ± 0.02 (0.51) | 0.12 | 0.10 | 0.15 | 0.02 | 0.08 | 0.04 |
| How much are you distressed by others not giving you proper credit for your achievements? | 0.43 ± 0.02 (0.49) | 0.14 | 0.15 | 0.09 | 0.02 | 0.07 | 0.03 |
| How much are you distressed by feeling that people will take advantage of you if you let them? | 0.52 ± 0.02 (0.51) | 0.08 | 0.11 | 0.06 | 0.02 | 0.04 | 0.07 |
| **Distress - Phobic Anxiety** | | | | | | | |
| How much are you distressed by feeling afraid in open spaces? | 0.30 ± 0.02 (0.40) | 0.14 | 0.17 | 0.12 | 0.09 | 0.10 | 0.08 |
| How much are you distressed by feeling afraid to travel on buses, subways, or trains? | 0.31 ± 0.02 (0.43) | 0.14 | 0.12 | 0.11 | 0.06 | 0.15 | 0.07 |
| How much are you distressed by having to avoid certain things, places, or activities because they frighten you? | 0.42 ± 0.02 (0.53) | 0.10 | 0.15 | 0.14 | 0.06 | 0.12 | 0.05 |
| How much are you distressed by feeling uneasy in crowds? | 0.45 ± 0.02 (0.55) | 0.08 | 0.08 | 0.10 | -0.03 | 0.07 | 0.05 |
| How much are you distressed by feeling nervous when you are left alone? | 0.32 ± 0.02 (0.44) | 0.13 | 0.18 | 0.14 | 0.02 | 0.14 | 0.09 |
| **Distress - Psychoticism** | | | | | | | |
| How much are you distressed by the idea that someone else can control your thoughts? | 0.31 ± 0.02 (0.46) | 0.16 | 0.20 | 0.11 | 0.08 | 0.16 | 0.09 |
| How much are you distressed by feeling lonely even when you are with people? | 0.43 ± 0.02 (0.53) | 0.12 | 0.14 | 0.10 | 0.04 | 0.10 | 0.01 |
| How much are you distressed by the idea that you should be punished for your sins? | 0.33 ± 0.02 (0.45) | 0.19 | 0.18 | 0.11 | 0.02 | 0.13 | 0.08 |
| How much are you distressed by never feeling close to another person? | 0.41 ± 0.02 (0.52) | 0.08 | 0.09 | 0.08 | 0.07 | 0.13 | -0.02 |

| | MAILA's correlation to ground truth (R) | | | | | | |
|---|---|---|---|---|---|---|---|
| **Question** | **Mean ± 95% CI (IQR)** | **baseline** | **follow -up** | **survey** | **game** | **touch-q** | **touch-d** |
| How much are you distressed by the idea that something is wrong with your mind? | 0.41 ± 0.02 (0.56) | 0.19 | 0.17 | 0.07 | 0.13 | 0.09 | 0.03 |
| **Distress - Somatization** | | | | | | | |
| How much are you distressed by faintness or dizziness? | 0.34 ± 0.02 (0.45) | 0.13 | 0.17 | 0.10 | 0.03 | 0.16 | 0.06 |
| How much are you distressed by pains in the heart or chest? | 0.34 ± 0.02 (0.50) | 0.12 | 0.11 | 0.09 | 0.00 | 0.06 | 0.05 |
| How much are you distressed by nausea or upset stomach? | 0.37 ± 0.02 (0.50) | 0.09 | 0.13 | 0.12 | 0.00 | 0.06 | 0.06 |
| How much are you distressed by trouble getting your breath? | 0.32 ± 0.02 (0.43) | 0.12 | 0.17 | 0.11 | 0.05 | 0.11 | 0.05 |
| How much are you distressed by hot or cold spells? | 0.31 ± 0.01 (0.40) | 0.17 | 0.17 | 0.15 | 0.10 | 0.12 | 0.03 |
| How much are you distressed by numbness or tingling in parts of your body? | 0.33 ± 0.02 (0.45) | 0.14 | 0.15 | 0.12 | 0.02 | 0.10 | 0.07 |
| How much are you distressed by feeling weak in parts of your body? | 0.41 ± 0.02 (0.53) | 0.12 | 0.12 | 0.03 | -0.01 | 0.10 | 0.02 |
| **Wellbeing - Emotional Wellbeing** | | | | | | | |
| To what extent do you feel happy? | 0.60 ± 0.01 (0.39) | 0.15 | 0.12 | 0.15 | 0.13 | 0.22 | 0.05 |
| To what extent do you feel interested in life? | 0.66 ± 0.01 (0.40) | 0.09 | 0.14 | 0.10 | 0.08 | 0.17 | 0.03 |
| To what extent do you feel satisfied with life? | 0.55 ± 0.02 (0.46) | 0.12 | 0.09 | 0.15 | 0.14 | 0.16 | 0.01 |
| **Wellbeing - Psychological Wellbeing** | | | | | | | |
| To what extent do you feel that you like most parts of your personality? | 0.62 ± 0.01 (0.38) | 0.05 | 0.07 | 0.09 | 0.05 | 0.16 | 0.05 |
| To what extent do you feel good at managing the responsibilities of your daily life? | 0.61 ± 0.01 (0.41) | 0.13 | 0.13 | 0.17 | 0.06 | 0.15 | 0.02 |
| To what extent do you feel that you have warm and trusting relationships with others? | 0.60 ± 0.01 (0.41) | 0.06 | 0.09 | 0.13 | 0.08 | 0.13 | -0.01 |
| To what extent do you feel that you have experiences that challenge you to grow and become a better person? | 0.62 ± 0.01 (0.35) | 0.06 | 0.04 | 0.17 | 0.04 | 0.12 | 0.02 |

| | | MAILA's correlation to ground truth (R) | | | | | |
|---|---|---|---|---|---|---|---|
| **Question** | **Mean ± 95% CI (IQR)** | **baseline** | **follow-up** | **survey** | **game** | **touch-q** | **touch-d** |
| To what extent do you feel confident to think or express your own ideas and opinions? | 0.64 ± 0.01 (0.38) | 0.08 | 0.12 | 0.12 | 0.04 | 0.13 | 0.00 |
| To what extent do you feel that your life has a sense of direction or meaning to it? | 0.56 ± 0.02 (0.51) | 0.08 | 0.12 | 0.09 | 0.03 | 0.16 | 0.05 |
| **Wellbeing - Social Wellbeing** | | | | | | | |
| To what extent do you feel that you have something important to contribute to society? | 0.59 ± 0.02 (0.47) | 0.11 | 0.13 | 0.06 | 0.09 | 0.15 | 0.04 |
| To what extent do you feel that you belong to a community (like a social group, or your neighborhood)? | 0.51 ± 0.02 (0.52) | 0.09 | 0.09 | 0.01 | 0.10 | 0.07 | 0.04 |
| To what extent do you feel that our society is a good place, or is becoming a better place, for all people? | 0.41 ± 0.02 (0.41) | 0.15 | 0.06 | 0.08 | 0.10 | 0.09 | 0.07 |
| To what extent do you feel that people are basically good? | 0.52 ± 0.01 (0.34) | 0.01 | 0.03 | 0.04 | -0.02 | 0.03 | 0.00 |
| To what extent do you feel that the way our society works makes sense to you? | 0.47 ± 0.01 (0.43) | 0.06 | 0.05 | 0.07 | 0.07 | 0.05 | 0.07 |

***Table S1. Predicting mental health from human–computer interactions.*** *For each questionnaire item, the average score is reported as mean ± 95% confidence interval (inter-quartile range). To the right, Spearman correlations (R) indicate the correspondence between predicted and true scores in the calibration cursor dataset (baseline, 5-fold cross-validation), follow-up data, a non-mental-health survey, and a gamified decision-making task. Follow-up models were trained on baseline recordings and applied without retraining. Survey and game models were calibrated on baseline and follow-up mental-health-task recordings and applied without further training to held-out alternate-task recordings from the same participants. The final two columns show correlations from two touch-based tasks, interface interaction (touch-q) and free-form drawing (touch-d), each evaluated using 5-fold cross-validation.*

## Table S2

| Question | R to ground truth |
|---|---|
| | **Body Awareness** |
| To what extent do you feel that strong lights or sounds affect your ability to focus? | -0.01 |
| To what extent do you feel that you can detect changes in your vision or hearing in different environments? | 0.05 |
| To what extent do you feel that you can distinguish between different textures or temperatures by touch? | 0.16 |
| To what extent do you feel that you can sense your body’s position in space during movement? | 0.00 |
| To what extent do you feel that you notice subtle bodily sensations (e.g., heartbeat, muscle tension)? | -0.02 |
| | **Civic** |
| To what extent do you feel that voting is important to you? | -0.01 |
| To what extent do you feel that your voice matters in society? | 0.04 |
| | **Cognition** |
| To what extent do you feel that you can detect subtle differences in colors? | 0.16 |
| To what extent do you feel that you can easily identify objects in cluttered scenes? | -0.09 |
| To what extent do you feel that you can recall specific events from two days ago? | -0.04 |
| To what extent do you feel that you easily recall names of people you meet? | 0.05 |
| To what extent do you feel that you remember people's faces after meeting them once? | -0.11 |
| | **Decision Making** |
| To what extent do you feel that uncertainty affects your decision-making? | -0.05 |
| To what extent do you feel that you can consider long-term outcomes when making choices? | 0.05 |
| To what extent do you feel that you enjoy solving complex logical problems? | 0.10 |
| To what extent do you feel that you prefer making decisions quickly rather than deliberating? | 0.01 |
| To what extent do you feel that you rely on intuition when making difficult decisions? | -0.03 |
| | **Economy** |
| To what extent do you feel that groceries are more expensive than last year? | -0.13 |
| To what extent do you feel that people work harder now than 10 years ago for the same housing? | -0.16 |
| To what extent do you feel that the economy has improved in the past year? | 0.04 |
| To what extent do you feel that you are paid fairly for your work? | 0.02 |
| To what extent do you feel that your income keeps up with the cost of living? | 0.03 |
| | **Head Impact** |
| To what extent do you feel that you can recall episodes of losing consciousness during or after sports? | 0.02 |
| To what extent do you feel that you've experienced head impacts during physical activities? | -0.05 |
| To what extent do you feel that you’ve noticed changes in your memory after repeated sports-related impacts? | 0.01 |
| To what extent do you feel that your past participation in contact sports has affected your physical coordination? | 0.11 |

| Question | R to ground truth |
| --- | --- |
| To what extent do you feel that your sports training emphasized head safety? | 0.05 |
| | **Motor Control** |
| To what extent do you feel that you can adjust your body movements in response to unexpected changes in your environment? | 0.04 |
| To what extent do you feel that you can coordinate both hands effectively for tasks like tying shoelaces or typing? | 0.17 |
| To what extent do you feel that you can keep your body still when needed (e.g., holding a posture or standing motionless)? | 0.19 |
| To what extent do you feel that you can maintain balance when walking on uneven surfaces? | 0.16 |
| To what extent do you feel that your movements are precise when doing tasks that require fine motor skills (e.g., writing, threading a needle)? | 0.13 |
| | **Politics** |
| To what extent do you feel that immigration strengthens the country? | -0.07 |
| To what extent do you feel that political news influences your daily decisions? | -0.06 |
| To what extent do you feel that public healthcare is important? | -0.01 |
| To what extent do you feel that the government should solve more social problems? | -0.03 |
| | **Society** |
| To what extent do you feel that climate change affects your everyday life? | 0.01 |
| To what extent do you feel that news media are trustworthy? | -0.09 |
| To what extent do you feel that people are treated equally regardless of race? | -0.15 |
| To what extent do you feel that public transport meets your daily needs? | 0.00 |
| To what extent do you feel that your cultural background shapes your identity? | 0.05 |
| To what extent do you feel that your education prepared you well for life? | 0.10 |
| | **Technology** |
| To what extent do you feel that fake news is easy to recognize? | 0.01 |
| To what extent do you feel that technology improves your quality of life? | 0.10 |
| To what extent do you feel that your online activity is private and secure? | -0.11 |
| | **Values** |
| To what extent do you feel that learning about other cultures enriches your perspective? | 0.06 |
| To what extent do you feel that religion plays a role in your life? | -0.05 |

***Table S2. Predicting responses to non-mental health items from human–computer interactions.*** *For each question from the non-psychological survey and gamified task, the table reports the Spearman correlation coefficient (R) between predicted and true item scores obtained within each dataset using 5-fold cross-validation. Dimensions correspond to thematic categories of items. Each group header indicates the corresponding dimension, and individual rows list the specific items within that domain. MAILA failed to predict the responses to the non-mental health survey ($R = 0.01 \pm 0.03$, $p = 0.28$), suggesting that cursor movements capture dynamic mental states associated with psychological distress and wellbeing, but not more stable self-assessments of abilities, attitudes, or beliefs, or response artifacts induced by our interface.*

## Table S3

| Prompt: Draw ... |
|---|
| **Digits** |
| the digits "8047" |
| the digits "9846" |
| the digits "1237" |
| the digits "5912" |
| the digits "0356" |
|  |
| **Objects** |
| a human face with glasses |
| a bow shooting an arrow |
| a spaceship |
| lightning coming out of a cloud |
| a key |
| a tent and a campfire |
| a traffic light |
| a fish in a fishbowl |
| a house with a chimney |
| a flower in a pot |
| a cat |
| a coffee mug |
| a mountain range with a sun |
| a hand with a wristwatch |
| a person riding a bike |

***Table S3. Predicting mental health from free-form digital behavior.*** *Participants were instructed to draw each prompt using a touchscreen interface. For readability, prompts are grouped by their underlying semantic category (e.g., objects, digits). Each prompt was shown once in random order across participants.*

## Table S4

| Agent and purpose | Information available | Core instruction | Output |
|---|---|---|---|
| User simulator (Claude-Sonnet-3.7): generate one everyday message | Shown: ground-truth item responses and 13-dimensional profile; construct definitions; one sampled scenario. Not shown: MAILA predictions, context condition, or profile-inference output. | “Write one natural first-person message that starts a conversation with a general AI assistant.”<br>“Let the scenario shape what you want help with and the private profile shape what you feel, notice, avoid, value, and struggle with.”<br>“Do not discuss mental health directly or reveal scores, questionnaire content, diagnoses, or the simulation.” | One short synthetic user message (typically 2–5 sentences). |
| Profile-inference agent (GPT-5.2): estimate the mental-health profile | Shown in all conditions: the synthetic message; dimension names, definitions, and item context. Condition-specific input: no MAILA context; MAILA predictions from the same participant; or MAILA predictions from another participant. Not shown: ground-truth profile or the user simulator’s private plan. | “Predict the participant’s 13 summary dimensions as population percentile ranks from 0 to 100.”<br>“Use only the transcript and auxiliary hint, when included.”<br>“If evidence is weak or ambiguous, choose around 50 rather than extremes.” | Thirteen numeric percentile-rank estimates, one per dimension, with no additional prose. |

***Table S4. Core prompts for the controlled synthetic LLM proof of concept.*** *The user simulator generated one everyday message from the private ground-truth profile and a sampled scenario. The profile-inference agent then estimated the profile from that message, with no MAILA context, matched MAILA context (out-of-fold cursor predictions for the same participant), or random MAILA context (out-of-fold cursor predictions from another participant). Messages and scenarios were drawn independently in each condition. Normalized error $e$ is the participant-level root mean squared difference between inferred and ground-truth percentile ranks across the 13 dimensions and is summarized as the mean and 95% CI. Quoted text reproduces the operative instructions in condensed form; implementation-only tool schemas and private planning fields are omitted for readability.*

## Table S5

| 1–24 | 25–48 | 49–72 | 73–96 | 97–120 |
|---|---|---|---|---|
| 1. A note to a child's teacher | 25. Drafting an info sheet for a small group | 49. Note to a school office | 73. Preparing for a home repair visit | 97. Starting a side project |
| 2. Adjusting a routine for the season | 26. Drafting feedback on a colleague's work | 50. Note to someone who helped them | 74. Preparing for a parent-teacher conference | 98. Starting to learn a language |
| 3. Asking for feedback on something they made | 27. Drafting feedback on a product | 51. Organizing a closet | 75. Preparing for a performance review | 99. Thinking about adopting a pet |
| 4. Balancing work and hobbies | 28. Farewell note to a departing colleague | 52. Organizing a recipe collection | 76. Preparing for a routine checkup | 100. Thinking through a commute change |
| 5. Budget conversation with a partner | 29. Fitting exercise into the week | 53. Organizing a small birthday gathering | 77. Preparing for an anniversary | 101. Thinking through a job offer |
| 6. Building a reading list | 30. Handling an upcoming social mixer | 54. Organizing a stack of paperwork | 78. Preparing for an upcoming wedding attendance | 102. Thinking through a request from a family member |
| 7. Building a study schedule | 31. Handling back-to-back commitments | 55. Organizing digital photos | 79. Preparing for jury duty | 103. Using an unexpected sum |
| 8. Celebrating a personal milestone | 32. Hosting a game night | 56. Planning a call to a distant relative | 80. Preparing questions for a routine appointment | 104. Using an unused space at home |
| 9. Changing a long-standing daily habit | 33. Hosting an out-of-town guest | 57. Planning a child's birthday | 81. Prioritizing social plans | 105. Using holiday leave |
| 10. Choosing a gift for someone | 34. Introducing oneself in a new group | 58. Planning a date night | 82. Questions for a financial decision | 106. Visit to an elderly relative |
| 11. Choosing a new book or show | 35. Letter to a local representative | 59. Planning a dinner party menu | 83. Questions for a mechanic visit | 107. Weekly grocery planning |
| 12. Choosing a new hobby to try | 36. Life update to distant friends | 60. Planning a joint project with a friend | 84. Redecorating a room | 108. Weighing two career paths |
| 13. Choosing a self-taught course | 37. Making a packing list | 61. Planning a move to a new place | 85. Reflecting on a tense meeting | 109. What to do with a free Saturday |
| 14. Choosing a vacation destination | 38. Message about rescheduling | 62. Planning a quiet evening alone | 86. Reflecting on the past year | 110. Writing a birthday message |
| 15. Choosing between two apartments | 39. Message asking for a favor | 63. Planning a small garden or houseplants | 87. Reply to a family group text | 111. Writing a condolence note |

| 1–24 | 25–48 | 49–72 | 73–96 | 97–120 |
|---|---|---|---|---|
| 16. Choosing between two educational programs | 40. Message for a work Slack channel | 64. Planning a small weekend trip | 88. Responding to a mildly ambiguous text | 112. Writing a dating profile bio |
| 17. Choosing between two event invitations | 41. Message to a former mentor | 65. Planning a video call with distant relatives | 89. Responding to unexpected news | 113. Writing a postcard |
| 18. Composing a resignation letter | 42. Message to a group of neighbors | 66. Planning a weekend of errands | 90. Response to family group chat friction | 114. Writing a reference for a friend |
| 19. Deciding on a volunteer commitment | 43. Message to a landlord | 67. Planning an outdoor day | 91. Setting up a cleaning schedule | 115. Writing a review for a local business |
| 20. Declining additional work | 44. Message to decline an invitation | 68. Planning meals for the week | 92. Setting up a recurring get-together | 116. Writing a short piece for a newsletter |
| 21. Designing a morning routine | 45. Message to reconnect with an old friend | 69. Preparing for a community meeting | 93. Setting up a shared calendar | 117. Writing a short professional bio |
| 22. Designing an evening wind-down | 46. Note to a book club or hobby group | 70. Preparing for a family visit | 94. Setting up next month's budget | 118. Writing a short speech for a gathering |
| 23. Downsizing possessions | 47. Note to a coworker about current workload | 71. Preparing for a hard conversation with a relative | 95. Shaping an open weekend | 119. Writing a short toast |
| 24. Drafting a complaint to a service provider | 48. Note to a friend in a life change | 72. Preparing for a holiday with relatives | 96. Short message to a neighbor | 120. Writing a thank-you note |

***Table S5. Scenario bank for the controlled synthetic LLM proof of concept.*** *The 120 broad, non-diagnostic assistant-use scenarios sampled during message generation are listed alphabetically across five columns.*

## Table S6

| Feature | Operational definition | Implementation notes |
|---|---|---|
| Average speed | Mean cursor/touch speed across samples. | Speed computed from successive differences in $(x_t, y_t)$ with $\Delta t = 1$; $s_t = \sqrt{v_x^2 + v_y^2}$. |
| Speed kurtosis | Kurtosis of the segment speed distribution. | Computed on $\{s_t\}$. |
| Jerk (speed change magnitude) | Mean absolute change in speed over time. | Discrete acceleration $a_t = \Delta s_t / \Delta t$; jerk is $\mathrm{mean}(\lvert a_t \rvert)$ (with $\Delta t = 1$). |
| Movement area | Bounding-box area covered by the trajectory. | $(\max x - \min x) \times (\max y - \min y)$. |
| Relative idle time | Fraction of samples with speed below a stationary threshold. | $\mathrm{mean}(s_t < \tau)$ with $\tau = 0.001$ (screen-normalized units/sample). |
| Path efficiency | Straight-line displacement divided by total path length. | Straight-line distance between first and last point divided by cumulative path length (0 if path length = 0). |
| Average turn angle | Mean absolute change in heading direction. | Heading $\theta_t = \mathrm{atan2}(v_y, v_x)$; angle changes $\Delta\theta_t$ wrapped to $(-\pi, \pi]$; report $\mathrm{mean}(\lvert \Delta\theta_t \rvert)$. |
| Tortuosity | Total path length divided by straight-line displacement. | Path length / straight-line distance (defined only if straight-line distance $> 0$; else 0). |
| Turn rate per distance | Mean absolute turning per unit distance traveled. | Average turn angle divided by total path length (0 if path length = 0). |
| Horizontal–vertical bias | Difference between horizontal and vertical motion magnitude. | $\mathbb{E}(\lvert v_x \rvert) - \mathbb{E}(\lvert v_y \rvert)$; positive values indicate more horizontal motion. |
| Speed entropy (spectral entropy) | Entropy of the normalized power spectrum of the speed trace. | FFT-based: compute power spectrum of $\{s_t\}$, normalize to probabilities, then compute entropy. |
| Speed fluctuation rate | Rate of sign changes in speed around its mean. | Count zero-crossings of $s_t - \bar{s}$ (sign changes relative to mean speed). |

***Table S6. Definitions of human-interpretable cursor/touch features.*** *The table defines the 12 handcrafted movement features used to interpret MAILA predictions. Features were computed for each trajectory segment and then averaged across segments within participant to obtain one participant-level feature vector; features were z-scored across participants prior to regression analyses.*

## Table S7

| | Status | |
|---|---|---|
| **Recommendation** | **Current efforts** | **Future priorities** |
| **Explainability** | | |
| Define the need and requirements for explainability with end users | Assessed mechanism of prediction, provided explanations of performance in terms of human-centered movement features | Develop interactive visualizations or summaries for non-experts |
| Evaluate explainability with end users (e.g., correctness, impact on users) | N/A | Conduct usability studies to assess how well explanations improve understanding |
| **Fairness** | | |
| Collect information on individuals' and data attributes | Collected demographics, self-reported mental health data, and hardware at the participant level | Expand data collection to include additional background information, e.g. electronic health records, additional dimensions of mental health, biomarkers (genetics, wearables, imaging) |
| Define any potential sources of bias from an early stage | Evaluated performance across demographic features available for participants recruited via an online experimental platform | Conduct targeted bias analyses for underrepresented (including clinical) populations; validate MAILA outside of online cohorts |
| Evaluate potential biases and, when needed, bias correction measures | Evaluated model stability across demographic groups, context, time, and input modality | Implement algorithmic fairness measures (e.g., re-weighting techniques) to actively mitigate bias |
| **General** | | |
| Define adequate evaluation plan (e.g., datasets, metrics, reference methods) | Defined evaluation protocol for cursor and touchscreen activity for regression and classification | Incorporate additional fairness, robustness, and real-world performance metrics |
| Engage interdisciplinary stakeholders throughout the AI lifecycle | N/A | Expand involvement to include ethicists, data privacy experts, and policymakers |
| Identify and comply with applicable AI regulatory requirements | N/A | Anticipate compliance plans aligned with AI standards such as GDPR, HIPAA, or ISO standards |
| Implement measures for data privacy and security | Emphasized anonymization, explored strategies for preventing unintended use (client-side scrambling) | Build a browser plugin for client-side scrambling |
| Implement measures to address identified AI risks | Discussed risk mitigation | Develop targeted strategies for mitigating potential misuse of human-computer interactions, starting with scrambling tools |
| Investigate and address application specific ethical issues | Acknowledged ethical concerns such as consent and transparency | Develop detailed guidelines for ethical data use and informed consent practices |

| Recommendation | Status<br>Current efforts | Future priorities |
|---|---|---|
| Investigate and address social and societal issues | Acknowledged ethical risks and societal implications | Conduct focus groups or interviews with key social groups to anticipate unintended consequences |
| **Robustness** | | |
| Define sources of data variation from an early stage | Conducted stress testing for noise, incomplete data, reduced training set size, and impoverished movement clusters | Validate in cohorts with known movement variation (e.g., movement disorders); Assess atypical movement patterns in people with movement disorders |
| Evaluate and optimize robustness against real world variations | N/A | Expand data collection to fully unconstrained computer use annotated with mental health labels |
| Train with representative real-world data | Collected data across various contexts intended to simulate everyday computer use, indirect validation on unlabeled naturalistic computer use | Expand data collection to fully unconstrained computer use annotated with mental health labels |
| **Traceability** | | |
| Define mechanisms for quality control of the AI inputs and outputs | Evaluated model performance using multiple metrics and targets (regression on inter-individual differences and classification of groups) | Implement ongoing quality control processes during deployment |
| Establish mechanisms for AI governance | N/A | Establish an advisory board to oversee ethical concerns and data management |
| Implement a logging system for usage recording | N/A | Develop secure logging protocols to track system performance and failures |
| Implement a risk management process throughout the AI lifecycle | Addressed ethical considerations regarding privacy and security, outlined scrambling as a way to mitigate unwanted digital profiling | Formalize a risk management framework, identifying potential failure points, build scrambler browser plugin & safety hardware |
| Implement a system for periodic auditing and updating | N/A | Develop procedures for continuous model updates based on evolving data |
| Provide documentation (e.g., technical, clinical) | Developed detailed methodology documentation for feature extraction, data analysis, and model development | Develop user-facing documentation for non-technical stakeholders, GitHub repo at the time of publication |
| **Universality** | | |
| Define intended clinical settings and cross setting variations | Tested generalization from models trained on the general population to populations with self-identified mental illness (depression & OCD) | Test cohorts with clinical diagnoses; define specific contexts for deployment (e.g., telehealth, digital wellbeing platforms) |

| Recommendation | Status<br>Current efforts | Future priorities |
|---|---|---|
| Evaluate and demonstrate local clinical validity | N/A | Conduct clinical trials in real-life healthcare settings |
| Evaluate using external datasets and/or multiple sites | Evaluated performance in several generalization datasets, applied trained models to external datasets | Evaluate model performance across multiple sites and diverse real-world conditions, e.g. gaming, naturalistic browsing, office work, coding, entertainment applications in large cohorts |
| Use community defined standards (e.g., clinical definitions, technical standards) | Used a novel questionnaire tool with favorable psychometric properties | Integrate structured interviews (e.g., SCID), expand to other self-report questionnaires, expand to predefined cohorts |
| **Usability** | | |
| Define intended use and user requirements from an early stage | Defined human-computer interactions as a scalable signal for mental health prediction | Develop specific deployment strategies for use in clinical or public health contexts |
| Establish mechanisms for human-AI interactions and oversight | N/A | Design user feedback mechanisms to improve model trustworthiness |
| Evaluate clinical utility and safety (e.g., effectiveness, harm, cost-benefit) | N/A | Conduct clinical safety and efficacy evaluations before deployment in clinical settings |
| Evaluate user experience and acceptance with independent end users | N/A | Conduct studies evaluating usability, interpretability, and trust |
| Provide training materials and activities (e.g., tutorials, hands-on sessions) | N/A | Develop educational content for clinicians, researchers, and end users |

***Table S7. Recommendations for responsible and transparent use of AI in mental health research, aligned with the FUTURE-AI framework.*** *Each row summarizes a key recommendation grouped by overarching category. The table outlines how the FUTURE-AI principles are currently addressed and highlights proposed next steps for advancing best practices.*